# Gravitational Waves from Cosmological Phase Transitions

by

**Moritz Breitbach**
born in Mainz, Germany

A thesis submitted in partial fulfillment of the requirements for the degree

*Master of Science in Physics*

to the Faculty of Physics, Mathematics and Computer Science
of Johannes Gutenberg University Mainz

April 24, 2018

1st advisor: Prof. Dr. Joachim Kopp
2nd advisor: Prof. Dr. Pedro Schwaller

I hereby declare that I have written the present thesis independently, without assistance from external parties and without use of other resources than those indicated. The ideas taken directly or indirectly from external sources are acknowledged in the text.

Mainz, April 24, 2018

Moritz Breitbach
PRISMA Cluster of Excellence
Mainz Institute for Theoretical Physics (MITP)
Johannes Gutenberg University Mainz
breitbach@uni-mainz.de

# I Introduction

In 1916, Albert Einstein predicted the existence of gravitational radiation as a consequence of his groundbreaking theory of general relativity [1, 2]. This radiation manifests itself as tiny ripples in the fabric of spacetime that propagate with the speed of light. Due to the weakness of gravity, it took more than a century of scientific progress to finally observe this kind of radiation: In 2015, gravitational waves originating from a binary black hole merger were directly detected by the LIGO observatory [3], a revolutionary discovery which was awarded the 2017 Nobel Prize of Physics. The detection capabilities of planned pulsar timing arrays and space-based observatories herald a new era of astronomical and cosmological research, complementary to current and future collider experiments. Massive binaries, inspirals, supernovae and spinning neutron stars will be astrophysical objects of interest. Additional to that, cosmological phenomena such as cosmic strings, the inflation and phase transitions are believed to generate a stochastic background. The observation or non-observation of such a background gives a direct probe of the very early universe, looking back in time much further than the CMB and involving energies beyond the reach of any earth-bound detector. It is hoped that these new prospects will shed light on some of the unresolved mysteries of modern cosmology and particle physics.

The present thesis focuses on cosmological phase transitions driven by the temperature dependence of the thermodynamical free energy density in the expanding and cooling universe. If such a transition occurs abruptly, i.e. if it is first-order, bubbles of the new phase nucleate, expand and finally collide. These collisions cause anisotropies acting as sources for gravitational radiation. After their production, the gravitational waves propagate through space undisturbed until they might be detected in the form of a redshifted stochastic background today. A prominent example of a cosmological phase transition is the breaking of electroweak symmetry by the Higgs mechanism, which is however a smooth crossover in the Standard Model. The transition can be rendered first-order by simply postulating a minimal set of additional particles. A variety of models beyond the Standard Model that try to explain baryogenesis or dark matter might thus be probed by a search for stochastic gravitational wave backgrounds. This thesis presents different models that feature dark matter candidates together with first-order phase transitions. Vev flip-flop models provide an appealing dark matter mechanism and feature a two-step phase transition at the electroweak scale, probable by future space-based interferometers. Furthermore, the dark photon model will be presented, which features a phase transition at sub-MeV scales and can be probed by pulsar timing arrays.

This document is structured as follows: Chapter II gives an overview over the theory and techniques that are relevant for the subsequent analysis. This comprises the basics of cosmology and the explicit calculation of the one-loop effective potential for a toy model, together with the phenomenon of false vacuum decay. Chapter III reviews the current status regarding gravitational waves in the context of cosmological phase transitions and contains new results in the form of several sensitivity plots. Chapters II and III are rather detailed, because they aim at providing a complete toolbox for a state-of-the-art analysis on gravitational waves for any given model at the one-loop level. In the course of this thesis, the machinery has been applied to different models with results being presented in Chapter IV.

## I.1 Notation

In this work, the usual conventions of the field are adopted, e.g. Einstein's summation convention and the Dirac slash notation to indicate Lorentz contraction with gamma matrices, i.e. $\not{k} \equiv k^\mu \gamma_\mu = g_{\mu\nu} k^\mu \gamma^\nu$. The Minkowski signature $(+,-,-,-)$, common in the field of particle physics, is used. Spatial vectors are printed bold $\boldsymbol{x}$ and their indices are Latin letters $\{i, j, \dots\}$, in contrast to spacetime vectors which are furnished with Greek letters $\{\mu, \nu, \dots\}$. Unless indicated otherwise, natural units $c = \hbar = k_\mathrm{B} = 1$ are employed such that spacial and temporal distances come with inverse energy units, whereas temperature, mass and momentum are given in energy units of $\mathrm{eV} \approx 1.6 \times 10^{-19}\,\mathrm{J}$.

# II Theory and Background

## II.1 Cosmology

The following overview is inspired by Kolb and Turner's standard reference [4] and Daniel Baumann's instructive lecture notes [5].

### II.1.1 The Geometry of Spacetime

On scales much larger than galaxy clusters, the structure of the universe is to a good approximation independent of position and direction, i.e. *homogeneous* and *isotropic*. Consequently, the universe can be described through a simple choice of geometry. 4-dimensional spacetime can be foliated into time-ordered slices of 3-dimensional space, each of which is maximally symmetric implying constant density and curvature throughout space. While the coordinates of a spacetime 4-vector $\mathrm{d}x^\mu = (\mathrm{d}t, \mathrm{d}\boldsymbol{x})$ depend on the choice of a coordinate system, the *invariant distance*

$$\mathrm{d}s^2 \equiv g_{\mu\nu}\mathrm{d}x^\mu \mathrm{d}x^\nu \equiv \mathrm{d}t^2 - \mathrm{d}l^2 \tag{II.1}$$

is identical for all non-accelerating observers. The *metric of spacetime* $g_{\mu\nu}$ is in the simplest case of flat Minkowski space given by

$$g_{\mu\nu} = \eta_{\mu\nu} \equiv \mathrm{diag}(1, -1, -1, -1). \tag{II.2}$$

According to general relativity, energy (and thus mass) curves spacetime, such that $g_{\mu\nu} = g_{\mu\nu}(t, \boldsymbol{x})$ in general. For a slice of maximally symmetric 3-space, we have to assign a constant curvature which can be either zero, positive or negative, as illustrated in Fig. II.1. When applying Pythagoras' theorem in positively (negatively) curved space, one notices that the hypotenuse of a triangle is shortened (lengthened) in comparison to the one in flat space, i.e.

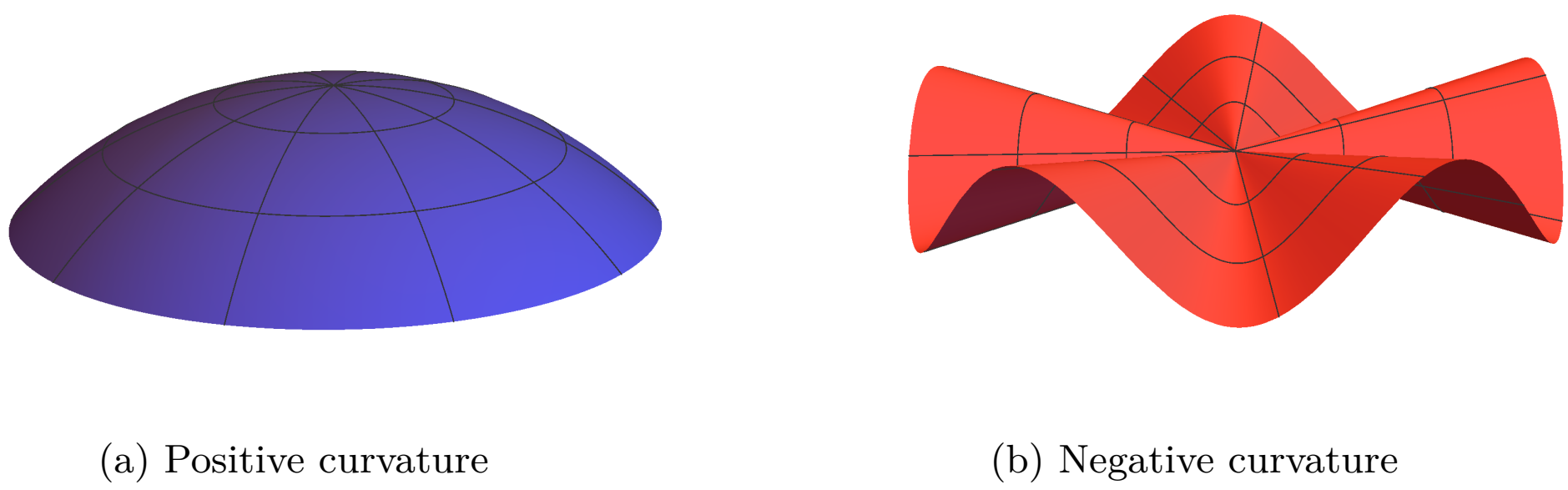

(a) Positive curvature

(b) Negative curvature

Figure II.1: Examples of 2-dimensional curved manifolds, embedded in 3 dimensions.

$$\mathrm{d}\boldsymbol{x}^2 = \mathrm{d}l^2 \mp \mathrm{d}u^2. \tag{II.3}$$

A possible embedding of a curved manifold in 4-dimensional space is realized by a 3-sphere (3-hyperboloid) described by

$$\boldsymbol{x}^2 \pm u^2 = \pm a^2 \tag{II.4}$$

where $u$ is the new direction the 3-manifold is curved into. $a$ defines the scale of the sphere (hyperboloid), so that the strength of curvature is $\sim 1/a$. After making the coordinates dimensionless by substituting $(\boldsymbol{x}, u) \to a(\boldsymbol{x}, u)$ and using (II.3) and (II.4), one can write

$$\mathrm{d}l^2 = a^2 \gamma_{ij} \mathrm{d}x^i \mathrm{d}x^j \tag{II.5}$$

with

$$\gamma_{ij} = \delta_{ij} + k \frac{x_i x_j}{1 - k\,|\boldsymbol{x}|^2}, \qquad k = \begin{cases} 0 & \text{flat} \\ +1 & \text{positive curvature} \\ -1 & \text{negative curvature} \end{cases}. \tag{II.6}$$

Returning to 4D-spacetime, the *scale factor* $a(t)$ now depends on time. This leads to the so-called *Friedmann-Robertson-Walker metric* (FRW) given by

$$\mathrm{d}s^2 = \mathrm{d}t^2 - a^2(t) \gamma_{ij} \mathrm{d}x^i \mathrm{d}x^j. \tag{II.7}$$

The now time-dependent scale $a(t)$ relates physical to dimensionless (*comoving*) coordinates by $\boldsymbol{x}_{\mathrm{phys}} = a(t)\boldsymbol{x}$. This gives rise to a modified physical velocity

$$\dot{\boldsymbol{x}}_{\mathrm{phys}} = a(t)\dot{\boldsymbol{x}} + H \boldsymbol{x}_{\mathrm{phys}} \tag{II.8}$$

with the time-dependent *Hubble parameter*

$$H \equiv \frac{\dot{a}}{a}. \tag{II.9}$$

Positive $H$ adds a velocity proportional to the distance of an object and therefore describes an expanding universe. A century ago, Edwin Hubble observed spectral lines of distant galaxies to appear redshifted [6]. This was the first evidence that we live in an expanding universe.

### II.1.2 Dynamics

*General relativity* describes the dynamics of the universe at large scales. It states that spacetime is not just the stage for gravitational interactions, but actually its mediator. One of the consequences of general relativity is the curvature of spacetime by energy. This insight is captured mathematically in *Einstein's field equations*

$$G_{\mu\nu} = 8\pi G\, T_{\mu\nu} \tag{II.10}$$

which relate local curvature to energy content. The former is encoded in the *Einstein tensor* $G_{\mu\nu}$ which is a function of metric $g_{\mu\nu}$ and its derivatives. The latter is represented by the *energy-momentum tensor* $T_{\mu\nu}$, describing energy, momentum and momentum flow at a given point in spacetime. The gravitational constant $G \approx 6.7 \times 10^{-11}\,\mathrm{m^3 kg^{-1} s^{-2}}$ is a fundamental constant of nature determining how much energy curves space and thereby the strength of gravity. In the FRW metric, matter can be described as a *perfect fluid* with

$$T_{\mu\nu} = (\rho + P)\dot{x}_\mu \dot{x}_\nu - P\, g_{\mu\nu} \tag{II.11}$$

where energy density $\rho$ and pressure $P$ are the only characterizing quantities. For a comoving observer, i.e. in the rest frame of the fluid, $\dot{x}_\mu = (1, 0, 0, 0)$ and

$$T_{\mu\nu} = \mathrm{diag}(\rho, P, P, P). \tag{II.12}$$

#### The continuity equation

In Minkowski space, $\rho$ and $P$ evolve with time following energy and momentum conservation

$$\partial_\mu T^{\mu\nu} = 0 \tag{II.13}$$

which can be separated into the *continuity equation* $\dot{\rho} = 0$ and the *Euler equation* $\partial_i P = 0$. In general relativity, going beyond Minkowski space, energy and momentum are not conserved in the conventional way: According to Neother's theorem, energy and momentum conversation is tied to translation invariance in time and space. A non-static universe with time- and space-dependent metric loses these symmetries. Differential geometry, the mathematical framework behind general relativity, introduces a *covariant derivative* $\nabla_\mu$ which generalizes the usual derivative for curved spacetime. The new operator is a linear combination of $\partial_\mu$ and additional terms which depend on the metric and its derivatives. It turns out that

$$\nabla_\mu T^{\mu\nu} = 0 \tag{II.14}$$

is a valid conservation law in curved spacetime, as it accounts for the energy and momentum that is mediated by gravity, i.e. by spacetime itself. The corresponding continuity equation in the FRW metric reads

$$\dot{\rho} + 3H(\rho + P) = 0. \tag{II.15}$$

With the above equation, one can examine how different types of fluids behave depending on scale factor $a(t)$. The possible forms of energy in the universe can be categorized in the following way:

**Matter** can be described as a non-relativistic gas, the energy density of which is dominated by the rest mass of its constituents. It has negligible pressure $P \ll \rho$, so the continuity equation yields

$$\rho \sim a^{-3}(t).$$

The energy density thus dilutes with increasing scale of a spacial volume.

**Radiation** or relativistic matter has negligible rest mass and a pressure $P = \frac{\rho}{3}$. The energy density therefore scales as

$$\rho \sim a^{-4}(t)$$

which accounts for dilution and redshift in an expanding universe.

**Dark energy** is a postulated energy of empty space and does per definition not dilute with increasing scale factor, thus

$$\rho = \text{const.}$$

This behavior is realized for negative pressure $P = -\rho$, which is clearly not a property of ordinary matter. Note that instead of adding a dark energy contribution $\sim g_{\mu\nu}$ to $T_{\mu\nu}$ one could move the new term to the other side of the Einstein field equations and call it *cosmological constant* $\Lambda$. This was originally part of Einstein's equations to explain a static universe [7], the common conjecture in the early 20th century. After the universe turned out to be expanding, $\Lambda$ was not needed any more. At the very end of the century, observations of distant supernovae [8] hinted at an accelerated expansion, explainable by a positive cosmological constant or equivalently dark energy.

### The Friedmann equations

The continuity equation gives information about the behavior of energy in expanding space. To infer the dynamics of the universe itself, we must employ the Einstein field equations (II.10). After calculating $G_{\mu\nu}$ for the FRW metric and inserting the stress-momentum tensor of a perfect fluid, one obtains the *Friedmann equations*

$$H = \frac{\dot{a}}{a} = \sqrt{\frac{8\pi G}{3}\rho - \frac{k}{a^2}}, \tag{II.16}$$

$$\frac{\ddot{a}}{a} = -\frac{4\pi G}{3}(\rho + 3P). \tag{II.17}$$

Assuming a flat universe, which is consistent with observations [9], and using today's Hubble parameter $H_0$, one can define today's *critical energy density* which satisfies the first Friedmann equation:

$$\rho_{\text{crit},0} \equiv \frac{3H_0^2}{8\pi G} \tag{II.18}$$

The relative composition of the universe can be expressed by dimensionless density parameters

$$\Omega_{i,0} \equiv \frac{\rho_{i,0}}{\rho_{\text{crit},0}} \tag{II.19}$$

where $i = \{\text{r}, \text{m}, \Lambda, \text{k}\}$ for radiation, matter, dark energy and curvature. The curvature part in (II.16) has been relabeled as if it would be an energy density, scaling as $\sim a^{-2}$. The Friedmann equation in terms of the different energy forms reads

$$H^2 \sim \frac{\Omega_\text{r}}{a^4} + \frac{\Omega_\text{m}}{a^3} + \frac{\Omega_\text{k}}{a^2} + \Omega_\Lambda \tag{II.20}$$

where both $H$ and $a$ are time-dependent.

### The $\Lambda$CDM model

The evolution of the universe is from a present day perspective best described by the *Lambda cold dark matter* ($\Lambda$CDM) model. It assumes the existence of dark energy, i.e. a positive $\Lambda$, and non-relativistic, often termed 'cold' dark matter. Based on numerous observations [10], the model suggests today's energy distribution to be

$$
\begin{aligned}
&\Omega_{\text{r},0} \approx 10^{-4} && \text{(radiation)}\\
&\Omega_{\text{b},0} \approx 0.05 && \text{(baryonic matter)}\\
&\Omega_{\text{d},0} \approx 0.27 && \text{(dark matter)}\\
&\Omega_{\Lambda,0} \approx 0.68 && \text{(dark energy)}\\
&\Omega_{\text{k},0} \lesssim 0.01 && \text{(curvature)}
\end{aligned}
$$

where baryonic and dark matter together make up $\Omega_{\text{m},0} \approx 0.32$. Using (II.20), one can extrapolate today's energy densities back to earlier times. The scale factor $a$ can be set to 1 for the present day, while values $< 1$ mark the past in our expanding universe. It becomes apparent that the universe was first radiation dominated, then went through an era of matter being most important and only very recently became dominated by dark energy (see Fig. II.2). The expansion started to accelerate in this last era, as the second Friedmann equation (II.17), yielding positive $\frac{\ddot{a}}{a}$ for $P < -\frac{1}{3}\rho$, tells us. Today's expansion rate $H_0$ could only be measured very roughly for many years. It was therefore parametrized as

$$H_0 = h \times 100\,\mathrm{km\,s^{-1}\,Mpc^{-1}} \tag{II.21}$$

with $h = 0.4 \sim 1$ and $1\,\mathrm{Mpc} \approx 3.1 \times 10^{22}\,\mathrm{m}$. One of the modern approaches is the observation of astronomical standard candles as done by the Planck Space Telescope [9] with the result $h \approx 0.68$. Alluding to the historical parametrization, dimensionless energy densities are often stated as $h^2\Omega$.

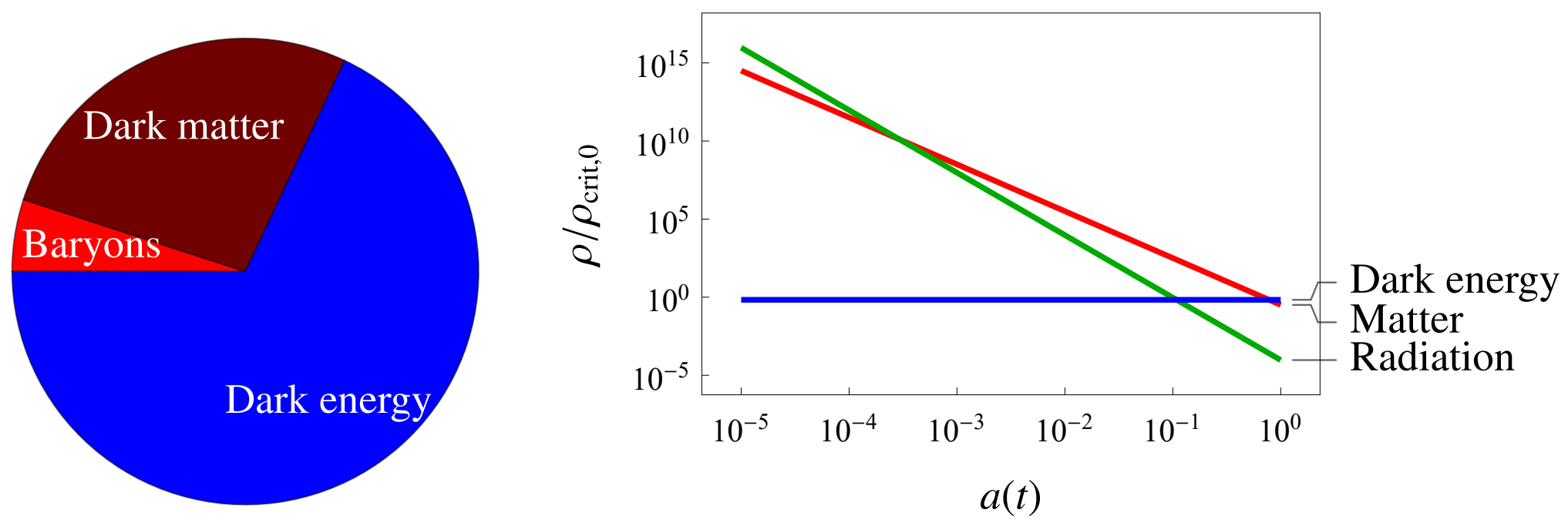

Figure II.2: Energy content of today's universe (left) and extrapolated back in time, i.e. for decreasing scale factor $a(t)$ (right).

### II.1.3 The History of our Universe

Extrapolating back towards earlier times, one eventually comes to a point of infinite temperature and density, the *singularity* at which space and time emerged. This event, referred to as the *Big Bang*, marks the emergence of our universe which is 13.8 billion years old [9].

#### Thermal equilibrium

After reheating, which will be explained further below, the universe was filled with a hot, dense particle plasma, which then started to cool down due to expansion of the universe. As long as all interacting constituents are in *thermal equilibrium*, one can assign a single temperature $T$ to the entire plasma. Since the decrease in temperature occurs monotonously, one can indicate a moment in time by stating the corresponding temperature. In thermal equilibrium, plasma particles maximize their entropy obeying the standard *phase space distribution functions*

$$f(p) = \frac{1}{e^{(E(p)-\mu)/T} \mp 1} \tag{II.22}$$

with $-$ for bosons and $+$ for fermions, energy $E(p) = \sqrt{m^2 + \boldsymbol{p}^2}$ and chemical potential $\mu$. We will neglect the latter, since it is not relevant at early times [4]. The *number density* $n$ and *energy density* $\rho$ of one particle *degree of freedom* (DOF) are found by integrating over phase space, i.e.

$$\begin{aligned} n &= \frac{1}{(2\pi)^3} \int \mathrm{d}^3 p \, f(p) \\ &= \frac{1}{2\pi^2} \int_0^\infty \mathrm{d}p \frac{p^2}{e^{\sqrt{m^2+p^2}/T} \mp 1} \end{aligned} \tag{II.23}$$

and

$$\begin{aligned} \rho &= \frac{1}{(2\pi)^3} \int \mathrm{d}^3 p \, f(p) E(p) \\ &= \frac{1}{2\pi^2} \int_0^\infty \mathrm{d}p \frac{p^2 \sqrt{m^2 + p^2}}{e^{\sqrt{m^2+p^2}/T} \mp 1}. \end{aligned} \tag{II.24}$$

These expressions evaluate to

$$n_{\mathrm{rel}} \approx \frac{6}{5\pi^2} T^3 \begin{cases} \times 1 & \text{(bosons)} \\ \times \frac{3}{4} & \text{(fermions)} \end{cases} \tag{II.25}$$

$$\rho_{\mathrm{rel}} = \frac{\pi^2}{30} T^4 \begin{cases} \times 1 & \text{(bosons)} \\ \times \frac{7}{8} & \text{(fermions)} \end{cases} \tag{II.26}$$

in the relativistic limit $T \gg m$ and yield

$$n_{\text{non-rel}} = \left( \frac{mT}{2\pi} \right)^{\frac{3}{2}} e^{-m/T} \tag{II.27}$$

$$\rho_{\text{non-rel}} = m \cdot n \tag{II.28}$$

for particles where the rest mass dominates, i.e. where $T \ll m$ and $E(p) \approx m$. In the latter case, a so-called *Boltzmann factor* $e^{-m/T}$ arises and suppresses contributions of heavy particles (see Fig. II.4). This suppression can be understood explicitly by looking at the underlying interaction processes. Consider for instance a process of annihilation and subsequent pair-production $a\bar{a} \rightleftarrows b\bar{b}$ which can proceed in both directions as long as $T > m_a, m_b$. As the temperature drops below one of the particle masses, e.g. $T < m_a$, the corresponding particles will due to lack of energy not be produced any longer. They will then be annihilated away via $a\bar{a} \to b\bar{b}$, which is reflected in the Boltzmann suppressing exponential.

The relativistic expressions for number and energy density are independent of particle masses. This allows us to simply add up all relativistic contributions

$$\rho_{\mathrm{rel}} = \sum_a \rho_{\mathrm{rel},a} = \frac{\pi^2}{30} g_{\mathrm{rel}} T^4$$

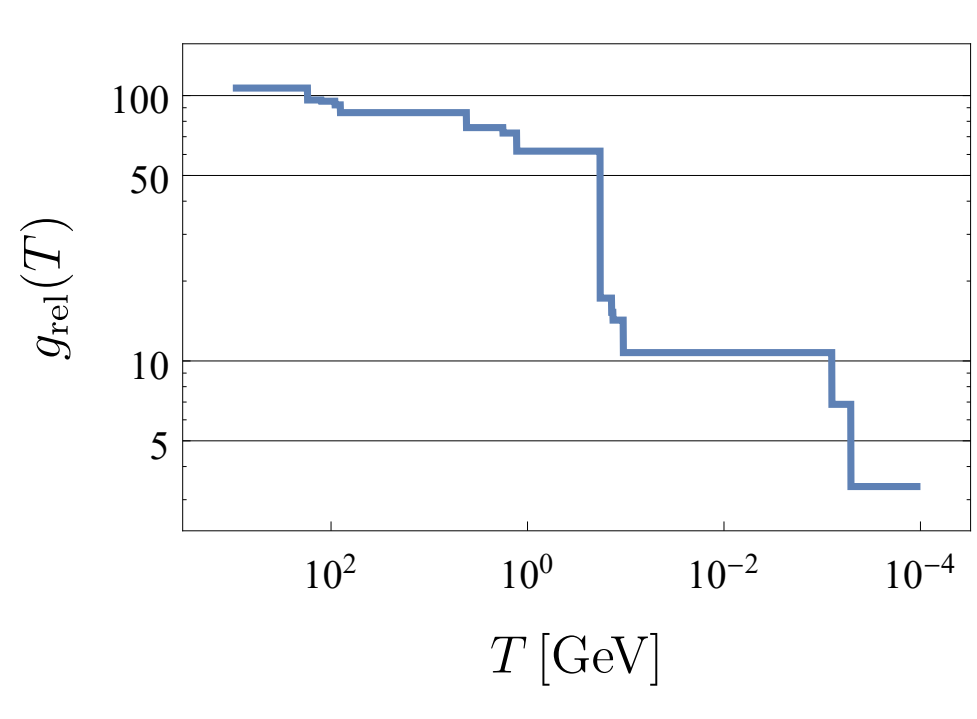

| Particle / Event | $T$ | $g_{\rm rel}$ |
|---|---|---|
| | | 106.75 |
| $t$ quark | 172 GeV | |
| | | 96.25 |
| Higgs boson | 125 GeV | |
| | | 95.25 |
| $Z$ boson | 91.2 GeV | |
| | | 92.25 |
| $W$ boson | 80.4 GeV | |
| | | 86.25 |
| $b$ quark | 4.18 GeV | |
| | | 75.75 |
| Tauon | 1.78 GeV | |
| | | 72.75 |
| $c$ quark | 1.29 GeV | |
| | | 61.75 |
| QCD transition | 150 MeV | |
| | | 17.25 |
| $\pi^\pm$ meson | 139 MeV | |
| | | 15.25 |
| $\pi^0$ meson | 135 MeV | |
| | | 14.25 |
| Muons | 106 MeV | |
| | | 10.75 |
| Electrons | 511 keV | |
| | | 3.363 |

Figure II.3: Temperature evolution of the relativistic DOF assuming *Standard Model* (SM) particle content. $g_{\rm rel}(T)$ is drawn here simplified as a step function, whereas in reality it would be smooth since the transition from relativistic to non-relativistic is not instantaneous. The big drop at $T \approx 150\,\text{MeV}$ is caused by the hadronization of the light quarks. Data taken from [11].

with the *relativistic degrees of freedom*

$$g_{\rm rel} \equiv \sum_b g_b + \frac{7}{8} \sum_f g_f. \tag{II.29}$$

When adding up DOF, particles and antiparticles count separately. Also the gauge charges (e.g. factor 3 for color charge) and helicity/spin DOF have to be taken into account. As the temperature of the universe declines, more and more particle species become non-relativistic, which implies that $g_{\rm rel}$ is a function of temperature (see Fig. II.3). Due to the Boltzmann suppression, one can assume $\rho \approx \rho_{\rm rel}$ when doing calculations concerning the radiation dominated era of the universe. Using this and setting $k = 0$ for a flat universe, the Hubble rate is given by the simple expression

$$H = \frac{1}{M_{\rm Pl}} \sqrt{\frac{\rho}{3}} = \frac{\pi T^2}{3 M_{\rm Pl}} \sqrt{\frac{g_{\rm rel}}{10}} \tag{II.30}$$

where the gravitational constant was absorbed into the reduced Planck mass $M_{\rm Pl} = (8\pi G)^{-1/2} \approx 2.4 \times 10^{18}\,\text{GeV}$.

Assuming the universe's expansion happens not too fast, i.e. adiabatic, entropy is conserved. The relativistic *entropy density* is

$$s = \frac{2\pi^2}{45} g_{\mathrm{rel},s} T^3 \tag{II.31}$$

and the conservation of entropy implies $sa^3 = \text{const.}$ Note that $g_{\mathrm{rel},s} = g_{\mathrm{rel}}$ as long as all relativistic species are part of the thermal bath. This conservation law implies a temperature increase whenever a species becomes Boltzmann suppressed, i.e. when $g_{\mathrm{rel},s}$ decreases. This energy for the rise in temperature is provided by the annihilation of the respective species.

### Particle freeze-out

Particles stay in equilibrium, i.e. are part of the thermal bath, due to annihilation and scattering with other particles in the plasma. The *interaction rate*

$$\Gamma \equiv n \langle \sigma v \rangle \tag{II.32}$$

is a function of the number density $n$ and the thermally averaged cross section $\langle \sigma v \rangle$ of the interacting particles. In an expanding universe, the Hubble rate $H$ competes with $\Gamma$. Thermal equilibrium is only possible as long the interaction time scale is shorter than the expansion time scale, i.e. if

$$\Gamma^{-1} < H^{-1}. \tag{II.33}$$

Otherwise, particles do not encounter each other frequently enough to maintain thermal equilibrium. By equating $\Gamma$ with $H$, which are both functions of $T$, one can identify the temperature $T_{\mathrm{f}}$ below which a particular particle species decouples from the thermal bath, a process named *freeze-out*. After decoupling, the abundance of a particle species is fixed. Depending on its mass, a species may or may not be relativistic at the time of freeze-out.

As an example, consider a simple 2-to-2 process with a heavy mediator, such that the thermally averaged cross section scales as

$$\langle \sigma v \rangle \sim \left| \; g \overset{M}{\rightsquigarrow} g \; \right|^2 \sim \frac{g^4}{M^4} T^2. \tag{II.34}$$

as a result of dimensional analysis and neglecting incoming and outgoing particle masses. Further assuming $n \sim T^3$ and $H \sim \frac{T^2}{M_{\mathrm{Pl}}}$ for a relativistic species, the freeze-

out condition becomes

$$
\begin{aligned}
1 \overset{!}{\sim} \frac{\Gamma}{H} &\sim \frac{T_{\mathrm{f}}^5 g^4/M^4}{T_{\mathrm{f}}^2/M_{\mathrm{Pl}}} \\
\Leftrightarrow \qquad T_{\mathrm{f}} &\sim \left(\frac{M^4}{g^4 M_{\mathrm{Pl}}}\right)^{\frac{1}{3}} .
\end{aligned}
\tag{II.35}
$$

It becomes apparent that a small coupling $g$ or a large mediator mass $M$ leads to an early freeze-out and vice versa. The above calculation can be applied to estimate the temperature of SM *neutrino decoupling*. Using $\frac{g^2}{M^2} \sim G_{\mathrm{F}} \approx 1.2 \times 10^{-5}\,\mathrm{GeV}^{-2}$, the freeze-out temperature evaluates to $T_{\mathrm{f}} \sim (G_{\mathrm{F}}^2 M_{\mathrm{Pl}})^{-1/3} \approx 1\,\mathrm{MeV}$. This is far above the current neutrino mass limit [12], implying that they froze out while being relativistic. This explains the anomalous behavior of $g_{\mathrm{rel}}$ after neutrino decoupling in Fig. II.3: The energy released by electron-positron annihilation at $T \sim m_e$, so shortly after neutrino freeze-out, goes predominantly into the photon thermal bath. The neutrino temperature $T_\nu$ in contrast decreases at an unchanged rate and does not experience the reheating. Entropy is conserved separately in each thermal bath, the temperature ratio after annihilation is hence given by the ratio of thermalized DOF

$$
\frac{T_\nu}{T} = \left(\frac{g_{\mathrm{rel}}^{\mathrm{th}}(T < m_e)}{g_{\mathrm{rel}}^{\mathrm{th}}(T \gtrsim m_e)}\right)^{1/3} = \left(\frac{2}{2 + \frac{7}{8} \cdot 4}\right)^{1/3} = \left(\frac{4}{11}\right)^{1/3} \tag{II.36}
$$

and the total number of relativistic DOF becomes

$$
g_{\mathrm{rel}}(T < m_e) = 2 + \frac{7}{8} \cdot 2 \cdot N_{\mathrm{eff}} \cdot \left(\frac{T_\nu}{T}\right)^4 \approx 3.363 \tag{II.37}
$$

with effective number of neutrino species $N_{\mathrm{eff}} \approx 3$.[1]

For the explicit determination of $n$ as a function of time, one has to employ the *Boltzmann equation*[2]

$$
\frac{\mathrm{d}n}{\mathrm{d}t} = \langle \sigma v \rangle \left(n^2 - n_{\mathrm{eq}}^2\right) - 3Hn \tag{II.38}
$$

with equilibrium number density $n_{\mathrm{eq}}$ given by (II.25) for relativistic and (II.27) for non-relativistic particles. This differential equation can be understood intuitively: The first term on the r.h.s. switches sign depending on whether $n$ is smaller or larger than the equilibrium density. This ensures, given a large enough cross section, that $n$ tracks $n_{\mathrm{eq}}$. The term involving $H$ accounts for the expansion of the universe and makes successful equilibrium tracking more difficult. For a better visualization, a

[1] The exact value is slightly larger because neutrino decoupling was not complete at the time of annihilation.

[2] This is the equation for a 2-to-2-processes under some simplifying assumptions such as neglecting the effects of Bose condensation and Fermi degeneracy and assuming the same distribution functions for all involved particles.

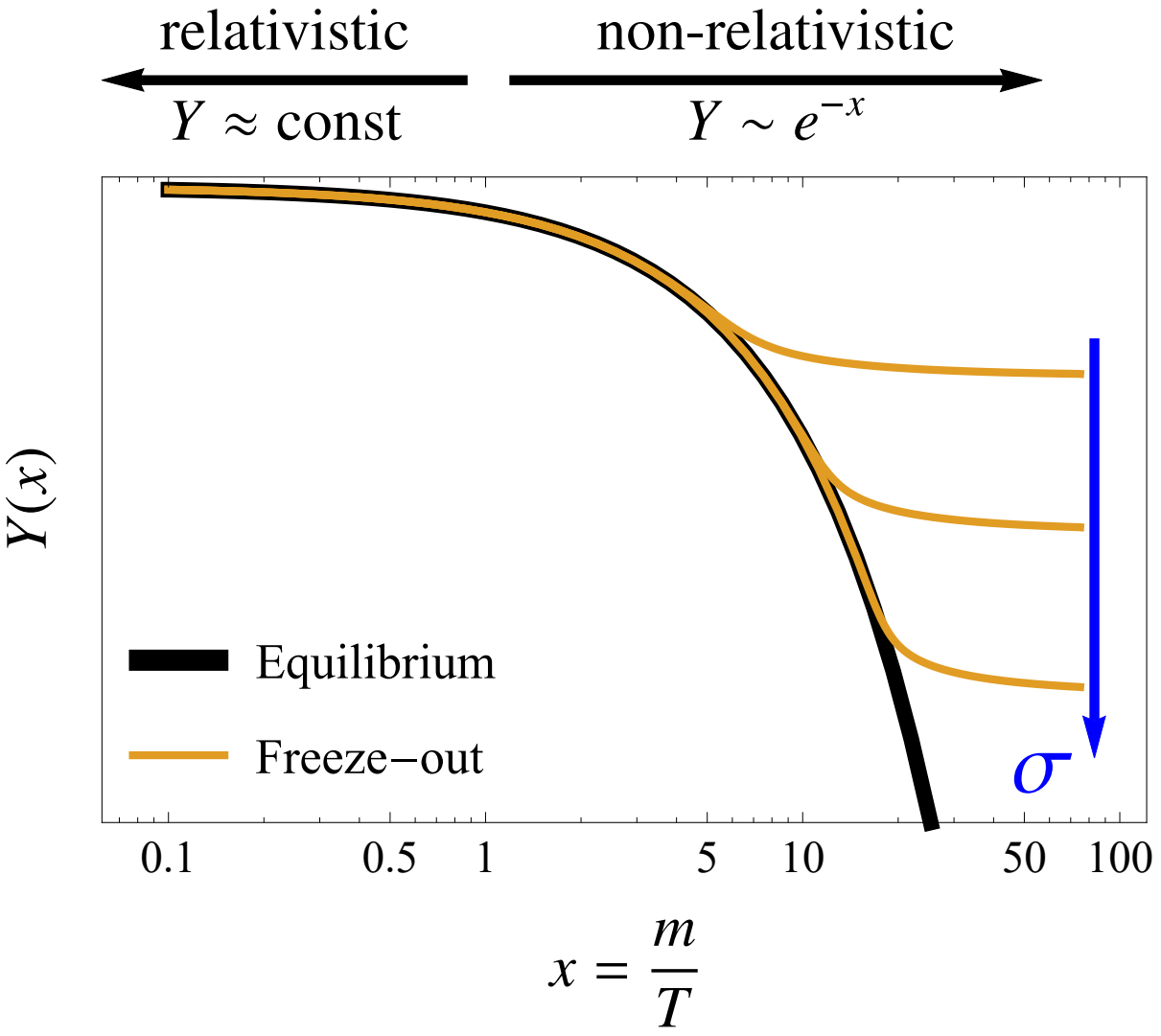

Figure II.4: Equilibrium yield (black) and schematic freeze-out curves (orange) for different cross sections.

convenient parameter transformation can be done by introducing the dimensionless parameters

$$x \equiv \frac{m}{T} \qquad \text{and } \textit{yield} \qquad Y \equiv \frac{n}{s} \tag{II.39}$$

with entropy density $s \sim T^3$. The conservation of entropy implies $s \sim a^{-3}(t)$, allowing us to convert between time and temperature with

$$\frac{1}{T}\frac{\mathrm{d}T}{\mathrm{d}t} = -\frac{1}{a}\frac{\mathrm{d}a}{\mathrm{d}t} = -H. \tag{II.40}$$

Possible freeze-out curves fulfilling the Boltzmann equation are schematically presented in Fig. II.4.

#### A timeline

The following list should serve as an overview of the most important events in the thermal history of our universe, beginning after the Big Bang. The entries are in chronological order with a brief explanation and the time and temperature they occur at.

**Inflation:** Observations of the sky confirm that the universe is indeed extremely homogeneous and isotropic. This seems surprising, since many points with large separation in the night sky have never been in causal contact. In other words, they are further apart from each other than light or any information could have traveled in the maximally available 13.8 billion years. At the same time, it requires an unnatural amount of fine-tuning to make the universe start off with exactly the same temperature in all causally disconnected patches. The most popular solution to this *horizon problem* is the idea of *cosmic inflation*, an era of accelerated expansion right after the Big Bang singularity. According to Friedmann's equations, this implies a shrinking comoving Hubble radius during this time, which in turn fixes the causality problem. A scalar *inflaton* field with a very specific potential can induce this period of inflation. The inflaton is assumed to decay into other particles after the inflationary period, a process called *reheating*. From this point on, the *Hot Big Bang* proceeds and the dynamics can be described by considering thermal equilibrium.

**Electroweak phase transition ($t \sim 10\,\mathrm{ps}$, $T \sim 150\,\mathrm{GeV}$):** The Higgs field assumes a *vacuum expectation value* (VEV) by spontaneous symmetry breaking and thereby gives masses to gauge bosons and fermions (see Section IV.1).

**Baryogenesis:** Matter and anti-matter are, apart from their opposing charges, similar and neither is considered more fundamental than the other. The world around us is however made of baryons instead of anti-baryons. For baryogenesis, the process inducing this asymmetry, three conditions have to be met according to Sakharov [13]: Departure from thermal equilibrium, baryon number violation as well as C- and CP-violation. The SM meets these conditions, but the effect would be too small [14]. Extensions of the SM featuring additional CP-violation and a strong first-order *electroweak phase transition* (EWPT) allow for *electroweak baryogenesis* (EWBG). For successful baryogenesis, the *phase transition* (PT)'s order parameter should be greater than unity and the expansion velocity of the broken phase bubbles must not be too large [15]. We will later see that in the context of the *gravitational waves* (GWs), a large order parameter is also desired but small bubble wall velocities are disfavored (see Sections II.2.6 and III.3.1).

**QCD phase transition ($t \sim 20\,\mathrm{\mu s}$, $T \sim 150\,\mathrm{MeV}$):** Quarks and gluons lose their asymptotic freedom and confine to hadrons, i.e. baryons and mesons.

**Neutrino decoupling ($t \sim 1\,\mathrm{s}$, $T \sim 1\,\mathrm{MeV}$):** Neutrinos, which couple to the SM only via the weak interaction, leave the thermal bath while still being relativistic.

**Electron-positron annihilation ($t \sim 6\,\mathrm{s}$, $T \sim 500\,\mathrm{keV}$):** Electrons and positrons become Boltzmann suppressed and heat up the photon bath while leaving the neutrino temperature unchanged.

**Big Bang nucleosynthesis ($t \sim 3\,\mathrm{min}$, $T \sim 100\,\mathrm{keV}$):** Baryons form the light elements helium, deuterium and lithium. This requires a net abundance of baryons in comparison to anti-baryons in the first place.

**Matter-radiation equality ($t \sim 60 \times 10^3\,\mathrm{yr}$, $T \sim 750\,\mathrm{meV}$):** Due to the different scaling, $\sim a^{-3}$ versus $\sim a^{-4}$, non-relativistic matter becomes the most dominant part of the energy content (see Fig. II.2).

**Recombination ($t \sim 380 \times 10^3\,\mathrm{yr}$, $T \sim 250\,\mathrm{meV}$):** As long as protons and electrons are ionized, photons cannot travel freely due to continuous scattering and the universe is opaque. After recombination of the ions into neutral hydrogen, light is able to spread through the whole universe. This light, forming the first 'snapshot' of the universe, is called *cosmic microwave background* (CMB). Its observation gives important insights about the early universe. Looking even further back in time might soon be possible by the observation of a neutrino background or stochastic GW spectra.

**The present day ($t \sim 13.8 \times 10^9\,\mathrm{yr}$, $T \sim 240\,\mu\mathrm{eV}$):** Homo sapiens on planet Earth has just begun to grasp the laws of nature and to explore the universe.

### II.1.4 Dark Matter

As already alluded to in Section II.1.2, only about 16% of the universe's matter content is visible to us. The unknown rest is called *dark matter* (DM) and couples either very weakly or not at all to the known particle spectrum, except through gravity. DM is still required to explain a series of observed cosmological and astrophysical phenomena.

#### Evidence

The first hints go back to 1884 when Lord Kelvin observed the *velocity dispersion* of stars in our galaxy and inferred, with help of the virial theorem, the Milky Way's mass. The result differed from the estimate of direct observation and he concluded that most of what makes up the galaxy has to be invisible dark matter, as Poincaré called it in 1906. Over the 20th century, these speculations were supported by increasingly precise observations of other galaxies. The spiral arms always seem to rotate faster than suggested by visible mass only. This phenomenon can be explained, if one assumes each galaxy to be embedded in the center of a giant spherical DM halo. Further evidence comes from the observation of distorted light passing through galaxies. This effect is called *gravitational lensing* and is due to the spacetime curvature induced by DM. Finally, DM also plays a huge role in *large scale structure formation*. Assuming ordinary matter alone, density perturbations in the early universe would have become washed out completely due to scattering with relativistic particles. DM however remains (mostly) unaffected by this effect and was therefore able to provide the gravitational wells required for the later formation of stars, galaxies and clusters. Today's arguably most compelling evidence for DM is the *bullet cluster*, two galaxy clusters in the process of collision, shown in Fig. II.5. Due to electromagnetic interaction, gas of ordinary matter is decelerated in course of the collision while DM passes through mostly undisturbed, as seen by the gravitational lensing effect.

#### Forms and production

Not much is known about the nature of DM and there are a number of possible candidates: From light *axions* or *sterile neutrinos*, to *weakly interacting massive particles* (WIMPs) in the electroweak mass range, to *massive astrophysical compact halo objects* (MACHOs). Even a candidate involving a hypothesized 5th dimension, the *Kaluza-Klein particle*, seems possible. Depending on their mass, the candidates can be separated into cold, warm and hot DM. The possible mass range is constrained from above by the observation of dwarf galaxies [17], giving the limit $m \lesssim 10 M_\odot \approx 10^{58}$ GeV. The lower limit $m \gtrsim 10^{-22}$ eV for scalar DM is given by the requirement of a bottom-up structure formation [18], while fermions have a much stronger constraint $m \gtrsim 100\,\text{eV}$ due to the Fermi exclusion principle [19]. These constraints rule out MACHOs and SM neutrinos from making up all the required DM alone [20].

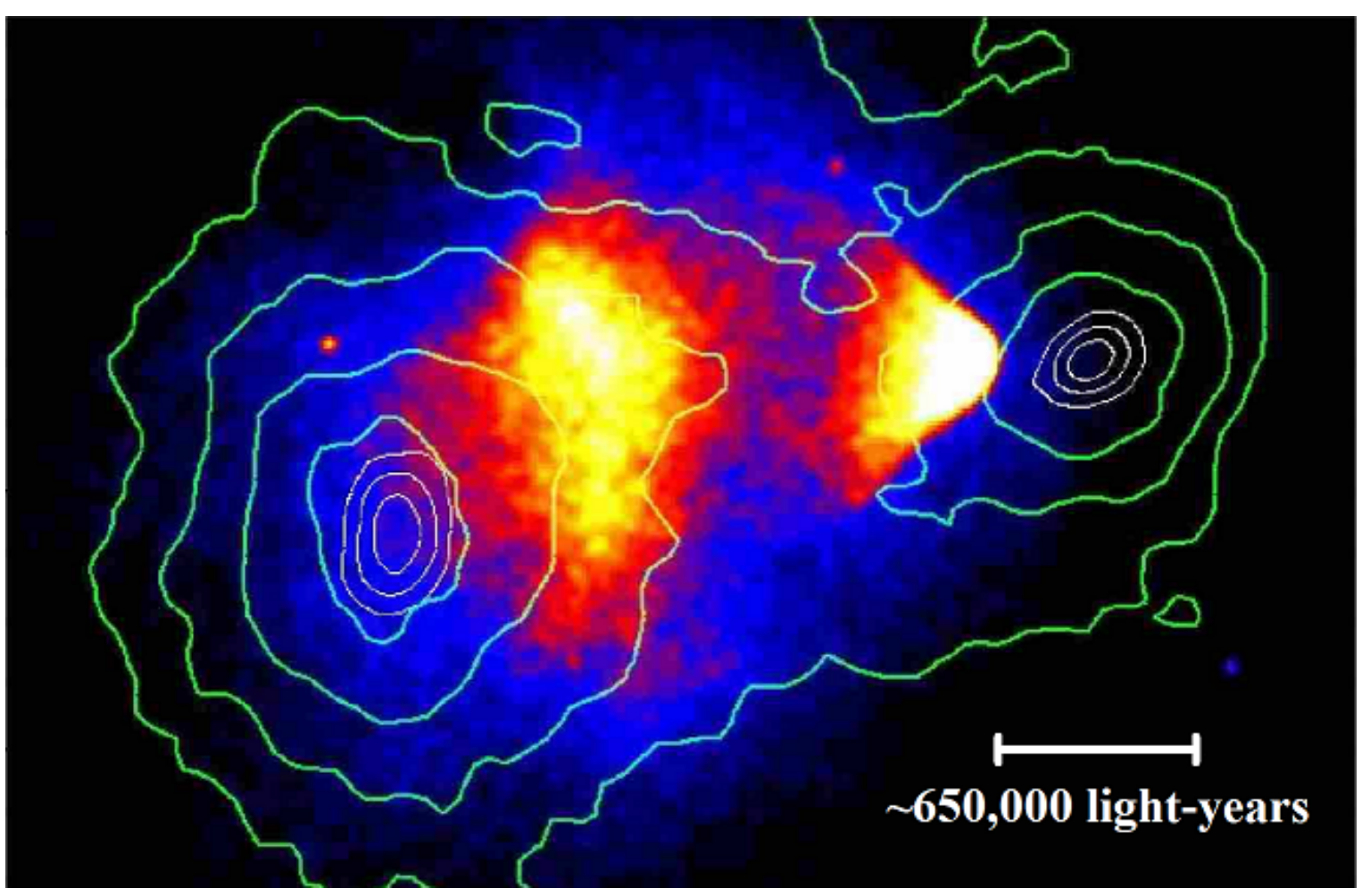

Figure II.5: Snapshot of the bullet cluster (1E0657-553) with green contours indicating mass (including DM). The visible dust clouds lag behind. Figure taken from [16].

In many proposed scenarios, the DM candidate has a small but non-vanishing coupling to the SM and is in thermal contact with it for a period of time. The limits in this case are much stronger: Perturbative unitarity dictates $m < \mathcal{O}(100\,\text{TeV})$ [21] while the lower limit is often stated as $m > \mathcal{O}(\text{MeV})$ due to constraints from BBN and the CMB [22]. The class of WIMP-like DM seems to be particularly promising. Depending on the model of interest, WIMP-like DM can be either thermalized after inflation and freeze out at later times or not be abundant until a certain process produces it, which would be a *freeze-in* scenario. In both cases, mass and coupling strength need to be tuned in the right way, in order to realize today's observed DM abundance. Applying the freeze-out condition (II.33) with a DM mass as input gives an estimate of the required cross-section. Surprisingly, assuming a weak scale DM mass of $\mathcal{O}(100\,\text{GeV})$ yields a coupling strength also comparable to those of weak SM interactions, a coincidence dubbed *WIMP miracle*. The freeze-out of WIMPs with $m > 10\,\text{GeV}$ typically starts at temperatures around $x = \frac{m}{T} \approx 20$ [23].

### Detection

Current DM searches mainly pursue the three approaches 'make it, shake it or break it' and rely on the assumption of particle DM with at least some tiny SM interaction. Direct detection experiments ('shake it') aim to observe recoils of atomic nuclei after being struck by DM that is passing through the detector medium. This approach requires a non-zero DM-SM cross-section and is based on the assumption of DM being abundant everywhere in the galaxy and not just in the center (as in case of MACHO DM). Indirect detection ('break it') refers to the attempt of observing DM self-annihilation or decay products coming from space. In this context, the most interesting sky regions are the ones with high DM densities, for example the center of our galaxy. Particle colliders such as the LEP or the LHC are also well suited for the search. DM production ('make it') in course of a particle collision could become noticeable in the form of missing energy.

None of the enormous efforts so far have led to a clear discovery of DM and its properties [24, 25, 26]. However, the parameter space spanned by mass and coupling strength becomes more and more constrained and model after model can be ruled out. It is to be hoped that DM will be identified at some point at all, since the possibility of non-interacting DM is always there.

## II.1.5 Gravitational Radiation

General relativity is based on the assumption of a dynamically curved spacetime. Energy curves spacetime locally, but still affects distant objects by its gravity. However, this effect at a distance cannot be instantaneous as special relativity imposes causality in form of a speed limit $c$ for the propagation of information. Going one step further, this suggests the possibility of curvature oscillations that travel trough space. The existence of this kind of radiation, namely GWs, is therefore a natural outcome of general relativity. In the following, a brief introduction about the nature of GWs, partly based on [27], shall be given.

### Linearized gravity

The Einstein field equations are non-linear and general solutions of them are difficult to obtain. In the context of GWs it is therefore useful to consider linearized gravity, i.e. small perturbations around the flat Minkowski metric:

$$g_{\mu\nu}(x) = \eta_{\mu\nu} + h_{\mu\nu}(x) \tag{II.41}$$

The smallness of $h_{\mu\nu}$ allows to drop all terms of quadratic or higher order which will lead to significant simplifications and furthermore allows raising and lowering indices with $\eta^{\mu\nu}$ or $\eta_{\mu\nu}$ respectively. The aim is now to describe the propagation of GWs in vacuum, where the energy-momentum tensor $T_{\mu\nu}$ vanishes. The field equations are

then

$$0 = G_{\mu\nu} \equiv R_{\mu\nu} - \frac{1}{2} g_{\mu\nu} R \tag{II.42}$$

with *curvature scalar* $R \equiv R^{\mu}_{\mu}$ and *Ricci tensor* $R_{\mu\nu} \equiv R^{\gamma}_{\mu\gamma\nu}$. The *Riemann curvature tensor* is defined as

$$R^{\mu}_{\nu\alpha\beta} \equiv \partial_\alpha \Gamma^{\mu}_{\nu\beta} - \partial_\beta \Gamma^{\mu}_{\nu\alpha} + \Gamma^{\mu}_{\alpha\rho} \Gamma^{\rho}_{\nu\beta} - \Gamma^{\rho}_{\nu\alpha} \Gamma^{\mu}_{\beta\rho}. \tag{II.43}$$

The *Christoffel symbols* $\Gamma$ are a tool of differential geometry to construct the covariant derivative and allow the comparison of two vectors or tensors at different points in spacetime. Coordinate transformations are the gauge transformations of general relativity, which makes the Christoffel symbols the analogue of gauge fields in *quantum field theory* (QFT). In terms of the metric, they are defined as

$$\Gamma^{\alpha}_{\mu\nu} \equiv \frac{1}{2} g^{\alpha\beta} (\partial_\mu g_{\beta\nu} + \partial_\nu g_{\beta\mu} - \partial_\beta g_{\mu\nu}). \tag{II.44}$$

After inserting all the above ingredients and doing some index-intensive algebra, one ends up with the linearized vacuum Einstein equations

$$0 = G_{\mu\nu} = \frac{1}{2} (\partial_\alpha \partial_\nu \bar{h}^{\alpha}_{\mu} + \partial_\alpha \partial_\mu \bar{h}^{\alpha}_{\nu} - \Box \bar{h}_{\mu\nu} - \eta_{\mu\nu} \partial_\alpha \partial_\beta \bar{h}^{\alpha\beta}) \tag{II.45}$$

with $\Box = \partial_\mu \partial^\mu$. Note that $\bar{h}_{\mu\nu} = h_{\mu\nu} - \frac{1}{2} \eta_{\mu\nu} h^{\alpha}_{\alpha}$ was introduced to shorten the expression. The curvature tensor and so the Einstein equations are gauge invariant, i.e. they are invariant under infinitesimal coordinate transformations

$$\begin{aligned} x^\mu &\to x^\mu + \xi^\mu(x), \\ h_{\mu\nu} &\to h_{\mu\nu} - \partial_\mu \xi_\nu - \partial_\nu \xi_\mu. \end{aligned} \tag{II.46}$$

$h_{\mu\nu}$, being symmetric as the metric itself, has 10 independent components. Choosing a coordinate system reduces the number of DOF by 4, leaving 6 functions for the actual geometry. One way of fixing the redundancy is Lorenz gauge, analogous to the gauge in electromagnetic theory, which imposes the condition

$$0 = \partial_\mu \bar{h}^{\mu}_{\nu} = \partial_\mu h^{\mu}_{\nu} - \frac{1}{2} \partial_\nu h^{\alpha}_{\alpha} \tag{II.47}$$

and simplifies the linearized field equations to

$$\Box \bar{h}_{\mu\nu} = 0. \tag{II.48}$$

This is a typical wave equation, solved most simply by plane waves

$$\bar{h}_{\mu\nu} = \mathrm{Re}[\epsilon_{\mu\nu} \exp(i k_\alpha x^\alpha)] \tag{II.49}$$

with $k_\mu k^\mu = 0$. GWs propagate with the speed of light, which is reflected in $k^\mu$ being lightlike.

### Transverse-traceless gauge

Imposing Lorenz gauge does not exhaust the gauge freedom of the metric. One can further restrict the metric perturbation to be purely spatial and traceless, i.e.

$$h_{0i} = h_{i0} = h^\mu_\mu = 0. \tag{II.50}$$

Note that in this gauge, there is no distinction between $h_{\mu\nu}$ and $\bar{h}_{\mu\nu}$. Applied to the plane wave ansatz (II.49), the additional constraints together with the Lorenz condition imply

$$\epsilon_{0\mu} = \epsilon_{\mu 0} = \epsilon^\mu_\mu = 0 \tag{II.51}$$

and

$$k^i \epsilon_{ij} = 0 \tag{II.52}$$

for the polarization tensor. The last equation makes the transverse nature of the waves explicit, hence the name *transverse-traceless gauge* (TT). It is certainly not necessary to pick this gauge, but it has the advantage of fixing the entire gauge freedom. Starting with 10 independent degrees of freedom in $\epsilon_{\mu\nu}$, the gauge conditions leave behind only *two physical polarization states* for GWs. Rotating the polarization tensor around the axis of propagation $k^\mu$ reveals that gravity is associated with a massless spin-2 field with helicity $\pm 2$.

To display the physical polarization states more explicitly, consider a GW of energy $E$ traveling in $z$-direction:

$$k^\mu = (E, 0, 0, E)$$

The requirement of spatial transversality leaves only 4 non-vanishing components for $\epsilon_{\mu\nu}$, which are further related due to tracelessness and symmetry. Finally, the metric perturbation becomes

$$h_{\mu\nu}(t, z) = \begin{pmatrix} 0 & 0 & 0 & 0 \\ 0 & \epsilon_{11} & \epsilon_{12} & 0 \\ 0 & \epsilon_{12} & -\epsilon_{11} & 0 \\ 0 & 0 & 0 & 0 \end{pmatrix} \cos[E(t - z)] \tag{II.53}$$

and can be split up into the polarization states

$$\epsilon_+ \sim \begin{pmatrix} 0 & 0 & 0 & 0 \\ 0 & -1 & 0 & 0 \\ 0 & 0 & 1 & 0 \\ 0 & 0 & 0 & 0 \end{pmatrix} \quad \text{and} \quad \epsilon_\times \sim \begin{pmatrix} 0 & 0 & 0 & 0 \\ 0 & 0 & 1 & 0 \\ 0 & 1 & 0 & 0 \\ 0 & 0 & 0 & 0 \end{pmatrix}.$$

#### Effect on test particles

As a consequence of the equivalence principle, local effects of gravity can be transformed away. Observing a single point particle is therefore not sufficient to make GWs visible. At least two freely floating test masses are required, which we consider to be at rest and placed on the $x$-axis separated by a distance $L_x$. A $+$-polarized GW propagating in $z$-direction is described by

$$g_{\mu\nu}(t,z) \sim \begin{pmatrix} 1 & 0 & 0 & 0 \\ 0 & -1-h_+(t,z) & 0 & 0 \\ 0 & 0 & -1+h_+(t,z) & 0 \\ 0 & 0 & 0 & -1 \end{pmatrix}$$

which implies the invariant line element to be

$$\mathrm{d}s^2 = \mathrm{d}t^2 - [1+h_+(t,z)]\mathrm{d}x^2 - [1-h_+(t,z)]\mathrm{d}y^2 - \mathrm{d}z^2. \tag{II.54}$$

The *proper distance*, which refers to a distance measured by a ruler, is determined by integrating the line element. The calculation restricted to the relevant subspace $(t,x)$ at $z=y=0$ yields

$$\begin{aligned} L_x(t) &= \int_0^{L_x} \sqrt{-\det g_{\mu\nu}}\mathrm{d}x \\ &= \sqrt{1+h_+(t)}L_x \\ &\approx \left[1+\frac{1}{2}h_+(t)\right] L_x. \end{aligned} \tag{II.55}$$

The same analysis can be done for two masses on the $y$-axis and the resulting relative displacements for both cases are

$$\begin{aligned} \frac{\Delta L_x(t)}{L_x} &= \frac{1}{2}h_+(t) = h\cos[E(t-z)], \\ \frac{\Delta L_y(t)}{L_y} &= -\frac{1}{2}h_+(t) = h\cos[E(t-z)+\pi] \end{aligned} \tag{II.56}$$

and imply a 180° phase-shifted oscillation in the $x$- and $y$-direction with *strain amplitude h*. The strain can be used to define the strength of a GW and is usually, due to the weakness of gravity, extremely small. One of the most violent processes in the universe, the coalescence of two black holes, causes a peak strain tiny as $\sim 10^{-21}$ on

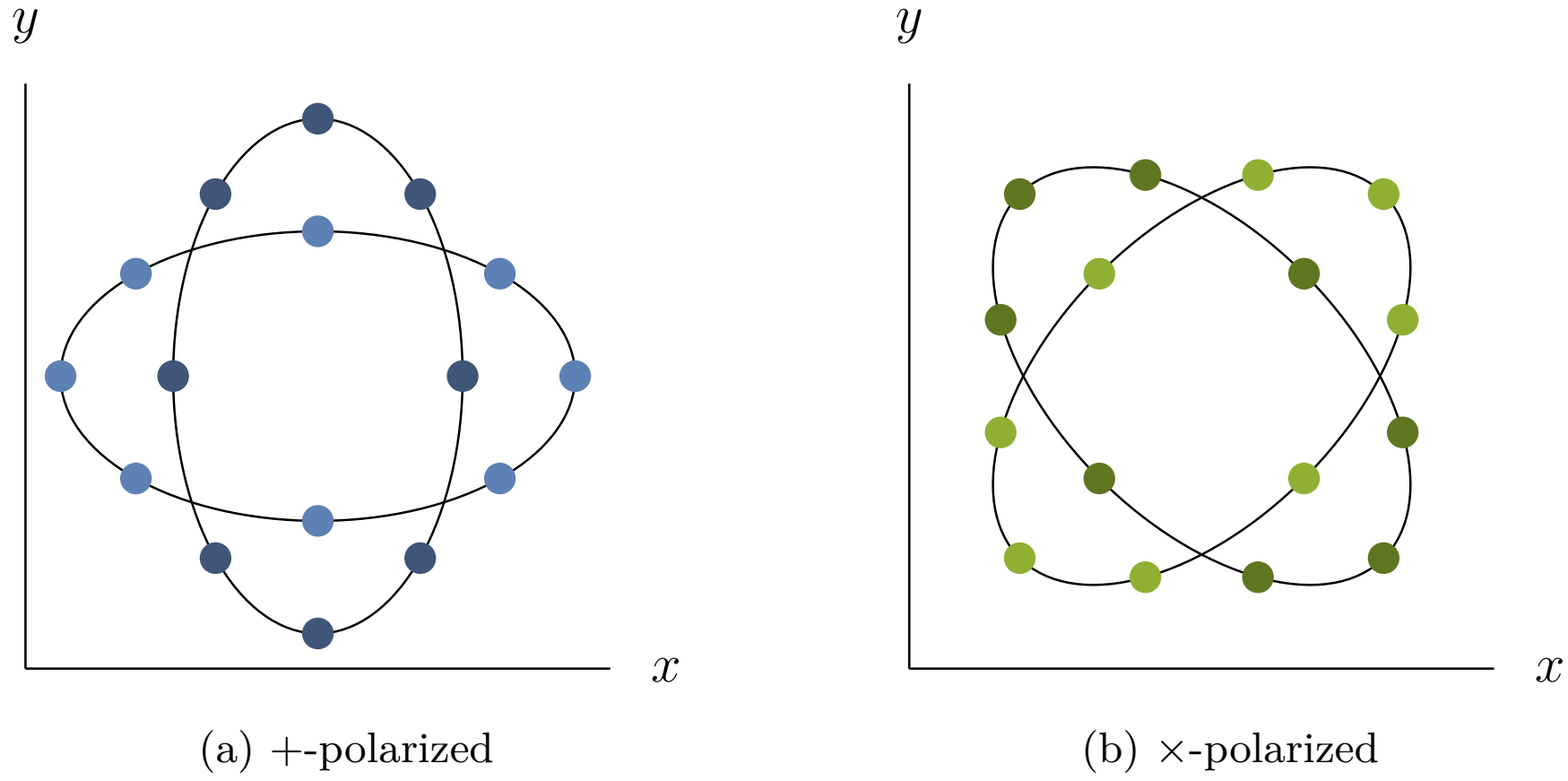

(a) +-polarized

(b) ×-polarized

Figure II.6: Displacement of test masses in the $x$-$y$-plane under the influence of polarized GWs in $z$-direction (strongly exaggerated).

Earth [3]. This is the ratio of a human hair's width to the distance between Alpha Centauri and us. The same exercise can be done for ×-polarized waves with particles sitting on axes rotated 45°. The oscillation patterns caused by the two polarizations are displayed in Fig. II.6.

### Sources

Assuming vacuum as propagation medium is appropriate in an almost empty universe. The production of GWs however requires a massive source. The linearized Einstein equations with source term

$$\Box\bar{h}_{\mu\nu} = -16\pi G\, T_{\mu\nu} \tag{II.57}$$

are solved by

$$\Box\bar{h}_{\mu\nu}(x) = 4G \int \mathrm{d}^3x' \frac{T_{\mu\nu}(t - |\boldsymbol{x} - \boldsymbol{x}'|\,, \boldsymbol{x}')}{|\boldsymbol{x} - \boldsymbol{x}'|}. \tag{II.58}$$

The usage of the *retarded time* $t - |\boldsymbol{x} - \boldsymbol{x}'|$ accounts for the time delay caused by the finite speed of causality. Assuming to be be far away from the source $|\boldsymbol{x} - \boldsymbol{x}'| \approx |\boldsymbol{x}| \equiv r$, considering the conservation law $\partial_\mu T^\mu_\nu = 0$ and imposing pure spatial polarization again leads us to the *quadrupole formula*

$$\bar{h}_{ij}(x) = \frac{2G}{r}\ddot{Q}_{ij}(t - r) \tag{II.59}$$

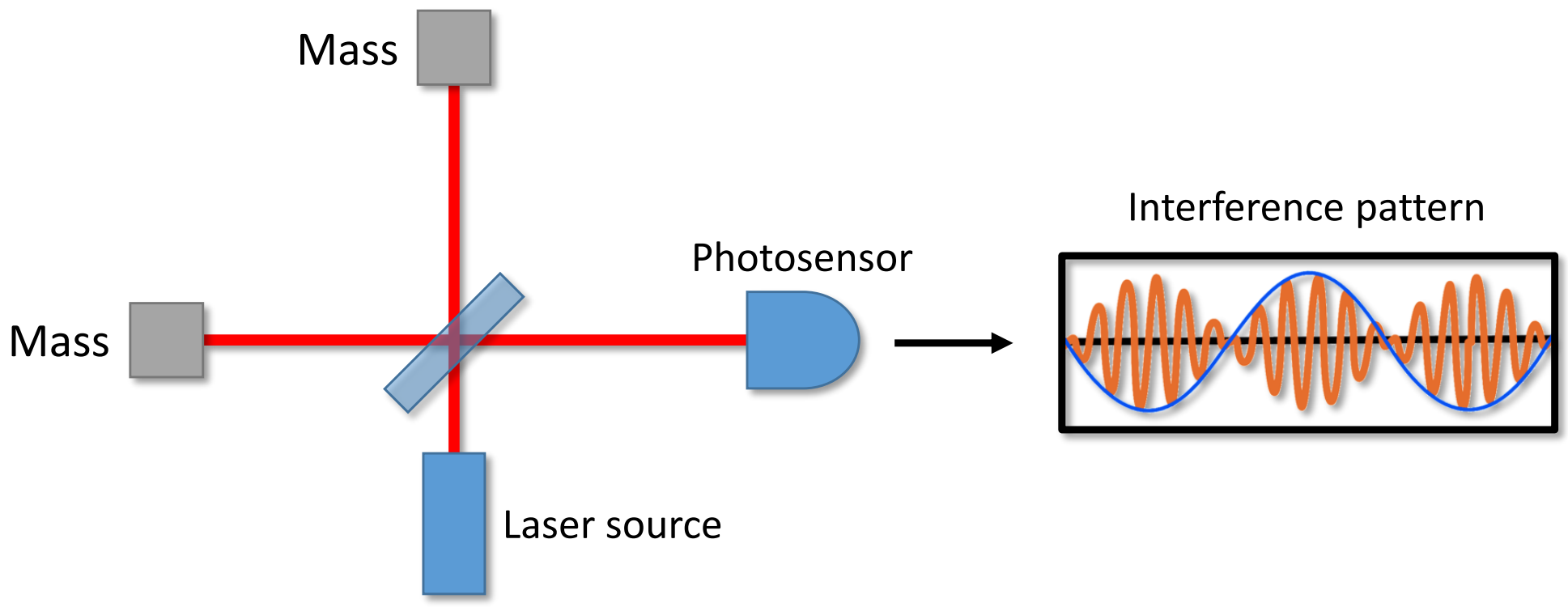

Figure II.7: Schematic of a Michelson interferometer.

featuring the second time derivative of the quadrupole tensor

$$Q_{ij}(t) = \int \mathrm{d}^3x' \, T_{00}(t, \boldsymbol{x}') x'_i x'_j. \tag{II.60}$$

This is an interesting outcome and reveals the nature of GW sources: They are time-dependent and *anisotropic* (non-spherical) motions of mass, where only the second and higher orders of the multipole expansion contribute. GWs carry energy and one can show that the corresponding energy-momentum tensor is given by

$$T^{\mathrm{GW}}_{\mu\nu} = \frac{1}{32\pi G} \left\langle \partial_\mu h^{\mathrm{TT}}_{ij} \partial_\nu h^{ij}_{\mathrm{TT}} \right\rangle \tag{II.61}$$

with time average $\langle \bullet \rangle$. By integrating the energy flow $T^{\mathrm{GW}}_{0i}$ over a spherical surface, the *luminosity*

$$L = \int \mathrm{d}A^i T^{\mathrm{GW}}_{0i} = \frac{G}{5} \left\langle \dddot{Q}^{\mathrm{TT}}_{ij} \dddot{Q}^{ij}_{\mathrm{TT}} \right\rangle \tag{II.62}$$

is obtained, i.e. the total amount of energy radiated by a quadrupole source per time. The TT representation $Q^{\mathrm{TT}}_{ij} = Q_{ij} - \frac{1}{3}\delta_{ij} Q^k_k$ is used at this point.

### Detection

The basic concept of GW observation is simple and just requires placing a ruler next to two test masses. This is basically what a Michelson interferometer does (see Fig. II.7): Monochromatic laser light is split up by a beam splitter into orthogonal directions. Both partial beams are then reflected by freely floating test masses at a certain distance. After passing through the splitter a second time, the two beams interfere with each other. A change in the interference pattern signals a changing interferometer arm length which can be caused by traversing GWs. The incredible smallness of strain $h$ makes it however extremely difficult to distinguish an actual signal from thermal, seis-mic or instrumental noise. It needed nothing less than the technically most advanced instruments available today and decades of work to finally make GWs visible.

Note that a single interferometer is only sensitive to one of the two polarization states and blind to any sources lying in its own plane. For this reason, and also to cancel local disturbances like environmental effects, it is helpful to operate multiple interferometers with varying orientation and at different places.

## II.2 The Effective Potential

The dynamics of a system described by QFT are determined by a Lagrangian density $\mathcal{L} \equiv \mathcal{L}_{\text{kin}} - V$ which consists of kinetic and potential terms. The potential $V$ gives rise to particle masses and field interactions, allowing particles to scatter and decay. A system's equations of motion are determined by the principle of least action $\delta S = 0$ with action $S \equiv \int \mathrm{d}^4 x \mathcal{L}$, which is in the static case (omitting field oscillations) equivalent to the minimization of potential $V$. The vacuum does therefore always occupy a minimum of the potential. The tree-level potential is not the end of the story since additional vertices arise at loop-level. The *effective potential* incorporates these corrections and will be schematically derived in the following, based on [28].

In a QFT with action $S[\phi] = \int \mathrm{d}^4 x \mathcal{L}[\phi]$ and quantized scalar field $\phi$,[3] the *partition function* is given by the amplitude for vacuum at past-infinity going to vacuum at future-infinity

$$\begin{aligned} Z[J] &\equiv \langle 0 | e^{-i(H+H_J)t} | 0 \rangle_J \\ &= \int \mathcal{D}\phi \exp\left\{ i \left[ S[\phi] + \int \mathrm{d}^4 x J \phi \right] \right\} \end{aligned} \tag{II.63}$$

where $\mathcal{D}\phi$ is the *path integral* measure which implies integration over $\phi$ separately at each spacetime point. $J$ is a classical (non-quantized) current that can source or destroy $\phi$ from vacuum and will be useful in the subsequent discussion. Diagrammatically, $Z[J]$ is the sum of all possible combinations of connected graphs $W$ to any loop order without external legs and in presence of $J$:

$$Z[J] = \sum_N \frac{1}{N!} (iW[J])^N = \exp\{iW[J]\} \tag{II.64}$$

$1/N!$ is a combinatorial factor for the permutation of vertices amongst connected graphs and $N = 0$ represents vacuum going to vacuum without any interaction taking place. $\phi$ can be written as functional derivative

$$\phi_J \equiv \frac{1}{Z[J]} \langle 0 | \phi | 0 \rangle_J = \frac{1}{i} \frac{\delta \log Z[J]}{\delta J} = \frac{\delta W[J]}{\delta J}, \tag{II.65}$$

where from now on the time-evolution exponential of (II.63) is not written out any

[3] We will consider only scalar potentials, because non-zero VEVs of fermion or vector boson fields would break Lorentz invariance of the vacuum.

more. This relation between $\phi$ and $J$ can be used to define the *effective action*

$$\Gamma[\phi] \equiv W[J] - \int \mathrm{d}^4x\, J_\phi \phi \tag{II.66}$$

as the Legendre transformation of $W[J]$. With (II.65) it is now easy to see that

$$\frac{\delta\Gamma[\phi]}{\delta\phi} = -J_\phi. \tag{II.67}$$

In the vacuum where $J = 0$, (II.67) becomes an equation of motion as it constrains the dynamics of $\phi$ to the minimum of $\Gamma$. This shows that $\Gamma$ does in fact behave like an action. Due to this extremal condition, the formalism describes a physical vacuum state, meaning that from now on $\phi$ has to be understood as a *classical background field*, i.e. a constant VEV around which the dynamic field is oscillating.[4]

To understand the actual meaning of $\Gamma$, consider a modified partition function with $\Gamma$ instead of $S$ as action

$$Z_\Gamma[J] = \int \mathcal{D}\phi \exp\left\{\frac{i}{\hbar}\left[\Gamma[\phi] + \int \mathrm{d}^4x J\phi\right]\right\} \tag{II.68}$$

where the $\hbar$ dependency is shown explicitly for now. Recall that a propagator is usually given by the inverse of the kinetic plus the mass term of the action, so it comes with $\hbar$. On the other hand, interaction terms as well as the source term vertex scale as $\hbar^{-1}$. A connected diagram thus has the dimension $\hbar^{\#\text{propagators}-\#\text{vertices}}$. Furthermore, the total number of loops in any connected diagram is given by

$$L = \#\text{propagators} - \#\text{vertices} + 1.$$

With this, the connected diagrams in $Z_\Gamma[J]$ can be expanded in loop order, i.e.

$$Z_\Gamma[J] = \exp\left\{i \sum_{L=0}^{\infty} \hbar^{L-1} W_\Gamma^{(L)}[J]\right\} \tag{II.69}$$

with $L$-loop contribution $W_\Gamma^{(L)}$. In the classical limit $\hbar \to 0$, the saddle point approximation can be applied to (II.68) by setting $\phi = \phi_J$. At the same time, $\hbar \to 0$ makes all terms involving loops ($L > 0$) small and only the tree-level term in (II.69) remains:

$$W_\Gamma^{(0)}[J] = \Gamma[\phi_J] + \int \mathrm{d}^4x\, J\phi_J \overset{\text{(II.66)}}{=} W[J] \tag{II.70}$$

This identity shows that all connected diagrams of a theory with action $S$, including all loop orders, can be obtained by considering only tree-level diagrams but with interaction vertices given by the effective action $\Gamma$ instead of $S$. This in turn implies that $\Gamma$ must contain all connected one-particle-irreducible Feynman diagrams with any possible number of external legs.

[4] In literature, the background field is sometimes labeled $\phi_\mathrm{c}$ or $\langle\phi\rangle$. We refrain from doing so, to make all equations more tidy and readable. Banking on the reader's ability to infer the meaning in a given context, dynamic fields and VEVs will therefore often be denoted by the same symbols throughout this work.

When writing the effective action as $\Gamma = \int \mathrm{d}^4x\, (\mathcal{L}_{\mathrm{kin}} - V_{\mathrm{eff}})$, the kinetic term $\mathcal{L}_{\mathrm{kin}}$ can be dropped by requiring the derivatives of $\phi$ to vanish. This is equivalent to the earlier mentioned conception of $\phi$ representing a classical background field instead of a dynamical one. As a consequence, the effective potential can be written as

$$V_{\mathrm{eff}}(\phi) = -\sum_{n=0}^{\infty} \frac{\phi^n}{n!} \Gamma^{(n)}(p=0) \tag{II.71}$$

$$= \Gamma^{(0)} + \Gamma^{(1)} + \Gamma^{(2)} + \Gamma^{(3)} + \dots$$

with $n$-point effective vertex $\Gamma^{(n)}$ in momentum space and all external momenta, i.e. derivatives in position space, set to zero. Each blob represents anything that can effectively happen at the corresponding vertex, including all higher loop orders. The zero-loop contribution of $V_{\mathrm{eff}}$ is, of course, nothing but the tree-level potential

$$V_{\mathrm{eff}}^{\text{0-loop}}(\phi) = V_{\mathrm{tree}}(\phi).$$

* * *

In the context of thermodynamics, the universe can (neglecting chemical potentials) be described as a *canonical ensemble*, i.e. a mechanical system of fixed volume and given temperature $T$. The effective potential resembles the *free energy density*

$$f = V_{\mathrm{eff}}(\phi, T). \tag{II.72}$$

The second law of thermodynamics causes entropy to increase, which is achieved by the minimization of $f$ in this ensemble. Other thermodynamic variables such as *pressure* $P$, *entropy density* $s$ and *energy density* $\rho$ can be derived by the usual relations

$$P = -f, \tag{II.73}$$

$$s = -\frac{\partial f}{\partial T}, \tag{II.74}$$

$$\rho = f + Ts = f - T\frac{\partial f}{\partial T}. \tag{II.75}$$

### II.2.1 One-Loop Contributions

S. Coleman and E. Weinberg were the first to derive the one-loop radiative corrections and to point out their ability to induce spontaneous symmetry breaking in some constellations [29]. As an illustrative toy example, consider a Lagrangian including a Dirac fermion $\psi$ and a complex scalar $\Phi = (\phi + i\varphi)/\sqrt{2}$ that are both charged under a $U(1)$ gauge symmetry with covariant derivative $D_\mu = \partial_\mu - igA_\mu$:

$$
\begin{aligned}
\mathcal{L} &= |D_\mu \Phi|^2 + \mu^2 |\Phi|^2 - \lambda |\Phi|^4 \\
&\quad + i\bar{\psi} \not{D} \psi - \frac{1}{2} y \Phi \bar{\psi}\psi + \text{h.c.} \\
&\supset \frac{1}{2}(\partial_\mu \phi)^2 + \frac{1}{2} g^2 \phi^2 A_\mu A^\mu + \frac{1}{2}\mu^2\phi^2 - \frac{1}{4}\lambda\phi^4 \\
&\quad + i\bar{\psi}\not{\partial}\psi + g\bar{\psi}\gamma^\mu\psi A_\mu - \frac{1}{\sqrt{2}} y \phi \bar{\psi}\psi
\end{aligned}
\tag{II.76}
$$

A VEV can always be rotated onto the real axis, we therefore focus only on the real component of $\Phi$. The following derivation of the one-loop effective potential partly follows [30].

### Scalar contribution

Each scalar diagram has $2n$ external legs, $n$ propagators and vertices and comes with symmetry factors $1/(n+n)$ (cyclic and anti-cyclic permutation of vertices) and $1/2^n$ (interchangeability of external legs at each vertex). The factor of 2 in front accounts for the complex loop scalar. The Feynman rules of the considered Lagrangian give factors of $-6i\lambda$ for each vertex and $i/(k^2 + \mu^2 + i\epsilon)$ for a propagator with momentum $k$. Setting all external momenta to zero, the contribution amounts to

$$
\begin{aligned}
V_{\text{eff}}^{\text{1-loop}}(\phi) &\supset \quad + \quad + \quad + \ldots \\
&= 2 \cdot i \sum_{n=1}^{\infty} \int \frac{\mathrm{d}^4 k}{(2\pi)^4} \frac{1}{2n} \left[ \frac{6\lambda\phi^2/2}{k^2 + \mu^2 + i\epsilon} \right]^n \\
&= \int \frac{\mathrm{d}^4 k_{\mathrm{E}}}{(2\pi)^4} \log\left\{ 1 + \frac{3\lambda\phi^2}{k_E^2 - \mu^2 - i\epsilon} \right\} .
\end{aligned}
\tag{II.77}
$$

To arrive at the last line, the energy component has been Wick rotated, i.e. $k_0 = ik_4$, $\mathrm{d}k_0 = i\mathrm{d}k_4$ and $k^2 = -k_E^2$. Before rewriting the result further, recall that $\phi$ actually represents a classical background field $\phi_{\mathrm{c}}$, i.e. a VEV. Despite not being relevant in the discussion so far, there is still a dynamical quantum field $\phi_{\mathrm{d}}$ which oscillates around the potential minimum. The frequency of this oscillation represents, according to the Klein-Gordon equation, the ground state energy i.e. the mass of the scalar particle. In other words, the mass is a measure for 'how parabolic' the potential is in its minimum, so it is identical to its second derivative. One can make the dynamic nature of the quantum field explicit for a moment by writing $\phi = \phi_{\mathrm{c}} + \phi_{\mathrm{d}}$ instead of just $\phi = \phi_{\mathrm{c}}$. The $\phi^4$ term in the tree-level potential then gives rise to a term $\sim \phi_{\mathrm{c}}^2 \phi_d^2$ which obviously contributes to the second derivative. The *field dependent mass*[5] is thus given by

$$m_\phi^2(\phi_c) = 3\lambda\phi_c^2 - \mu^2.$$

Inserting into (II.77) and dropping the subscript 'c' again, we arrive at

$$V_{\text{eff}}^{\text{1-loop}}(\phi) \supset 2 \cdot \frac{1}{2} \int \frac{\mathrm{d}^4 k_{\mathrm{E}}}{(2\pi)^4} \log\left\{k_{\mathrm{E}}^2 + m_\phi^2(\phi)\right\}. \tag{II.78}$$

Note that an *infrared* (IR) divergent, field independent term has been dropped in the last step.

#### Fermion contribution

The one-loop contribution with $2n$ internal fermion propagators $i\not{k}/(k^2 + i\epsilon)$ has $2n$ external legs and the same number of vertices, each of which contributes $-iy/\sqrt{2}$. When dealing with fermion loops, the spinor indices are contracted by an overall trace and an extra minus sign arises. Since traces over any odd numbers of $\gamma$-matrices (one is contained in each $\not{k}$) vanish, again only diagrams with an even number of propagators and external legs contribute. As a result, the fermion contribution reads

$$\begin{aligned} V_{\text{eff}}^{\text{1-loop}}(\phi) &\supset \quad + \quad + \dots \\ &= -i \sum_{n=1}^{\infty} \int \frac{\mathrm{d}^4 k}{(2\pi)^4} \frac{1}{2n} \mathrm{Tr}\left\{ \left[ \frac{y\phi\not{k}/\sqrt{2}}{k^2 + i\epsilon} \right]^{2n} \right\} \\ &= -4 \cdot \frac{1}{2} \int \frac{\mathrm{d}^4 k_{\mathrm{E}}}{(2\pi)^4} \log\left\{ 1 + \frac{y^2\phi^2/2}{k_{\mathrm{E}}^2} \right\} \\ &\supset -4 \cdot \frac{1}{2} \int \frac{\mathrm{d}^4 k_{\mathrm{E}}}{(2\pi)^4} \log\left\{k_{\mathrm{E}}^2 + m_\psi^2(\phi)\right\} \end{aligned} \tag{II.79}$$

where the identity $\not{k}\not{k} = k^2 I_{4\times4}$ for Dirac fermions as well as the field dependent Yukawa mass $m_\psi(\phi) = \frac{y}{\sqrt{2}}\phi$ was used.

[5] 'VEV dependent mass' would be a more correct denotation, but the usual jargon is being followed here.

### Gauge boson contribution

In Landau gauge, where ghosts decouple from the scalar field, the gauge boson propagator is given by

$$\Delta^{\mu\nu}(k) = \frac{-i\left(g^{\mu\nu} - \frac{k^\mu k^\nu}{k^2}\right)}{k^2 + i\epsilon} \tag{II.80}$$

and is equipped with the property

$$\mathrm{Tr}\left[(\Delta^{\mu\nu}(k))^n\right] = \left(\frac{-i}{k^2 + i\epsilon}\right)^n \underbrace{\left(g^\mu_\mu - \frac{k^\mu k_\mu}{k^2}\right)}_{=3} . \tag{II.81}$$

The coupling between $\phi$ and gauge boson $A^\mu$ gives a factor of $2ig^2$ for each vertex with $g$ being the gauge coupling constant. The gauge boson contribution is

$$\begin{aligned} V_{\text{eff}}^{\text{1-loop}}(\phi) \supset \; & \ldots + \ldots + \ldots + \ldots \\ &= 3 \cdot i \sum_{n=1}^{\infty} \int \frac{\mathrm{d}^4 k}{(2\pi)^4} \frac{1}{2n} \left[\frac{2g^2\phi^2/2}{k^2 + i\epsilon}\right]^n \\ &\supset 3 \cdot \frac{1}{2} \int \frac{\mathrm{d}^4 k_E}{(2\pi)^4} \log\left\{k_E^2 + m_A^2(\phi)\right\} \end{aligned} \tag{II.82}$$

with $m_A^2(\phi) = g^2\phi^2$. The Lorentz indices of the closed gauge boson loop were contracted by a trace.

Besides vanishing ghost diagrams, working in Landau gauge entails yet another advantage: One-loop contributions containing gauge bosons and scalars at the same time do not contribute. They vanish, due to their derivative couplings [29], for example

$$k_\mu \quad k_\nu \quad = 0$$

which can easily be seen by $k_\mu \Delta^{\mu\nu}(k) k_\nu = 0$.

* * *

The resulting expressions of the three contributions are similar, with a minus sign for fermions. We can thus combine them by summing over all particles $a$ with $n_a$ DOF, field dependent masses $m_a(\phi)$ and $\eta_a = +1$ $(-1)$ for bosons (fermions):

$$V_{\text{eff}}^{\text{1-loop}}(\phi) = \sum_a \frac{\eta_a n_a}{2} \int \frac{\mathrm{d}^4 k_{\mathrm{E}}}{(2\pi)^4} \log\left\{k_{\mathrm{E}}^2 + m_a^2(\phi)\right\} \tag{II.83}$$

Note that the explicit calculations above already contain multiple DOF: 2 (complex scalar), 4 (Dirac fermion) and 3 (massive vector) respectively.[6]

Before replacing the infinite sums by logarithms, the diagrams with 2 and 4 external legs were *ultraviolet* (UV) divergent. This divergence appears again in the logarithm now and can be treated with the technique of *dimensional regularization* [31]. In doing so, the dimensionality 4 of the integral is shifted to $d = 4 - \epsilon$ which encapsulates the singularities into terms $\sim \frac{1}{\epsilon}$. The result of the calculation is the regularized one-loop potential [32]

$$V_{\text{eff}}^{\text{1-loop}}(\phi) = \sum_a \eta_a n_a \frac{m_a^4(\phi)}{64\pi^2}\left[\log\frac{m_a^2(\phi)}{\Lambda^2} - C_a - C_{\text{UV}}\right] \tag{II.84}$$

with

$$\begin{aligned} C_a &= \begin{cases} {}^3/_2 & \text{(scalars and fermions)} \\ {}^5/_6 & \text{(gauge bosons)} \end{cases}, \\ C_{\text{UV}} &= \frac{2}{\epsilon} - \text{const} + \mathcal{O}(\epsilon) \end{aligned}$$

and *renormalization scale* $\Lambda$, which is usually chosen to be the largest tree-level VEV or mass of a given model. Note that the choice of $\Lambda$ does not influence any physics, it just affects the effective strength of the quartic term in the Lagrangian and can be absorbed by a redefinition of $\lambda$ for any theory.

To be able to numerically add up the different contributions to the potential, infinities will be dropped right from the beginning and *finite counterterms* $V_{\text{ct}}$ are added. They fix the freedom of choosing the constant term in $C_{\text{UV}}$ arbitrarily. Imposing the *renormalization conditions*

$$\begin{aligned} 0 &\overset{!}{=} \left.\frac{\partial\left[V_{\text{eff}}^{\text{1-loop}}(\phi) + V_{\text{ct}}(\phi)\right]}{\partial\phi}\right|_{\phi=\phi_c} && (\textit{tadpole}) \\ 0 &\overset{!}{=} \left.\frac{\partial^2\left[V_{\text{eff}}^{\text{1-loop}}(\phi) + V_{\text{ct}}(\phi)\right]}{\partial\phi^2}\right|_{\phi=\phi_c} && (\textit{self energy}) \end{aligned} \tag{II.85}$$

determines the couplings of $V_{\text{ct}}$ in a way which ensures that the potential minimum and the mass of $\phi$ stay unchanged with respect to the tree-level potential.

### II.2.2 Thermal Quantum Field Theory

Particle interactions at colliders are well described by conventional quantum field theory, which pictures particles as freely arriving from and leaving to spatial infinity, interacting only at a single point in space and time. In presence of an energetic background, for example in the hot and dense plasma of the early universe, thermal effects have to be taken into account. A very comprehensible introduction to this topic is given in [33] and will serve as an inspiration for the following discussion.

[6] Three DOF have to be counted even for massless gauge bosons like the photon, since they receive mass corrections in thermal environments.

In a canonical ensemble with Hamiltonian $H$ and energy eigenstates $|n\rangle$, the expectation value of an operator $A$ is given by a thermally weighted sum

$$\langle A\rangle_T \equiv \frac{1}{Z}\sum_n \langle n|\, e^{-\frac{H}{T}} A \,|n\rangle = \frac{1}{Z}\mathrm{Tr}\left[e^{-\frac{H}{T}} A\right] \tag{II.86}$$

with the partition function

$$Z = \sum_n e^{-\frac{E_n}{T}}$$

as normalization factor. To find the formal connection between thermodynamics and QFT, a thermal two-point function is rearranged by cyclic permutation of the trace argument

$$\begin{aligned}
\langle \phi(\boldsymbol{y},t)\phi(\boldsymbol{x},0)\rangle_T &= \frac{1}{Z}\mathrm{Tr}\left[e^{-\frac{H}{T}}\phi(\boldsymbol{y},t)\phi(\boldsymbol{x},0)\right] \\
&= \frac{1}{Z}\mathrm{Tr}\left[e^{-\frac{H}{T}}\phi(\boldsymbol{y},t)e^{-iH(-\frac{i}{T})}\phi(\boldsymbol{x},-iT^{-1})e^{iH(-\frac{i}{T})}\right] \\
&= \frac{1}{Z}\mathrm{Tr}\left[e^{-\frac{H}{T}}\phi(\boldsymbol{x},-iT^{-1})\phi(\boldsymbol{y},t)\right] \\
&= \left\langle \phi(\boldsymbol{x},-iT^{-1})\phi(\boldsymbol{y},t)\right\rangle_T
\end{aligned} \tag{II.87}$$

where the quantum mechanical time evolution $\phi(t) = e^{iHt}\phi(0)e^{-iHt}$ was used. The discovered equation is called *Kubo-Martin-Schwinger relation* (KMS) and requires the bosonic (fermionic) field $\phi$ to be symmetric (anti-symmetric) and cyclic in time, with periodicity $-iT^{-1}$:

$$\begin{aligned}
&\phi(\boldsymbol{x},0) = \pm\phi(\boldsymbol{x},-iT^{-1}) \\
\Leftrightarrow \qquad &\phi(\boldsymbol{x},0) = \pm\phi(\boldsymbol{x},T^{-1})
\end{aligned} \tag{II.88}$$

We now moved to imaginary time by performing a Wick rotation $t \to \tau = it$, as usually also done in the path integral formalism. By identifying imaginary time $\tau$ with inverse temperature $T^{-1}$, the formal relation between QFT and thermodynamics is revealed.

The cyclic condition of (II.88) turns any continuous Fourier integral involving $\phi$ into a discrete sum. The Fourier transform of any $n$-point function thus becomes

$$f(\boldsymbol{x},\tau) = \int \frac{\mathrm{d}k_0}{2\pi} f(\boldsymbol{x},k_0)e^{-k_0\tau} \to iT\sum_n f(\boldsymbol{x},i\omega_n)e^{-i\omega_n\tau} \tag{II.89}$$

with *Matsubara frequencies*

$$\omega_n = \begin{cases} 2n\pi T & \text{(bosons)} \\ (2n+1)\pi T & \text{(fermions)} \end{cases}$$

and $n \in \mathbb{Z}$. The prefactor $T$ ensures correct normalization and dimensionality of the Fourier sum. With this we have a prescription for quantum loop corrections at finite temperature: Replace the $k_0$ part of a loop integral by an infinite sum and employ imaginary Matsubara frequencies $i\omega_n$ as energy $k_0$. In Euclidean momentum space, the replacement rule reads $k_4 \to \omega_n$ and the $i$ in front of (II.89) becomes redundant. Section A.1 in the Appendix gives a recipe for the evaluation of Matsubara sums.

### II.2.3 The Thermal Effective Potential

The prescription found in the last section can be used to evaluate the effective potential at finite temperature. Applying it to the one-loop result (II.83) yields

$$V_{\text{eff}}^{\text{1-loop}}(\phi, T) = \eta \frac{T}{2} \sum_n \int \frac{\mathrm{d}^3 k}{(2\pi)^3} \log\left\{\omega_n^2 + |\boldsymbol{k}|^2 + m^2(\phi)\right\} \tag{II.90}$$

where for simplicity only one DOF is considered, with $\eta = +1$ in the bosonic and $\eta = -1$ in the fermionic case. To simplify the Matsubara sum, the derivative

$$\frac{\mathrm{d}V_{\text{eff}}^{\text{1-loop}}(\phi, T)}{\mathrm{d}m^2(\phi)} = \eta \frac{T}{2} \sum_n \int \frac{\mathrm{d}^3 k}{(2\pi)^3} \frac{1}{\omega^2 + \omega_n^2} \tag{II.91}$$

can be taken, where $\omega^2 \equiv |\boldsymbol{k}|^2 + m^2(\phi)$. Applying (A.6) yields

$$\frac{\mathrm{d}V_{\text{eff}}^{\text{1-loop}}(\phi, T)}{\mathrm{d}m^2(\phi)} = \eta \frac{1}{2} \int \frac{\mathrm{d}^3 k}{(2\pi)^3} \left[\frac{1}{2\omega} + \frac{1}{\omega} \frac{\eta}{e^{\omega/T} - \eta}\right] \tag{II.92}$$

and after integration

$$\begin{aligned} V_{\text{eff}}^{\text{1-loop}}(\phi, T) &= \eta \int \frac{\mathrm{d}^3 k}{(2\pi)^3} \left[\frac{\omega}{2} + T \log\left\{1 - \eta e^{-\omega/T}\right\}\right] \\ &= V_{\text{eff}}^{\text{1-loop}}(\phi, T = 0) + V_{\text{eff}}^{\text{1-loop}}(\phi, T > 0) \end{aligned} \tag{II.93}$$

the potential splits up into a zero-temperature and a temperature dependent part.

#### Temperature independent part

The temperature independent *Coleman-Weinberg potential* is equivalent to the earlier derived zero-temperature result (II.83):

$$V_{\text{CW}}(\phi) \equiv V_{\text{eff}}^{\text{1-loop}}(\phi, T = 0) = \eta \int \frac{\mathrm{d}^3 k}{(2\pi)^3} \frac{\omega}{2} = \frac{\eta}{2} \int \frac{\mathrm{d}^4 k_{\text{E}}}{(2\pi)^4} \log\left\{k_{\text{E}}^2 + m_a^2(\phi)\right\} \tag{II.94}$$

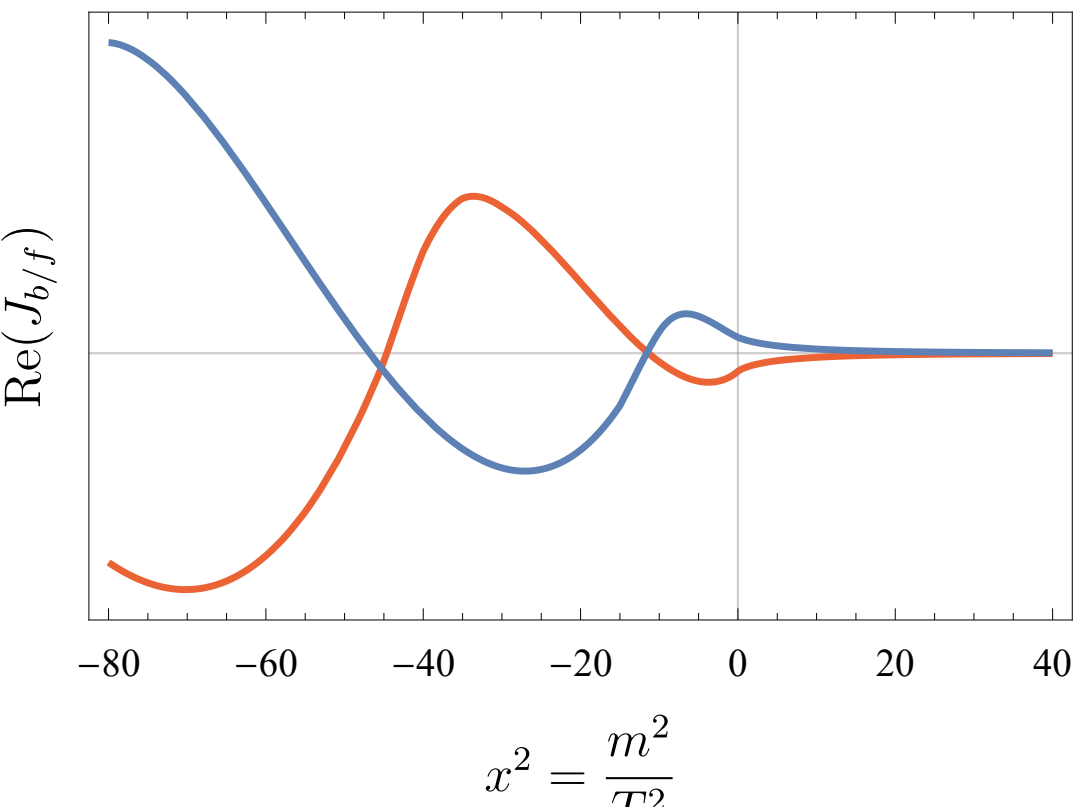

Figure II.8: Thermal functions $J_b(x^2)$ in red and $J_f(x^2)$ in blue.

This equivalence is based on the identity

$$\frac{\omega}{2} = \int_{-i\infty}^{i\infty} \frac{\mathrm{d}k}{2\pi} \log\{k^2 + \omega^2\}$$

which is easy to verify by taking the derivative with respect to $\omega$ and applying the residue theorem.

### Finite temperature part

The temperature dependent part can be rewritten as

$$\begin{aligned} V_T(\phi, T) \equiv V_{\text{eff}}^{\text{1-loop}}(\phi, T > 0) &= \eta T^4 \int_0^\infty \frac{\mathrm{d}y\, y^2}{2\pi^2} \log\left\{1 - \eta \exp\left[-\sqrt{y^2 + \frac{m^2(\phi)}{T^2}}\right]\right\} \\ &= \eta \frac{T^4}{2\pi^2} J_{b/f}\left(\frac{m^2(\phi)}{T^2}\right) \end{aligned} \tag{II.95}$$

by defining $y \equiv \frac{k}{T}$ and the thermal functions

$$J_{b/f}(x^2) \equiv \int_0^\infty \mathrm{d}y\, y^2 \log\left\{1 - \eta \exp\left[-\sqrt{y^2 + x^2}\right]\right\} \tag{II.96}$$

for bosonic ($b$) and fermionic ($f$) contributions. When evaluating the integral numerically, one has to take care of the poles arising for $x^2 \equiv \frac{m^2}{T^2} \leq 0$. Boson fields can in fact have a negative squared mass, namely if the effective potential is evaluated at a point where the system is unstable, e.g. a Mexican hat potential at its center.[7] The thermal functions are exponentially suppressed if $T \ll m(\phi)$ and oscillate for $x^2 < 0$,

[7] The hypothetical particle corresponding to this unstable vacuum configuration is referred to as *tachyon*, as it would travel faster than light.

as can be seen in Fig. II.8. The high-temperature expansion is given by [34]

$$
\begin{aligned}
J_b\left(x^2\right) &\overset{x\to 0}{\approx} -\frac{\pi^4}{45}+\frac{\pi^2}{12}x^2-\frac{\pi}{6}x^3-\frac{1}{32}x^4\log\left(x^2\right)+\text{const},\\
J_f\left(x^2\right) &\overset{x\to 0}{\approx} \frac{7\pi^4}{360}-\frac{\pi^2}{24}x^2-\frac{1}{32}x^4\log\left(x^2\right)+\text{const}.
\end{aligned}
\tag{II.97}
$$

When considering this approximation only up to second order, the thermal one-loop potential becomes

$$
V_T(\phi,T)\approx T^2\left[\sum_b\frac{n_b}{24}m_b^2(\phi)+\sum_f\frac{n_f}{48}m_f^2(\phi)\right]
\tag{II.98}
$$

with $n_b$ ($n_f$) bosonic (fermionic) DOF.

### II.2.4 Symmetry Restoration in the Early Universe

Whilst thermal effects were completely negligible for most of our universe's history, in the very first fractions of seconds after reheating they played an important role. In this hot era, the effective potential was qualitatively different and the universe not in the same vacuum state as today. To see this schematically, consider $V_{\text{tree}}$ and the high-temperature approximation of $V_T$ up to order $\sim x^3$. By noting that usually $m^2(\phi)\sim\phi^2$, one can parametrize the effective potential as

$$
\begin{aligned}
V_{\text{eff}}(\phi,T) &\approx V_{\text{tree}}(\phi)+V_T(\phi,T)\\
&=\left(AT^2-\frac{1}{2}\mu^2\right)\phi^2-BT|\phi|^3+\frac{1}{4}\lambda^4\phi^4.
\end{aligned}
\tag{II.99}
$$

For very high temperatures, the term $\sim A$ dominates and makes the potential parabolic. At this stage, the universe is in a state where $\phi$ has a VEV of zero. As the universe expands and thereby cools down, this dominance abates and the tree-level potential becomes more and more important. If $\mu^2$ is positive, a minimum at non-zero $\phi$ appears and represents the energetically favorable state. The universe undergoes a PT and $\phi$ acquires a non-zero VEV. When talking about the SM where $\phi$ represents the Higgs field, this PT is the breaking of electroweak symmetry. More generally speaking, any PT in which a gauged scalar field acquires a VEV is breaking the underlying gauge symmetry.

The coefficients $A$ and $B$ are composed of several coupling constants and are model dependent. Recall that of both approximated thermal functions $J_{b/f}$, only the one for bosons comes with a term $\sim x^3$. The term $\sim B$ in (II.99) thus vanishes if no bosonic DOF couple to $\phi$. A cubic term strongly affects the nature of the PT as shown in the following two scenarios. The potential minimum $\phi_0$ and critical temperature

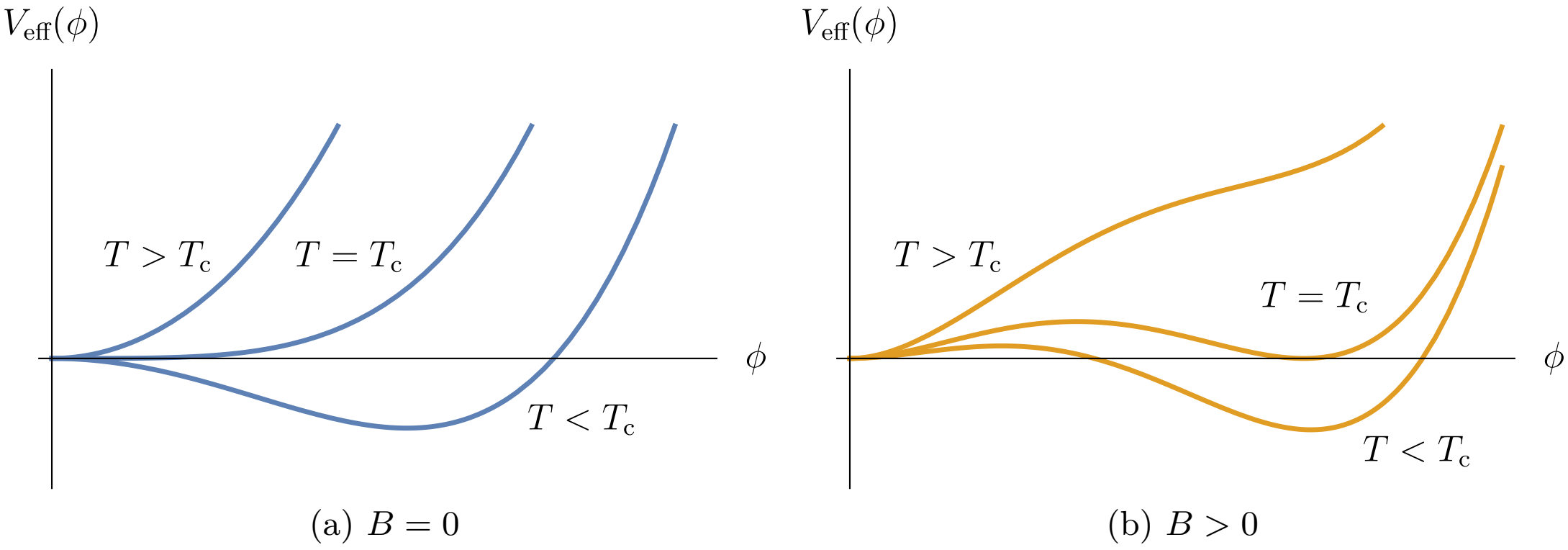

Figure II.9: Schematic plots of the effective potential at different temperatures.

$T_c$ will be determined by

$$\left.\frac{\partial V_{\text{eff}}}{\partial \phi}\right|_{\phi=\phi_{\text{min}}} \stackrel{!}{=} 0 \tag{II.100}$$

and

$$V_{\text{eff}}(\phi_{\text{min}}, T_c) \stackrel{!}{=} V(0, T_c). \tag{II.101}$$

#### Scenario $B = 0$, no barrier

Without cubic terms, no barrier hinders $\phi$ from acquiring a VEV immediately when the *critical temperature* $T_c$ is reached (Fig. II.9, left). This continuous change of the VEV marks a PT of second or higher order. Critical temperature and potential minimum at $T < T_c$ are given by

$$T_c = \frac{\mu}{\sqrt{2A}}, \tag{II.102}$$

$$\phi_{\text{min}} = \sqrt{\frac{\mu^2 - 2AT^2}{\lambda}}. \tag{II.103}$$

#### Scenario $B > 0$, boson couplings induce a barrier

In this scenario $\phi$ is couples to bosonic DOF and the thermal potential includes a cubic term. This ingredient allows for two distinct minima, separated by a potential barrier (Fig. II.9, right). The critical temperature is now the point at which both minima become degenerate:

$$T_{\mathrm{c}} = \frac{\mu}{\sqrt{2(A - B^2/\lambda)}} \tag{II.104}$$

$$\phi_{\min} = \frac{1}{2\lambda}\left[3BT + \sqrt{T^2(9B^2 - 8A\lambda) + 4\lambda\mu^2}\right] \tag{II.105}$$

Unlike before, the universe is first stuck in the $\phi = 0$ minimum until it will eventually tunnel through the barrier and nucleate bubbles of broken phase at $T_{\mathrm{n}} < T_{\mathrm{c}}$. This discontinuous change of the *order parameter* $\frac{\phi}{T}$ marks a *first-order phase transition.*

### II.2.5 Thermal Mass Corrections

In thermal environments, particles receive mass corrections at loop order, so-called *Debye masses*. The corresponding diagrams at one-loop order are referred to as *hard thermal loops*. The calculation of the leading Debye masses for the considered abelian toy Lagrangian (II.76) is carried out in Section A.2 of the Appendix. The results for the different contributions are listed in the following. The scalar Debye masses amount to

$$\approx 2 \cdot \lambda \frac{T^2}{4} = 2 \cdot \frac{\partial^2 m_\phi^2(\phi)}{\partial \phi^2}\frac{T^2}{24}, \tag{II.106}$$

$$\approx 4 \cdot y^2 \frac{T^2}{48} = 4 \cdot \frac{\partial^2 m_\psi^2(\phi)}{\partial \phi^2}\frac{T^2}{48}, \tag{II.107}$$

$$\approx 3 \cdot g \frac{T^2}{12} = 3 \cdot \frac{\partial^2 m_A^2(\phi)}{\partial \phi^2}\frac{T^2}{24}. \tag{II.108}$$

By comparing the results to the high-temperature expansion of the thermal potential (II.98), it can bee seen that the Debye mass of a scalar particle is simply given by

$$\Pi_\phi(T) = \frac{\partial^2 V_T(\phi, T)}{\partial \phi^2}. \tag{II.109}$$

The calculation of the gauge boson Debye masses yields

$$+ \quad \approx \begin{cases} 2 \cdot g^2 \frac{T^2}{6} & \text{(longitudinal)} \\ 0 & \text{(transverse)} \end{cases}, \tag{II.110}$$

$$\approx \begin{cases} 4 \cdot g^2 \frac{T^2}{12} & \text{(longitudinal)} \\ 0 & \text{(transverse)} \end{cases} . \tag{II.111}$$

Fermions also receive thermal masses, but they are not relevant for the effective potential as we will see later. Note that the above stated thermal mass corrections already contain the DOF of the loop particles. The considered toy model contains a 4-component Dirac fermion with $U(1)$ charge 1. To generalize the result for chiral theories like the SM, where left- and right-handed fermions carry different hypercharges, the 4-component Dirac spinor can be split up into two chiral Weyl spinors each of which contributes only 2 DOF. The generalized results summarize to

$$\Pi^{\mathrm{L}}_{U(1)} = g_1^2 T^2 \left[ \frac{1}{6} \sum_s n_s Y_s^2 + \frac{1}{12} \sum_f n_f Y_f^2 \right] \tag{II.112}$$

with $U(1)$ coupling $g_1$ and charges $Y$. $s$ $(f)$ is running over all scalar (chiral fermion) DOF. To generalize the result to non-abelian gauge theories with local $SU(N)$ symmetries, one has to take additional diagrams involving gauge boson self-couplings and ghosts into account. Gauge boson loops yield traces over pairs of generators $T^i$ with $i = 1 \dots (N^2 - 1)$. The result reads [35]

$$\Pi^{\mathrm{L}}_{SU(N)} = g_N^2 T^2 \left[ \frac{N}{3} + \frac{1}{6} \sum_s n_s C(r_s) + \frac{1}{12} \sum_f n_f C(r_f) \right] , \tag{II.113}$$

where $C(r)$ is the characteristic constant of a gauge group representation $r$ with generators $T^i$, defined by $C(r)\delta^{ij} = \mathrm{Tr}[T^i T^j]$, and $g_N$ is the gauge coupling.

### II.2.6 Breakdown of Perturbativity and Daisy Resummation

The coexistence of two different scales, mass parameter $\mu$ and temperature $T$, signals a breakdown of perturbation theory in the regime $T \gg \mu$ [34]. For a simplified demonstration of this breakdown we will consider only the $\lambda$ interaction vertex in the following. The $T$-dependence of a diagram can simply be determined by dimensional analysis, i.e. by considering the *superficial degree of divergence*

$$D = 4\#\text{loops} - 2\#\text{boson prop.} - \#\text{ferm. prop.}$$

The rule is, that a diagram scales as $\sim T^D$ but is always at least linear in $T$ due to the Matsubara frequency sum prefactor. The hard thermal loop has $D = 2$ and therefore

scales as

$$\Pi_{\text{hard}} \equiv \quad \sim \lambda T^2, \tag{II.114}$$

which is confirmed by the explicit calculation in Section A.2 of the Appendix. When adding loops on top of the main loop, the number of propagators increases. This results in $D < 2$ which implies a linear $T$ dependence. Each attached loop contributes a factor $\sim \lambda T^2$ and an appropriate power of the mass parameter $\mu$ must be multiplied to the denominator to fix dimensionality. One ends up with

$$\begin{aligned}\Pi_{\text{daisy}} \equiv \quad &\sim \frac{1}{\mu^{2N-3}}(\lambda T)(\lambda T^2)^{N-1} \\ &= \lambda^N \frac{T^{2N-1}}{\mu^{2N-3}}\end{aligned} \tag{II.115}$$

for the $N$-loop *daisy diagram* with $N-1$ petals. At the critical temperature $T_{\text{c}} \sim \frac{\mu}{\sqrt{\lambda}}$, the negative tree-level mass term $-\mu^2$ and the one-loop contribution $\sim \lambda T^2$ cancel each other. Close to a PT we thus have

$$\alpha \equiv \lambda \frac{T^2}{\mu^2} \sim 1$$

and the scaling can be rewritten as

$$\Pi_{\text{daisy}} \sim \alpha^{N-3/2} \lambda^{3/2} T^2 \sim \lambda^{3/2} T^2. \tag{II.116}$$

This shows that corrections do not recede with increasing loop order, but make a contribution $\sim \lambda^{3/2}$. This is in conflict with perturbation theory where diagrams of higher loop order are usually suppressed. The origin of this breakdown can also be seen in a more explicit calculation:

$$\Pi_{\text{daisy}} \sim \underbrace{T \sum_n \int_0^\infty \frac{\mathrm{d}k}{2\pi^2} \frac{k^2}{(\omega_n^2 + k^2)^{N-1}}}_{\text{main loop}} \times \underbrace{\left[ T \sum_n \int_0^\infty \frac{\mathrm{d}k}{2\pi^2} \frac{k^2}{\omega_n^2 + k^2} \right]^{N-1}}_{\text{petals}} \tag{II.117}$$

In the high-temperature regime where masses are negligible, the main loop contribution is IR divergent for any loop order $N > 2$. This breakdown of perturbativity occurs only for bosonic DOF for which the Matsubara frequencies vanish at $n = 0$. This justifies that we did not worry about fermion Debye masses in the first place.

The daisy diagrams can be resummed by adding up propagators with increasing number of attached loops. The infinite sum is a geometric series and thus yields

which is algebraically written

$$
\begin{aligned}
&\frac{1}{p^2 - m^2} + \frac{\Pi}{(p^2 - m^2)^2} + \frac{\Pi^2}{(p^2 - m^2)^3} + \dots \\
&\quad = \frac{1}{p^2 - m^2} \sum_{i=0}^{\infty} \left[ \frac{\Pi}{p^2 - m^2} \right]^i \\
&\quad = \frac{1}{p^2 - m^2} \frac{1}{1 - \frac{\Pi}{p^2 - m^2}} \\
&\quad = \frac{1}{p^2 - m^2 - \Pi}
\end{aligned}
\tag{II.118}
$$

with a *dressed propagator* as result. The infinite sum of diagrams hence boils down to the simple replacement

$$
m^2(\phi) \to m^2(\phi) + \Pi(T)
$$

in the effective potential, which dresses each of the loop propagators. Recalling the UV divergence of the Coleman-Weinberg potential, it becomes apparent that $T$-dependent counterterms would now be required. This seems to be against physical intuition, because it connects a theory's UV regime to its IR dynamics where the ring diagrams dominate [32]. To avoid this, the mass shift can be restricted to the zero Matsubara mode which gives the most dominant contribution anyway. After the replacement, the thermal one-loop potential reads

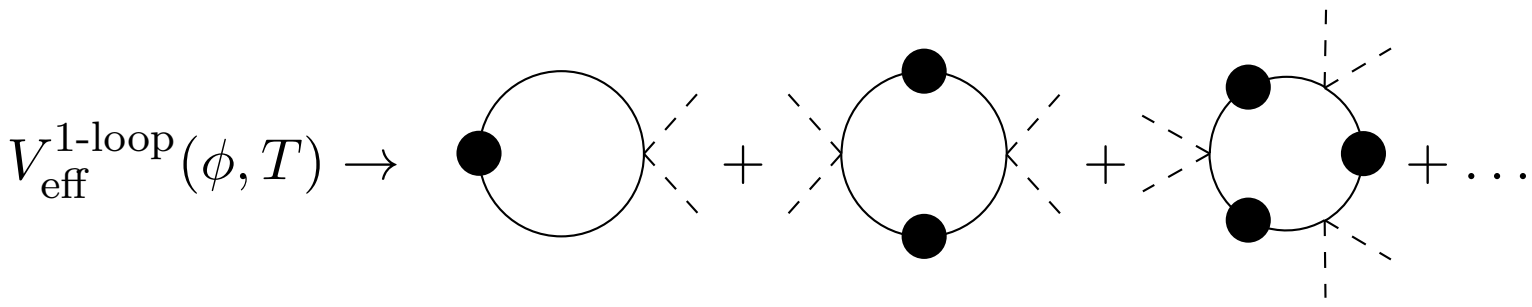

$$
\begin{aligned}
&= \frac{T}{2}\sum_{n\neq 0}\int \frac{\mathrm{d}^3k}{(2\pi)^3}\log\left\{\omega_n^2+|\boldsymbol{k}|^2+m^2(\phi)\right\} \\
&\quad + \frac{T}{2}\int \frac{\mathrm{d}^3k}{(2\pi)^3}\log\left\{|\boldsymbol{k}|^2+m^2(\phi)+\Pi(T)\right\} \\
&= \frac{T}{2}\sum_{n}\int \frac{\mathrm{d}^3k}{(2\pi)^3}\log\left\{\omega_n^2+|\boldsymbol{k}|^2+m^2(\phi)\right\} \\
&\quad + \frac{T}{2}\int \frac{\mathrm{d}^3k}{(2\pi)^3}\log\left\{1+\frac{\Pi(T)}{|\boldsymbol{k}|^2+m^2(\phi)}\right\} \\
&= V_{\text{eff}}^{\text{1-loop}}(\phi,T)+V_{\text{daisy}}(\phi,T)
\end{aligned}
\tag{II.119}
$$

with solid lines symbolizing scalars or gauge bosons. A new term has been separated from the rest of the potential and can be further rewritten as [36]

$$
V_{\text{daisy}}(\phi,T) = -\frac{T}{12\pi}\left[\left(m^2(\phi)+\Pi(T)\right)^{3/2}-\left(m^2(\phi)\right)^{3/2}\right]. \tag{II.120}
$$

Note the minus sign in front of the new contribution, revealing a competition between $V_{\text{daisy}}$ and $V_T$. This can prevent symmetry restoration and change the phase structure, depending on the examined model. Corrections of higher order than the daisy contribution are not included in the analyses of this work, but are briefly reviewed in Section A.3 of the Appendix.

#### Cancellation of cubic terms

The above derived formula for $V_{\text{daisy}}$ has an important implication on the considerations of Section II.2.4, where the outcome was that cubic terms in $V_T$ induce a barrier and can thus render a PT first-order. Expanding in the high-temperature regime, i.e.

$$
\begin{aligned}
V_{\text{daisy}} &= -\frac{T}{12\pi}\left[\left(m^2(\phi)+\Pi\right)^{3/2}-\left(m^2(\phi)\right)^{3/2}\right] \\
&\approx -\frac{T^4}{12\pi}\left[\left(\frac{\Pi}{T^2}\right)^{3/2}+\frac{3}{2}\sqrt{\frac{\Pi}{T^2}}\frac{m^2(\phi)}{T^2}-\left(\frac{m^2(\phi)}{T^2}\right)^{3/2}\right] \\
&\supset \frac{T^4}{12\pi}\left(\frac{m^2(\phi)}{T^2}\right)^{3/2} \\
&\sim \frac{T^4}{12\pi}\left(\frac{\phi^2}{T^2}\right)^{3/2},
\end{aligned}
\tag{II.121}
$$

explicitly reveals the occurrence of a $\phi^3$ term in $V_{\text{daisy}}$. Compared to the high-temperature approximation of $V_T$, this term has the same prefactor with a relative minus sign. This implies a cancellation of the thermally induced barrier. There is however still hope to achieve a first-order PT from thermal effects: The cancellation is not complete for gauge bosons since their transverse polarization modes do not receive thermal mass corrections and thus give no contribution to $V_{\text{daisy}}$. Hence from this point of view, scalar fields with gauge charges are required if a first-order PT is desired.

This level of analysis is however not always sufficient to make reliable predictions about the nature of a PT. Firstly, a thermally induced barrier can always be overpowered by tree-level effects, making the barrier small in comparison. Secondly, the barrier which arose when plugging $m^2(\phi) \sim \phi^2$ into the thermal function $J_b$ can only play a noticeable role if $\phi^2 \sim m^2(\phi) \gtrsim \Pi(T) \sim T^2$ at the critical temperature $T_c$ [34]. In the SM, this condition is related to the general requirement for perturbativity: The order parameter in the broken phase immediately after the PT must be larger than unity [37], i.e.

$$\frac{\phi_c}{T_c} \equiv \frac{\phi_{\min}(T_c)}{T_c} \gtrsim 1. \tag{II.122}$$

### II.2.7 Summary

All the derived contributions to the effective potential can be summarized to

$$V_{\text{eff}}(\phi, T) \approx V_{\text{tree}}(\phi) + V_{\text{CW}}(\phi) + V_{\text{ct}}(\phi) + V_T(\phi, T) + V_{\text{daisy}}(\phi, T) \tag{II.123}$$

with

$$V_{\text{CW}}(\phi) = \sum_a \eta_a n_a \frac{m_a^4(\phi)}{64\pi^2} \left[ \log \frac{m_a^2(\phi)}{\Lambda^2} - C_a \right], \tag{II.124}$$

$$V_T(\phi, T) = \frac{T^4}{2\pi^2} \sum_a \eta_a n_a J_{b/f}\left( \frac{m_a^2(\phi)}{T^2} \right) \tag{II.125}$$

$$\approx T^2 \left[ \sum_b \frac{n_b}{24} m_b^2(\phi) + \sum_f \frac{n_f}{48} m_f^2(\phi) \right], \tag{II.126}$$

$$V_{\text{daisy}}(\phi, T) = -\frac{T}{12\pi} \sum_b n_b^{\text{L}} \left[ \left( m^2(\phi) + \Pi(T) \right)_b^{3/2} - \left( m^2(\phi) \right)_b^{3/2} \right] \tag{II.127}$$

and

$$
\begin{aligned}
n_a &= \#\text{DOF}, \\
n_{b/f} &= \#\text{DOF (bosonic/fermionic)}, \\
n_b^{\mathrm{L}} &= \#\text{DOF (longitudinal bosonic)}, \\
\eta_a &= \begin{cases} +1 & \text{(scalars)} \\ -1 & \text{(fermions)} \end{cases}, \\
\Lambda &= \text{renormalization scale}, \\
C_a &= \begin{cases} {}^3/_2 & \text{(scalars and fermions)} \\ {}^5/_6 & \text{(gauge bosons)} \end{cases}, \\
J_{b/f}\left(x^2\right) &= \int_0^\infty \mathrm{d}y\, y^2 \log\left\{1 \mp \exp\left[-\sqrt{y^2 + x^2}\right]\right\}.
\end{aligned}
$$

When counting DOF, the massless Goldstones have to be counted in addition to the longitudinal gauge boson modes.[8] Note that in the notation here, unlike in some other literature, $n$ is meant to be positive also for fermions and the minus is provided by $\eta$. The expression $(m^2(\phi) + \Pi(T))_b$ in the daisy potential has to be understood as the $b$-th eigenvalue of the full mass matrix. We should keep in mind that the effective potential suffers from gauge-dependence, as already discovered 40 years ago [38]. It was shown more recently that the one-loop potential, formulated in the right way, becomes gauge-independent at least at its minima [39].

## II.3 False Vacuum Decay

As already pointed out in the previous sections, the potential can develop multiple minima with barriers in between. This allows the universe to be trapped in a potential minimum which was formerly a global minimum but at some point turned into a local one. In such a situation of *supercooling*,[9] the *false vacuum* can decay by tunneling to the global minimum. The new phase is then referred to as *true vacuum* or *broken phase* as the acquired VEV usually breaks a gauge symmetry.[10]

The process of false vacuum decay has its analogy in conventional thermodynamics: A liquid in a superheated fluid phase enters the vapor phase by nucleating bubbles. The potential energy of a bubble with radius $r$ is proportional to $-r^3$, whereby the surface tension goes with $+r^2$. Bubble growth will thus only occur for radii above a certain threshold. Below the threshold, the bubbles shrink to nothingness and the PT cannot proceed [40]. In the cosmological analogue, bubbles are spacial volumes with non-zero VEV which nucleate at random places in the universe and subsequently expand (see Fig. II.10).

[8] Despite the Goldstone equivalence theorem, this is no double counting. A reasoning is given in [32].

[9] The term 'supercooling' is sometimes used to specifically emphasize a strongly delayed decay of the false vacuum with $T_\mathrm{n} \ll T_\mathrm{c}$. This can lead to a phase of vacuum domination and cause a 'mini inflation'.

[10] The phrase 'broken phase' will be used to describe the energetically favorable state, even when talking about PTs without symmetry breaking.

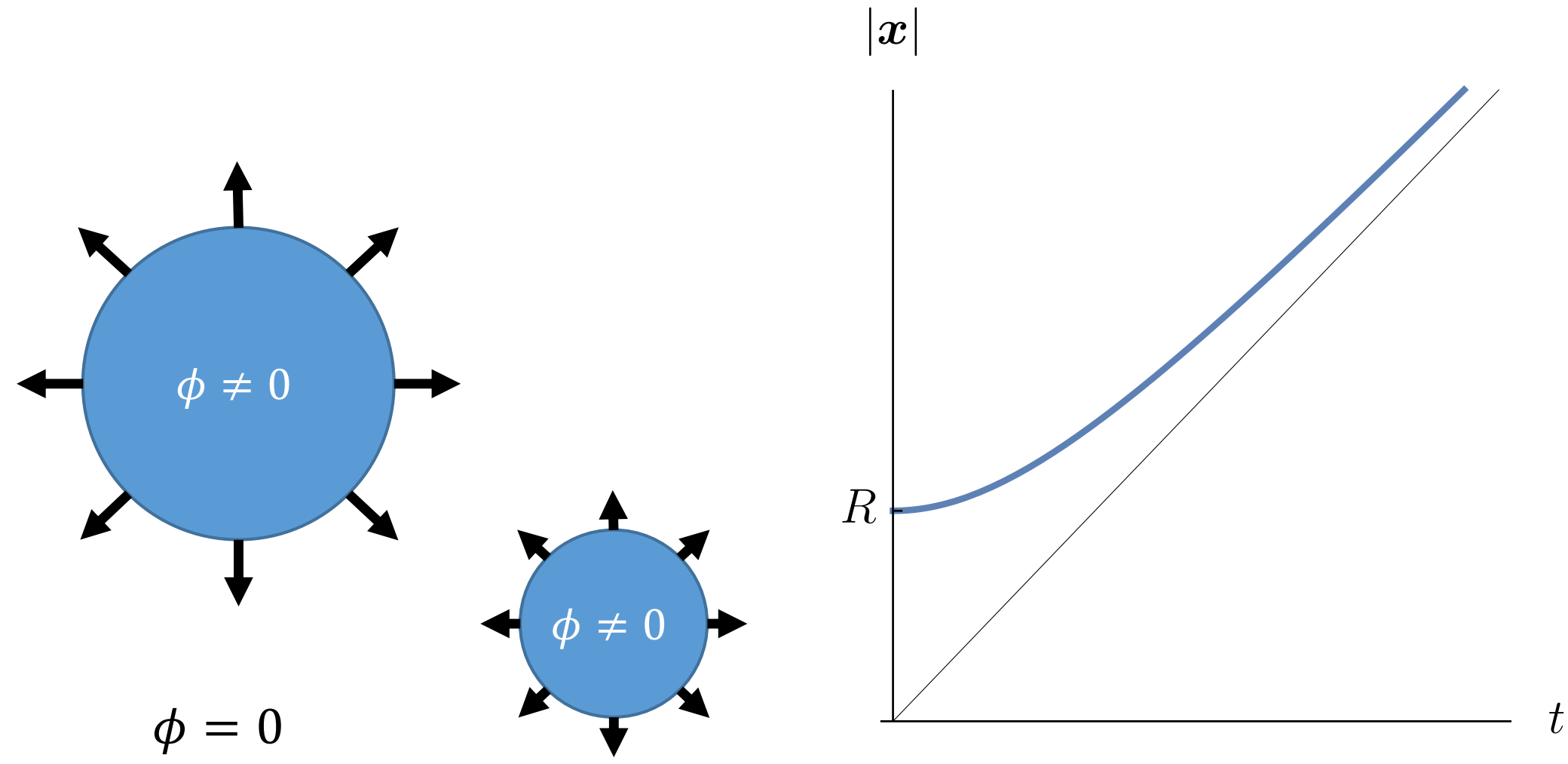

Figure II.10: Bubbles of broken phase nucleate and expand.

Figure II.11: The dynamics of a bubble wall describe a space-time hyperbola.

### II.3.1 Bubble Formation

In the following, consider a potential with two minima, one at $\phi = 0$ and another at $\phi > 0$. The *nucleation rate* per unit volume for cosmological bubbles at zero temperature is given by

$$\Gamma = \mathcal{A} e^{-S[\phi]} \tag{II.128}$$

with $\mathcal{A}$ of energy dimension 4 and *Euclidean action*

$$S[\phi] = \int \mathrm{d}^4 x_\mathrm{E} \left[ \frac{1}{2} \left( \frac{\mathrm{d}\phi}{\mathrm{d}\tau} \right)^2 + \frac{1}{2} (\nabla \phi)^2 + V_\mathrm{eff}(\phi) \right] . \tag{II.129}$$

Applying the principle of extremal action yields the Klein-Gordon equation in presence of a classical potential

$$\frac{\mathrm{d}^2 \phi}{\mathrm{d}\tau^2} + \Delta \phi = \frac{\mathrm{d}V_\mathrm{eff}}{\mathrm{d}\phi} \equiv V'_\mathrm{eff}(\phi) \tag{II.130}$$

with boundary conditions $\phi(\varrho \to \infty) \to 0$ and $\phi'(\varrho = 0) = 0$, where $\varrho \equiv \sqrt{\tau^2 + |\boldsymbol{x}|^2}$. The $\phi$ in (II.128) is understood to be a solution of (II.130) and thus represents the shape of a nucleated bubble, i.e. the VEV profile of field $\phi$ as function of space and time. It turns out that the solutions which minimize the action are typically $O(4)$ symmetric [40]. The equation of motion can thus be simplified to

$$\frac{\mathrm{d}^2\phi}{\mathrm{d}\varrho^2} + \frac{3}{\varrho}\frac{\mathrm{d}\phi}{\mathrm{d}\varrho} = V'_{\mathrm{eff}}(\phi). \tag{II.131}$$

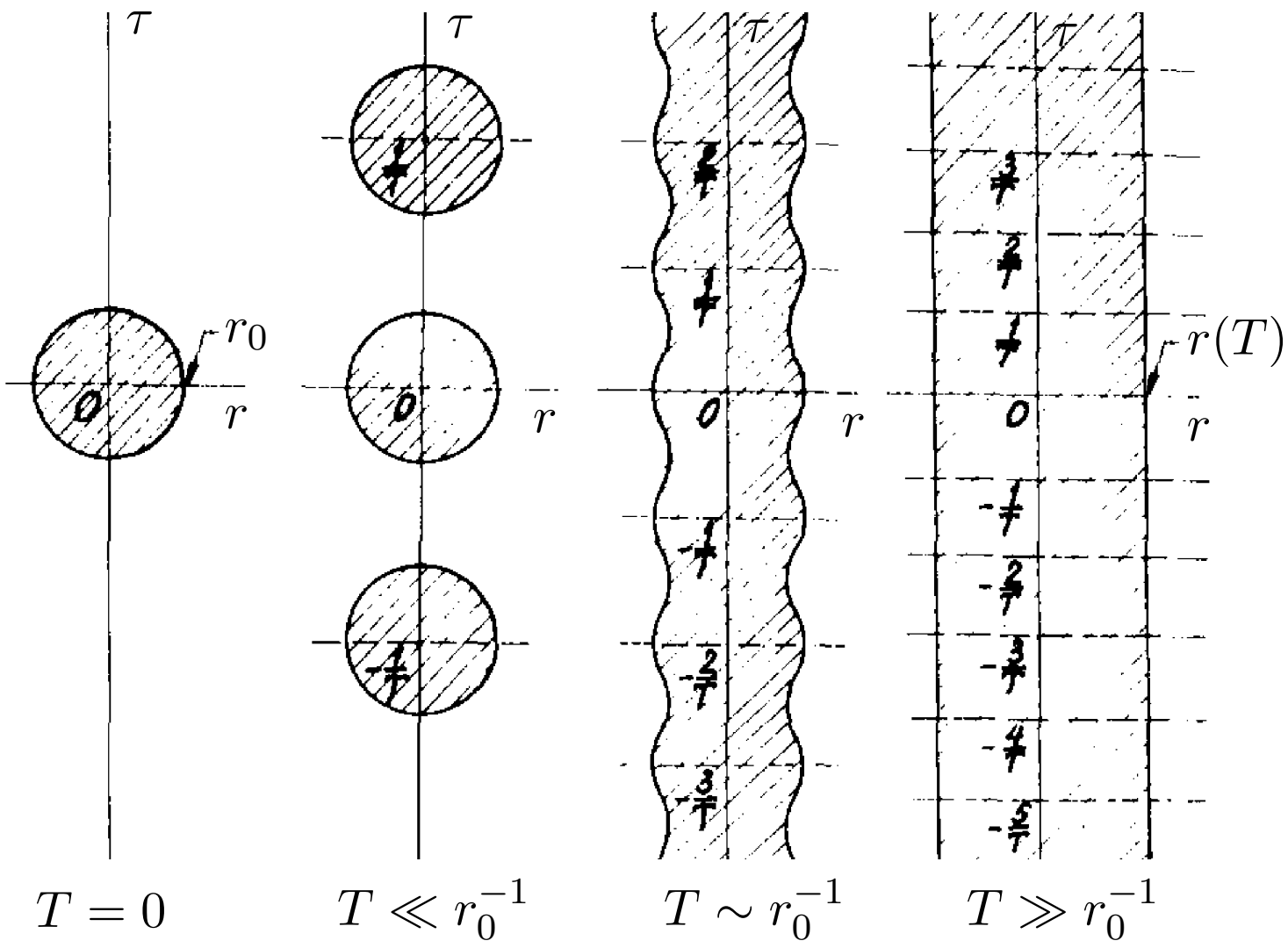

Figure II.12: Bubble solution in the $r$-$\tau$-plane for different temperatures. Figure taken from [42].

The so-called *bounce solution* of (II.131) can be obtained analytically under certain circumstances [41], but is usually calculated numerically. An attempt of applying the under-/overshoot method is presented in Section A.4 of the Appendix. Multiple solutions can coexist, e.g. for a potential with more than two minima. In this case, the solution with smallest action dominates the nucleation rate.

The radius of a bubble be can be defined, for instance, by the points in its wall profile where half of the final VEV is reached. A $O(4)$ symmetric solution implies that these points form a 4-dimensional sphere of radius $\rho = R$ in Euclidean space. The radius in 3-space is hence given by

$$r \equiv |\boldsymbol{x}| = \sqrt{R^2 + t^2} \tag{II.132}$$

which describes a hyperbola in Minkowski spacetime (see Fig. II.11). From this we can infer that bubbles have an initial radius of $R$ and expand subsequently, with bubble walls approaching the speed of light.

### II.3.2 Thermal Tunneling

To make the step towards thermal field theory, recall the KMS relation (II.88) which states that fields are periodic in imaginary time, with periodicity $T^{-1}$. Fig. II.12 pictures how temperature affects the bounce solution: The bubble profile does repeat itself on the $\tau$-axis in intervals of $T^{-1}$. With increasing temperature, the bubbles of radius $r_0$ move closer to each other in the $r$-$\tau$-plane until they start to merge at $T = r_0^{-1}$. For very high temperatures, the solution becomes a cylinder, the spacial cross section of which is a $O(3)$-symmetric bubble with radius $r(T) > r_0$. As a consequence, the imaginary time integral reduces to $T^{-1}$ and we are left with [42]

$$S[\phi, T] = \frac{S_3[\phi, T]}{T} = \frac{1}{T} \int \mathrm{d}^3 x \left[ \frac{1}{2} (\nabla\phi)^2 + V_{\text{eff}}(\phi, T) \right]. \tag{II.133}$$

Demanding stationary action yields the $O(3)$-symmetric bounce equation

$$\frac{\mathrm{d}^2\phi}{\mathrm{d}r^2} + \frac{2}{r}\frac{\mathrm{d}\phi}{\mathrm{d}r} = V'_{\text{eff}}(\phi, T) \tag{II.134}$$

with boundary conditions $\phi(r \to \infty) \to 0$ and $\phi'(r = 0) = 0$ and the *thermal bubble nucleation rate* is given by

$$\Gamma(T) = \mathcal{A}(T) e^{-S_3[\phi, T]/T} \tag{II.135}$$

where $\mathcal{A}(T) \sim T^4$ for dimensional reasons [42].

## II.4 Used Software

For the calculations and analyses in this work, special software packages were employed. Since much of the effort for this thesis went into applying, modifying and extending these programs, it seems sensible to briefly introduce them.

### II.4.1 CosmoTransitions

`CosmoTransitions` [43] is a `Python` package allowing the numerical analysis of cosmological PTs driven by the effective potentials of scalar fields at finite-temperature. In the physics community, it is a widely used software when it comes to studies of baryogenesis or GWs in the context of first-order PTs.

Before running the tool, one has to implement the model of interest in the form of `Python` code. The needed ingredients are the tree-level potential, the particle mass spectrum, Debye masses and some additional information such as the reheating temperature, underlying scalar symmetries and so forth. `CosmoTransitions` calculates the effective finite-temperature one-loop potential by using the same formulae that were derived in Section II.2. The program then determines all possible phases by tracking the minima while scanning over temperature. In a next step, it tracks the VEV starting from $T = 0$ and going up to the defined maximum temperature, i.e. the reheating temperature. Whenever multiple phases coexist in a certain temperature range, the nucleation criterion is continuously evaluated to find the exact point $T_\text{n}$ of a PT. To do so, the program repeatedly solves the bounce equation (II.134) numerically and uses the solution to calculate the action according to (II.133). In a multidimensional scalar field space, `CosmoTransitions` is capable of minimizing the action as well. In this case, the tunneling path is divided into a number of linear segments and is deformed iteratively.

#### Own extensions

To allow for a GW analysis, the calculation of the characteristic parameters had to be implemented. Furthermore, additional modules were added, e.g. for the automation of model parameter scans or for the output of *trace files* including information about the thermal evolution of VEVs and particle masses. The added code is designed to work with any given model and also allows for a convenient exploitation of multi-core processing capabilities.

### II.4.2 SARAH

The `Mathematica` package `SARAH` [44] is a tool for SUSY but also non-SUSY model building. After entering only the very basic features of a model such as gauge symmetries, representations, particle content and the Lagrangian, `SARAH` is able to calculate vertices, mass matrices, tadpole equations, self-energies and two-loop RGEs. A variety of `SARAH` sub-packages allow the generation of output for other tools (`FeynArts`, `SPheno`, `Vevacious`,...) and further analysis.

#### Own extensions

Since the implementation of models with lengthy mass spectra in `CosmoTransitions` is tedious and error prone, it seems helpful to have a tool for that. `SARAH` already delivers most of the information needed as input for `CosmoTransitions`, it just has to be brought into the right form. A draft of a `SARAH` extension which outputs `CosmoTransitions` files for arbitrary models has been created in the course of this work. It is planned to improve and refine the software in the future.

# III Gravitational Waves from Bubble Collisions

The decay of the false vacuum ends after the whole universe arrived in the new, energetically favorable state. As a consequence, the nucleated bubbles of broken phase have to collide and merge at some point. This is an anisotropic process capable of sourcing GWs. In order to characterize the resulting stochastic GW spectrum, it is useful to introduce four characteristic parameters.

## III.1 Characteristic Parameters

### III.1.1 Nucleation Temperature $T_\mathrm{n}$

The *nucleation temperature* $T_\mathrm{n}$ marks the onset of the first-order PT. As soon as this temperature is reached by the cooling universe, emerging bubbles successfully expand instead of collapsing immediately. To determine $T_\mathrm{n}$ numerically, one has to take into account the expansion of the universe which competes with the bubble nucleation rate. Nucleation proceeds as soon as the integrated number of bubbles generated in a Hubble volume $H^{-3}$ reaches one, i.e. [45]

$$\begin{aligned} 1 &\sim \int_{-\infty}^{t_n} \Gamma(t) H^{-3} \mathrm{d}t \\ &= \int_{T_\mathrm{n}}^{\infty} \Gamma(T) H^{-4} \frac{\mathrm{d}T}{T} \\ &= \int_{T_\mathrm{n}}^{\infty} \left( \frac{3}{\pi} \sqrt{\frac{30}{g}} \frac{M_\mathrm{P}}{T} \right)^4 e^{-S_3/T} \frac{\mathrm{d}T}{T} \end{aligned} \tag{III.1}$$

where the thermal bubble nucleation rate (II.135) with $\mathcal{A}(T) \sim T^4$ was used. Furthermore, the first Friedmann equation $H = \sqrt{\rho}/(3M_\mathrm{Pl})$ together with $\rho = \pi^2 g^2 T^4/30$ was employed, assuming a flat and radiation dominated universe at the time of bubble nucleation. Evaluating (III.1) for $g \approx 100$, as in the SM, yields the *nucleation condition* [30]

$$\frac{S_3}{T_\mathrm{n}} \sim 140 - 4 \log \left[ \frac{T_\mathrm{n}}{100\,\mathrm{GeV}} \right]. \tag{III.2}$$

The simpler approach of just demanding $\Gamma(T_\mathrm{n}) H_\mathrm{n}^{-4} \sim 1$ is sometimes presented in the literature and yields a comparable result.

The nucleation temperature $T_{\mathrm{n}}$ does not necessarily have to coincide with the temperature at bubble collision $T_*$. In an epoch of vacuum domination, e.g. due to sizable supercooling, the decay of the false vacuum leads to a reheating of the universe resulting in $T_* \gg T_{\mathrm{n}}$ [46]. This scenario will only be briefly considered in Section III.3.3, but as it is not applicable to the models in this work, we will assume $T_* = T_{\mathrm{n}}$ from this point on.

### III.1.2 Phase Transition Strength $\alpha$

A measure for the strength of a PT is naturally given by the amount of released energy. The *latent heat density* is the energy liberated during the PT and is given by the difference in the free energy density $-\Delta f > 0$ minus the energy used for the change of entropy density $\Delta s > 0$, i.e. [47]

$$\begin{aligned} \epsilon \equiv -\Delta\rho &= -\Delta f - T_{\mathrm{n}}\Delta s \\ &= -\Delta V_{\mathrm{eff}} + T_{\mathrm{n}} \left.\frac{\partial V_{\mathrm{eff}}}{\partial T}\right|_{T_{\mathrm{n}}} > 0. \end{aligned} \tag{III.3}$$

The liberated energy is partly injected into the surrounding plasma, while the remainder goes into the acceleration of the bubble walls. To make this quantity comparable between different scenarios, it can be normalized with respect to the total energy density of the unbroken phase at the time of collision, i.e.

$$\alpha \equiv \frac{\epsilon}{\rho_{\mathrm{rel}}} \tag{III.4}$$

where the assumption of a radiation-dominated universe was made. The relevance of this parameter for the resulting GWs is obvious, as it relates to the amount of energy available for their production.

### III.1.3 Inverse Time Scale $\beta$

Shortly before and during the PT, the variation in the bubble nucleation rate (II.135) is mostly attributable to the change of action $S$ with time [48]. This quantity is representing a measure for the *inverse time scale* of the transition and is defined as

$$\beta \equiv - \left.\frac{\mathrm{d}S}{\mathrm{d}t}\right|_{t_{\mathrm{n}}} \approx \left.\frac{1}{\Gamma}\frac{\mathrm{d}\Gamma}{\mathrm{d}t}\right|_{t_{\mathrm{n}}} \tag{III.5}$$

such that

$$\Gamma(t) \approx e^{\beta t}.$$

In cosmology, it is often more meaningful to express time scales in proportion to Hubble time $H^{-1}$. This factor arises naturally when switching from time to temperature.

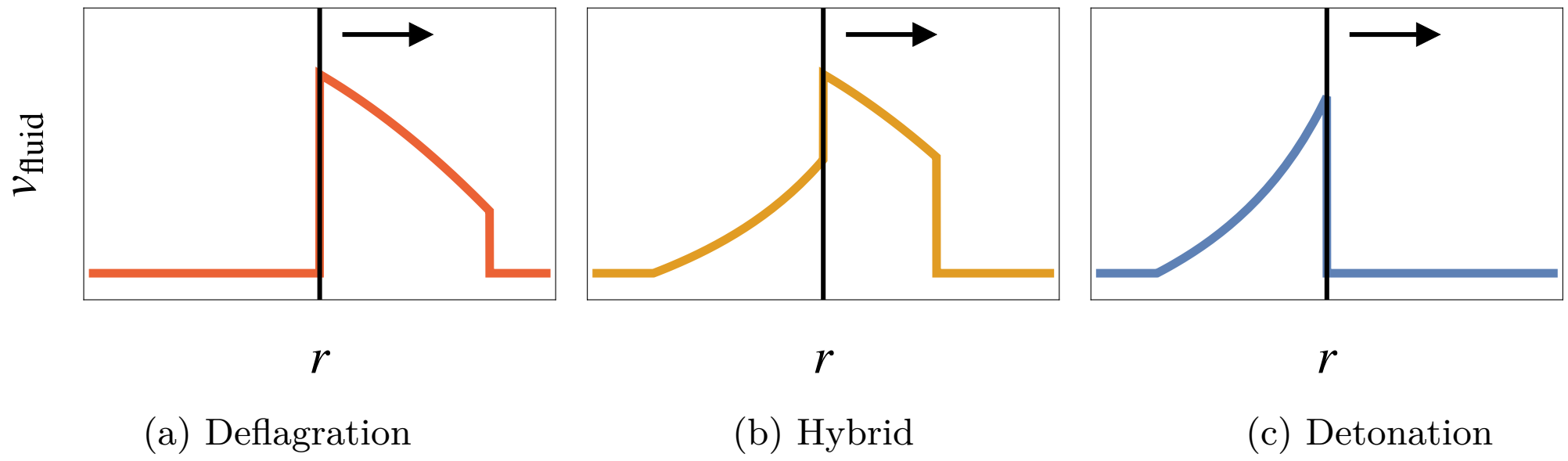

(a) Deflagration (b) Hybrid (c) Detonation

Figure III.1: Schematic illustration of the fluid velocity profile in three different regimes. The wall of the expanding bubble, marked by the black line, is moving towards higher radii $r$.

One can thus write

$$\frac{\beta}{H} = T_{\text{n}} \frac{\mathrm{d}}{\mathrm{d}T} \left( \frac{S_3}{T} \right) \Big|_{T_{\text{n}}} \tag{III.6}$$

which is dimensionless and depends on the shape of $V_{\text{eff}}$ at the time of nucleation. Two derived quantities can be inferred from the time scale $\beta^{-1}$ together with the bubble wall velocity $v_{\text{w}}$: $R \sim v_{\text{w}}\beta^{-1}$ is the approximate bubble radius at the time of collision and $\Gamma\beta^{-1}$ is the spatial bubble density.

### III.1.4 Bubble Wall Velocity $v_{\text{w}}$

The *bubble wall velocity* $v_{\text{w}}$ is perhaps the most difficult parameter one has to determine. It depends firstly on the liberated energy, measured by $\alpha$, since the acceleration of bubble walls requires energy. Secondly, the coupling between bubble wall (i.e. the scalar field) and surrounding particle plasma plays a significant role, due to the friction between the two. Note that this is highly model dependent. In scenarios with friction, part of the bubble wall's energy goes into the bulk motion of the plasma, dragging the particles along. For the analysis of this process, the surrounding particles are typically described as a relativistic fluid, in which density waves travel with velocity $c_{\text{s}} = 1/\sqrt{3} \approx 0.58$ [49].

Depending on $v_{\text{w}}$, the behavior of the fluid with respect to the bubble wall (see Fig. III.1) can be categorized as follows [47]. In the case of subsonic bubble walls, i.e. $v_{\text{w}} < c_{\text{s}}$, a shock-front of increased density develops in the plasma and precedes the bubble wall. After the PT front has passed in this *deflagration* scenario, particles are at rest again. For supersonic bubble walls, the underlying process is a *detonation*. The wall now travels faster than the density waves, preventing a shock front from building up. Particles in the fluid are at rest when the bubble wall hits and accelerates them. Behind the wall, the particles slow down again and form a rarefaction wave. For $v_{\text{w}} \gtrsim c_{\text{s}}$, a *hybrid* combination of both types can exist, featuring a shock-front and a rarefaction wave.

A common treatment for the determination of $v_\mathrm{w}$ is to assume a bubble expansion behavior similar to the one of a regular chemical combustion. This so-called *Chapman-Jouguet detonation* [50] marks the boundary between the detonating and the hybrid regime. It was shown however, that this assumption is not always realistic for cosmological PTs, especially in the case of small supercooling [51, 52]. A more advanced approach is to consider the model dependent couplings, calculate the resulting friction coefficients and solve Boltzmann-type equations [47]. If a PT is strong or if it proceeds during a vacuum-dominated epoch, bubble walls can and will reach luminal velocities $v_\mathrm{w} \sim 1$ [53].

Whilst subsonic wall velocities are attractive for baryogenesis as they give enough time to build up baryon asymmetry, higher velocities are favorable for a strong GW spectrum because they result in more violent collisions. The focus of the subsequent analyses will be on the physics behind the PT and not on the bubble wall dynamics. Following the agenda of considering what is optimistically possible, $v_\mathrm{w} \sim 1$ will be assumed in the remainder of this work.

## III.2 Bubble Wall Speed Limit

In the vacuum, the energy liberated by the PT goes completely into the acceleration of the PT front, i.e. the bubble wall velocity approaches light speed and its relativistic $\gamma$-factor grows with propagation distance. Bubble walls with this behavior belong to the *runaway regime* [54]. If, on the other hand, the involved scalar field is coupled to the surrounding plasma, collisions with the bubble wall act like a friction term. The total forward driving pressure is then given by

$$P_\mathrm{tot} = -\Delta V_\mathrm{tree} - P_\mathrm{fric} \tag{III.7}$$

where $\Delta V_\mathrm{tree} = V_\mathrm{tree}(\phi_1) - V_\mathrm{tree}(\phi_0) < 0$ is the difference in the tree-level potential between unbroken and broken phase with $\phi_0 < \phi_1$.

In the following, consider the bubble wall as a plane moving through the plasma in the $z$-direction. When passing a particle $a$ with relative momentum $p_z$, the wall feels a retaining force $F_a$ due to the changing scalar field $\phi$. Using energy and momentum conservation in the $x, y$-direction

$$0 = \frac{\mathrm{d}E^2}{\mathrm{d}t} = 2p_z \frac{\mathrm{d}p_z}{\mathrm{d}t} + \frac{\mathrm{d}m^2}{\mathrm{d}t} \tag{III.8}$$

and $\frac{\mathrm{d}z}{\mathrm{d}t} = \frac{\gamma m v_z}{\gamma m} = \frac{p_z}{E}$, this force can be expressed as

$$F_a = -\frac{\mathrm{d}p_z}{\mathrm{d}t} = \frac{1}{2p_z}\frac{\mathrm{d}m_a^2}{\mathrm{d}t} = \frac{1}{2E_a}\frac{\mathrm{d}m_a^2}{\mathrm{d}z} = \frac{1}{2E_a}\frac{\mathrm{d}m_a^2}{\mathrm{d}\phi}\frac{\mathrm{d}\phi}{\mathrm{d}z}. \tag{III.9}$$

To obtain the total force per wall unit area, i.e. $P_\mathrm{fric}$, one has to integrate over the wall width, sum over all particles DOF $n_a$ coupling to $\phi$ and integrate over their momentum distributions $f_a$ [54]:

$$
\begin{aligned}
P_{\text{fric}} &= \frac{1}{A}\int \mathrm{d}^3x \sum_a n_a \int \frac{\mathrm{d}^3p}{(2\pi)^3} F_a\, f_a(\boldsymbol{p}, z) \\
&= \int \mathrm{d}z \frac{\mathrm{d}\phi}{\mathrm{d}z} \sum_a n_a \frac{\mathrm{d}m_a^2}{\mathrm{d}\phi} \int \frac{\mathrm{d}p\, p^2}{(2\pi)^2 E_a} f_a(p, z)
\end{aligned}
\tag{III.10}
$$

In the limit $\gamma \gg 1$, the kinetic energy of plasma particles relative to the bubble wall is almost always high enough to pass the wall, i.e. no reflections take place. Furthermore, a highly relativistic bubble wall appears Lorentz contracted by a factor $\sim \gamma^{-1}$ to the surrounding plasma. One can assume all particles to be in equilibrium because the wall approaches faster than any signal that could be received by the plasma in advance. In this limit, effectively all particles pass the bubble wall in one direction. With these approximations, the friction is directly given by the total momentum change

$$
P_{\text{fric}} \approx \sum_a n_a \int \frac{\mathrm{d}^3p}{(2\pi)^3} \left[p_{z,\text{in}} - p_{z,\text{out}}\right] f_{a,\text{in}}(\boldsymbol{p}) \tag{III.11}
$$

where 'in' ('out') labels quantities concerning incoming (outgoing) particles. Considering energy conservation again and neglecting the rest masses, i.e.

$$
\sqrt{p_{z,\text{in}}^2 + m^2(\phi_0)} = \sqrt{p_{z,\text{out}}^2 + m^2(\phi_1)} \tag{III.12}
$$

$$
\Leftrightarrow \qquad p_{z,\text{in}} - p_{z,\text{out}} \approx \frac{m^2(\phi_1) - m^2(\phi_0)}{2E}, \tag{III.13}
$$

the friction becomes

$$
\begin{aligned}
P_{\text{fric}} &\approx \sum_a n_a \left[m_a^2(\phi_1) - m_a^2(\phi_0)\right] \int \frac{\mathrm{d}^3p}{(2\pi)^3 2E_a} f_{a,\text{in}}(\boldsymbol{p}) \\
&\approx T^2 \left[\sum_b \frac{n_b}{24} m_b^2(\phi) + \sum_f \frac{n_f}{48} m_f^2(\phi)\right]_{\phi_0}^{\phi_1} \\
&\approx \Delta V_T .
\end{aligned}
\tag{III.14}
$$

The result is equivalent to the change of the thermal one-loop potential $V_T$ in the $T \gg m$ expansion. As a consequence, the total forward pushing pressure is given by the change of the effective potential:[1]

$$
P_{\text{tot}} \approx -\Delta V_{\text{tree}} - \Delta V_T \approx -\Delta V_{\text{eff}} \tag{III.15}
$$

This makes sense, because if we assume a bubble to nucleate at the critical temperature $T_{\text{c}}$ where $\Delta V_{\text{eff}} = 0$ by definition, we expect the forces in both directions to balance and the bubble wall to stay at rest. The result also matches a basic relation of thermodynamics which equates pressure with the negative free energy density for the canonical ensemble.

[1] We are working in the high-temperature expansion, the temperature independent Coleman-Weinberg contribution can therefore safely be neglected.

The particle density scales as $\sim \gamma$ due to Lorentz contraction, while $\Delta p = \frac{\Delta m^2}{2E}$ scales as $\sim \gamma^{-1}$. Together, this results in a finite large-$\gamma$ limit of $P_{\text{fric}}$ and the bubbles will always run away if they become relativistic. One can determine whether a PT features runaway bubbles or not by comparing the latent heat $\epsilon$ to the friction term: If

$$\epsilon > P_{\text{fric}} \approx \Delta V_T,$$

one would expect the bubble walls to run away [47]. The released energy in that case is enough to overcome the friction and further accelerate the wall. This condition can be expressed in terms of the PT strength parameter $\alpha$ by normalizing for the total relativistic energy density at the time of the transition. The *runaway condition* thus reads

$$\alpha \equiv \frac{\epsilon}{\rho_{\text{rel}}} > \frac{\Delta V_T}{\rho_{\text{rel}}} \equiv \alpha_\infty \tag{III.16}$$

with

$$\alpha_\infty \equiv \frac{30}{\pi^2 g T_{\text{n}}^2} \left[ \sum_b \frac{n_b}{24} \Delta m_b^2 + \sum_f \frac{n_f}{48} \Delta m_f^2 \right]. \tag{III.17}$$

This expression is meant to sum over the mass changes of all physical particle DOF, including the longitudinal gauge boson modes but not the Goldstones [47]. Note that in the earlier derived effective potential, both had to be included.

#### Friction due to transition radiation

Besides the changing momentum of the plasma particles, as explained in the previous section, there is additional friction caused by so-called *transition radiation*, which may limit the acceleration of bubble walls. Transition radiation receives contributions from any splitting process in which the coupling strength or the participating masses change during the PT. Pictorially speaking, transition radiation occurs because the 'radiation clouds' on both sides of the bubble wall are not the same such that the 'difference' simply radiates away.

As shown in [53], the main contribution to the radiation induced friction comes from gauge bosons with phase-dependent masses and scales as $\sim \gamma g^2 m T^3$ with gauge coupling $g$ and gauge boson mass $m$ in the broken phase. Since the friction term now increases linearly with $\gamma$, the bubble wall will stop to accelerate at $\gamma \sim \frac{4\pi}{g^2} < \infty$. Under these circumstances, the runaway regime is thus ruled out. As a consequence, bubble walls can still be close to $v_{\text{w}} \sim 1$ but do not become ultra-relativistic, meaning the energy stored in the bubble walls is limited. Since the derivative couplings of gauge bosons make up the main contribution to transition radiation, PTs involving gauge singlet scalars are not affected by the above arguments and can still live in the runaway regime.

## III.3 The Gravitational-Wave Spectrum

### III.3.1 Scaling Estimate

In order to quantify the GW spectrum, a scale independent quantity is best suited. We therefore use the relative energy density of the produced GWs, denoted as

$$\Omega_{\mathrm{GW}} \equiv \frac{\rho_{\mathrm{GW}}}{\rho_{\mathrm{crit}}}.$$

In the following we will investigate how his this quantity scales with the parameters introduced in the preceding sections, based on [48]. As shown in Section II.1.5, the radiated power or luminosity of a GW source is given by

$$L = \frac{G}{5} \left\langle (\dddot{Q}_{ij}^{\mathrm{TT}})^2 \right\rangle. \tag{III.18}$$

With the definition of the quadrupole tensor (II.60) one can determine the dimension of its third time derivative to be

$$\dddot{Q}_{ij}^{\mathrm{TT}} \sim \frac{\text{mass} \times \text{length}^2}{\text{time}^3} \sim \frac{\text{kinetic energy}}{\text{time}} \tag{III.19}$$

such that luminosity takes the qualitative form $L \sim G\dot{E}_{\mathrm{kin}}^2$. The relevant time scale is $\tau = \beta^{-1}$ while the length scale is given by the bubble radius $R = v_{\mathrm{w}}\beta^{-1}$ at the time of collision. The kinetic energy here refers to energy in the bubble wall and in the bulk motion of the fluid, as opposed to the energy that just heats up the plasma. In other words, not all of a PT's latent heat density $\epsilon$ is available for GWs. We therefore introduce an *efficiency factor* $\kappa \leq 1$ such that the kinetic energy is given by

$$E_{\mathrm{kin}} = \kappa\, \epsilon\, V \sim \kappa\, \alpha \rho_{\mathrm{rel}}\, v_{\mathrm{w}}^3 \beta^{-3} \tag{III.20}$$

with volume $V \sim R^3$. Due to dimensionality $\frac{\mathrm{d}}{\mathrm{d}t} \sim \frac{\mathrm{d}}{\mathrm{d}(\beta^{-1})}$ such that the time derivative of the kinetic energy is

$$\dot{E}_{\mathrm{kin}} \sim \kappa\, \alpha \rho_{\mathrm{rel}}\, v_{\mathrm{w}}^3 \beta^{-2} \tag{III.21}$$

implying

$$L \sim G(\kappa\, \alpha \rho_{\mathrm{rel}}\, v_{\mathrm{w}}^3 \beta^{-2})^2. \tag{III.22}$$

The GW energy density can be written as

$$\rho_{\mathrm{GW}} = \frac{E_{\mathrm{GW}}}{V} \sim \frac{L\tau}{R^3} \sim \frac{L\beta^{-1}}{v_{\mathrm{w}}^3 \beta^{-3}}. \tag{III.23}$$

Together with the Friedmann equation $G \sim \frac{H^2}{\rho_{\mathrm{rel}}}$ and $\rho_{\mathrm{crit}} = \rho_{\mathrm{rel}} + \epsilon = (1+\alpha)\rho_{\mathrm{rel}}$ we have

$$\begin{aligned} \Omega_{\mathrm{GW}} \equiv \frac{\rho_{\mathrm{GW}}}{\rho_{\mathrm{crit}}} &\sim \frac{1}{\rho_{\mathrm{crit}}} \frac{H^2}{\rho_{\mathrm{rel}}} \kappa^2 \alpha^2 \rho_{\mathrm{rel}}^2 v_{\mathrm{w}}^3 \beta^{-2} \\ &= \left( \frac{\kappa\alpha}{1+\alpha} \right)^2 \left( \frac{H}{\beta} \right)^2 v_{\mathrm{w}}^3 \end{aligned} \tag{III.24}$$

which hints at strong signals for large $\alpha$ and small $\frac{\beta}{H}$, i.e. for violent and long-lasting PTs. A careful analysis of the PT's energy budget yields an efficiency factor of

$$\kappa(\alpha) = \frac{\alpha}{0.73 + 0.083\sqrt{\alpha} + \alpha} \tag{III.25}$$

for luminal wall velocities [47].

Finally, a statement about the relative energy density *today* should be made, since this is the one we can eventually measure. To account for the redshift due to the expansion of the universe since production, consider energy conservation $\rho_{\mathrm{GW},0} = (a/a_0)^4 \rho_{\mathrm{GW}}$ and $\rho_{\mathrm{crit},0} \sim (H_0/H)^2 \rho_{\mathrm{crit}}$ due to the Friedmann equation, with subscript '0' marking today's quantities. We further take the conservation of entropy $sa^3 \sim g_{\mathrm{rel},s} T^3 a^3$ into account.[2] Redshifting the relative energy density yields [55]

$$\begin{aligned} \Omega_{\mathrm{GW}} &\to \left( \frac{a}{a_0} \right)^4 \left( \frac{H}{H_0} \right)^2 \Omega_{\mathrm{GW}} \\ &= \left( \frac{T_0}{T} \right)^4 \left( \frac{g_{\mathrm{rel},s,0}}{g_{\mathrm{rel},s}} \right)^{4/3} \frac{\pi^2 T^4 g_{\mathrm{rel}}}{90 M_{\mathrm{pl}}^2 H_0^2} \Omega_{\mathrm{GW}} \\ &\approx 1.66 \times 10^{-5} h^{-2} \times \left( \frac{100}{g_{\mathrm{rel}}} \right)^{1/3} \Omega_{\mathrm{GW}} \end{aligned} \tag{III.26}$$

where $H_0 = h \times 2.13 \times 10^{-42}$ GeV, $T_0 = 2.43 \times 10^{-13}$ GeV was used and $H$ is given by the flat, radiation dominated Friedmann equation. The relativistic entropy DOF amount to $g_{\mathrm{rel},s,0} = 3.91$ today [11]. Remarkably, the result is not temperature dependent, except for small adjustments due to $g_{\mathrm{rel}}(T)$. Analogously to $\Omega_{\mathrm{GW}}$, the peak frequency $f_{\mathrm{p}}$ of a given spectrum also experiences a redshift

$$f_{\mathrm{p}}' = \frac{H'}{H} f_{\mathrm{p}} \tag{III.27}$$

[2] Entropy and energy DOF are not equal if decoupled relativistic species exist (like the neutrinos at late times). We assume here that this is not the case at the time of collision and thus $g_{\mathrm{rel},s} = g_{\mathrm{rel}}$.

with shifted Hubble rate

$$\begin{aligned} H' &= \frac{a}{a_0} H = \frac{T_0}{T} \left( \frac{g_{\mathrm{rel},s,0}}{g_{\mathrm{rel}}} \right)^{1/3} H \\ &\approx 16.5\,\mu\mathrm{Hz} \times \left( \frac{T}{100\,\mathrm{GeV}} \right) \left( \frac{g_{\mathrm{rel}}}{100} \right)^{1/6} . \end{aligned} \tag{III.28}$$

### III.3.2 Contributions

With a reasonable estimate at hand, the different processes involved in the production of GWs will now be introduced. The last section estimated the total relative energy density $\Omega_{\mathrm{GW}}$. To examine the spectrum as a function of frequency, we will consider the quantity[3]

$$\Omega_{\mathrm{GW}}(f) \equiv \frac{\mathrm{d}\Omega_{\mathrm{GW}}}{\mathrm{d}\log f} .$$

The exact spectral shapes, peak frequencies and the scaling with the characteristic parameters is obtained by combinations of analytical and numerical methods as well as simulations. The results indicate that a general spectrum can be parameterized as

$$h^2 \Omega_{\mathrm{GW}}(f) = \Omega_0 \left( \frac{100}{g_{\mathrm{rel}}} \right)^{1/3} \Delta \left( \frac{\kappa \alpha}{1+\alpha} \right)^{a} \left( \frac{H}{\beta} \right)^{b} s(f) \tag{III.29}$$

where $\Omega_0$, $\kappa$, $a$, $b$, velocity factor $\Delta$ and power law spectral shape $s(f)$ differ for the three different contributions [46], which are explained in the following:

**Scalar contribution** ($\Omega_\phi$) refers to the initial collision of the bubble walls, i.e. of the scalar field gradient itself. The spectral properties are usually obtained in the *envelope approximation*, which assumes sizable interactions only at the intersection points and a quick dispersion after collision.

**Sound waves** ($\Omega_{\mathrm{sw}}$) or 'density waves' are induced as the bubble wall passes through the plasma and collide at some later point as well. Compared to the scalar contribution, the effect of colliding sound waves lasts longer and is therefore enhanced by a factor of $\frac{\beta}{H}$. Note that it currently remains unclear whether these results apply for $\alpha > 0.1$ [46].

**Turbulence** ($\Omega_{\mathrm{turb}}$) or, due to the ionized plasma, *magnetohydrodynamic turbulence* (MHD) occurs after sound wave collisions and is powered by the same energy budget. Turbulences last for several Hubble times, hence the occurrence of $H'$ in the spectral shape $s(f)$.

The parameters for the respective contributions are listed in Tab. III.1.

[3] Note that in literature, as well as in this work, the frequency dependence is usually implied and just $\Omega_{\mathrm{GW}}$ is written.

| | $\Omega_\phi$ | $\Omega_{\rm sw}$ | $\Omega_{\rm turb}$ |
|---|---|---|---|
| $\Omega_0$ | $1.67 \times 10^{-5}$ | $2.65 \times 10^{-6}$ | $3.35 \times 10^{-4}$ |
| $a$ | 2 | 2 | $\frac{3}{2}$ |
| $b$ | 2 | 1 | 1 |
| $\Delta$ | $\frac{0.11 v_{\rm w}^3}{0.42+v_{\rm w}^2}$ | $v_{\rm w}$ | $v_{\rm w}$ |
| $f_{\rm p}$ | $\frac{0.62\beta}{1.8-0.1v_{\rm w}+v_{\rm w}^2}$ | $\frac{2\beta}{\sqrt{3}v_{\rm w}}$ | $\frac{3.5\beta}{2v_{\rm w}}$ |
| $s(f)$ | $\frac{3.8(f/f'_{\rm p})^{2.8}}{1+2.8(f/f'_{\rm p})^{3.8}}$ | $(f/f'_{\rm p})^3 \left(\frac{7}{4+3(f/f'_{\rm p})^2}\right)^{7/2}$ | $\frac{(f/f'_{\rm p})^3}{(1+f/f'_{\rm p})^{11/3}(1+8\pi f/H')}$ |
| Reference | [56] | [57] | [58] |

Table III.1: Spectrum parameters for the three GW contributions.

### III.3.3 Three Bubble Scenarios

Depending on the environment in which bubbles emerge and on the characterizing parameters, there are three possible bubble scenarios which differ in the composition of their characteristic spectra.

#### Scenario 1: Non-runaway bubbles

The particle plasma exerts friction on the expanding bubble walls which can hinder them from reaching ultra-relativistic velocities. As pointed out in Section III.2, this is the case for $\alpha \leq \alpha_\infty$. In this regime, the energy in the bubble walls themselves is negligible and with it the contribution $\Omega_\phi$. The full spectrum is thus given by

$$\Omega_{\rm GW} = \Omega_{\rm sw} + \Omega_{\rm turb}$$

as shown in Fig. III.2. The efficiency factor for the sound wave contribution is $\kappa_{\rm sw} = \kappa(\alpha)$, as given by (III.25), while the one for turbulence is $\kappa_{\rm turb} = \varepsilon_{\rm turb}\kappa_{\rm sw}$. It turns out that turbulence is only a subordinate effect with $\varepsilon_{\rm turb} = 0.05 \sim 0.1$ [57]. The optimistic value of 0.1 will be used in this work. Note that non-runaway bubbles can still reach $v_{\rm w} \sim 1$ if the PT is not too weak, the only difference being a limited relativistic $\gamma$-factor [53].

#### Scenario 2: Runaway bubbles in plasma

The PT is now assumed to still take place in a particle plasma, but with less friction than in Scenario 1. This is the case for strong PTs where $\alpha > \alpha_\infty$ such that the bubble walls accelerate continuously and become ultra-relativistic. The scalar field collisions

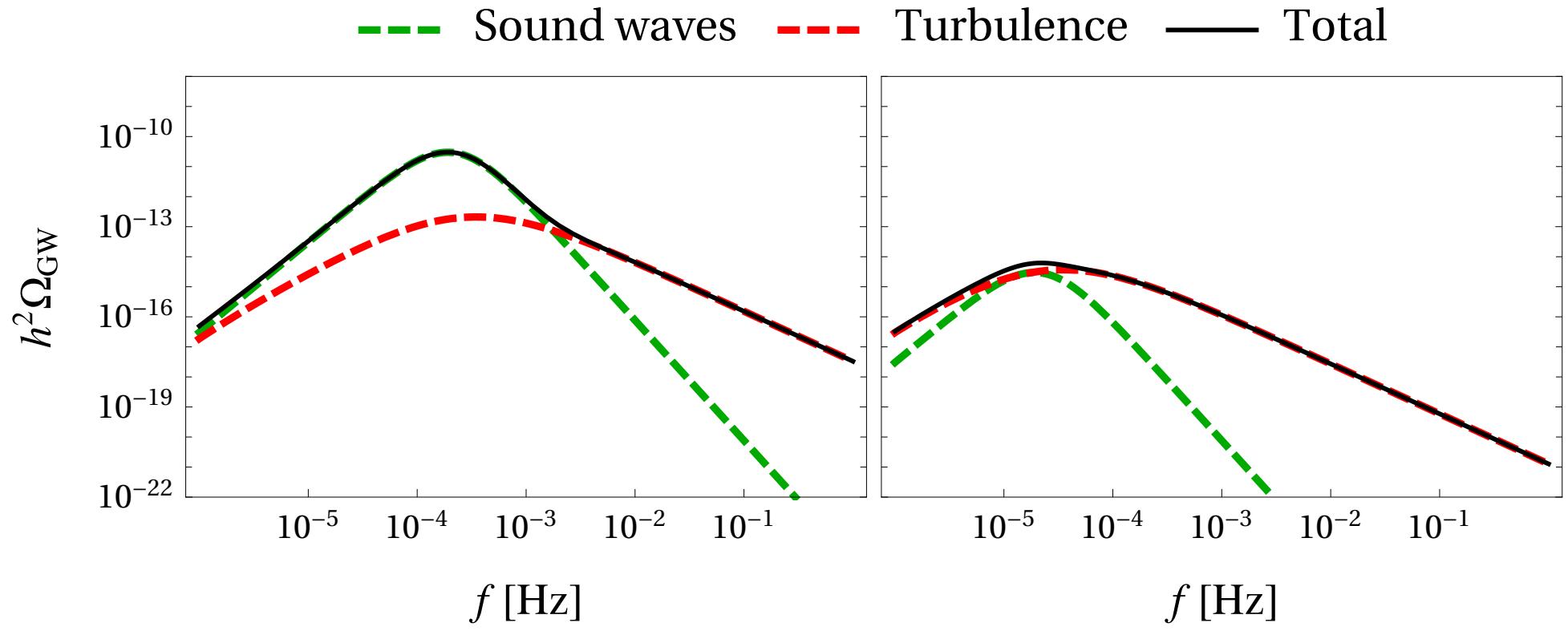

Figure III.2: Non-runaway spectrum for $T_\mathrm{n} = 100\,\mathrm{GeV}$, $g_\mathrm{rel} = 100$ and $\alpha = 0.1$, $\frac{\beta}{H} = 10$ (left), $\alpha = 0.001$, $\frac{\beta}{H} = 1$ (right). The contribution $\Omega_\mathrm{turb}$ becomes more dominant for small $\alpha$ or $\frac{\beta}{H}$.

are now relevant, leading to a spectrum with contributions

$$\Omega_\mathrm{GW} = \Omega_\phi + \Omega_\mathrm{sw} + \Omega_\mathrm{turb}.$$

The initial energy budget splits up into heat, bulk motion of the fluid and acceleration of the bubble walls. The fraction of the released energy that goes into heat and bulk motion is limited by $\frac{\alpha_\infty}{\alpha}$. The fraction $1 - \frac{\alpha_\infty}{\alpha}$ is a surplus energy which was not available in Scenario 1 and goes completely into the acceleration of the bubble walls. The efficiencies are therefore

$$\kappa_\mathrm{sw} = \frac{\alpha_\infty}{\alpha}\kappa(\alpha_\infty), \tag{III.30}$$

$$\kappa_\phi = 1 - \frac{\alpha_\infty}{\alpha}, \tag{III.31}$$

which implies a dominant (negligible) scalar contribution for $\alpha_\infty \ll \alpha$ ($\alpha_\infty \approx \alpha$). A double bump signature can be achieved with a certain choice of parameters (see Fig. III.3).

As mentioned in Section III.2, there is probably a speed limit on bubble walls involving gauged scalar fields. This would rule out the runaway regime and make this scenario obsolete. However, the effort of investigating to which extent those results really apply to the considered models in this work has not been made. Furthermore, there are several recent publications, [59, 60, 61, 62] just to name few, in which this scenario is still taken into account. We will therefore include this scenario in the subsequent analyses.

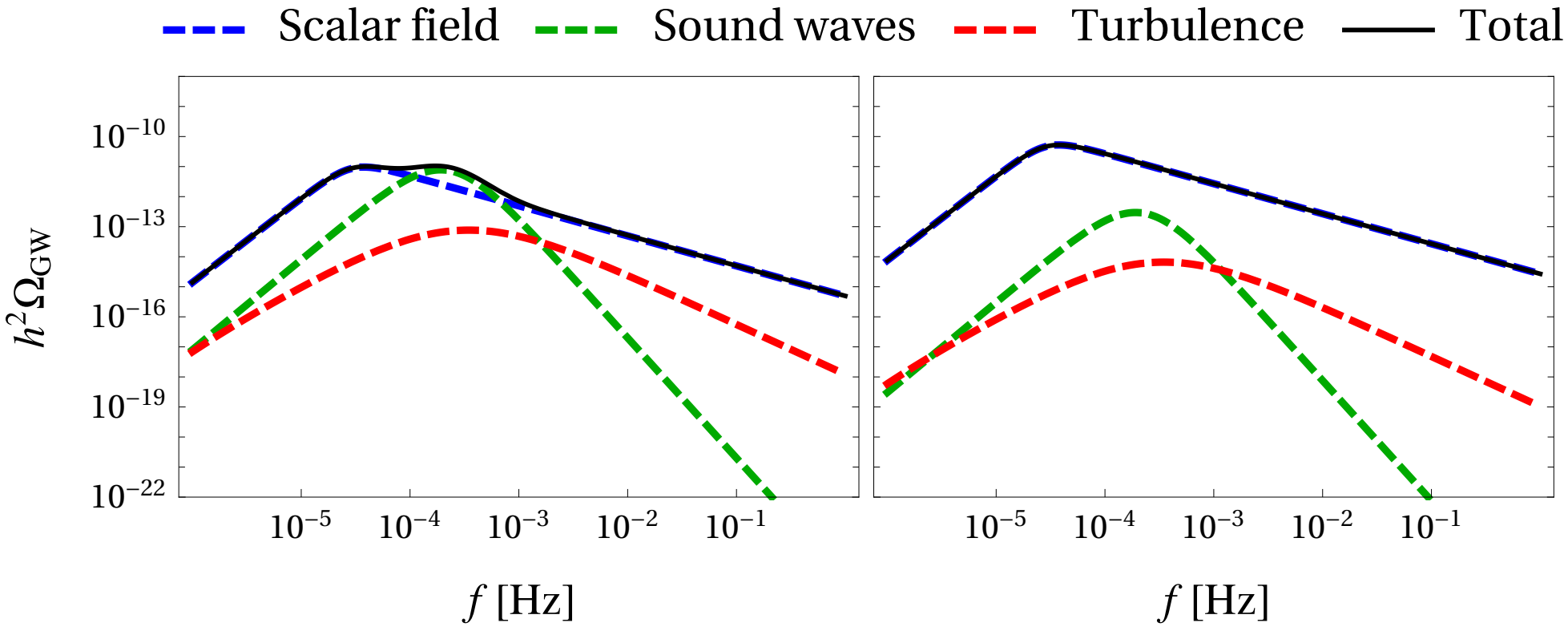

Figure III.3: Runaway spectrum for $T_{\rm n} = 100\,\text{GeV}$, $g_{\rm rel} = 100$, $\alpha = 0.1$, $\frac{\beta}{H} = 10$ and $\alpha_\infty = 0.7\alpha$ (left), $\alpha_\infty = 0.3\alpha$ (right). The contribution $\Omega_\phi$ would become less dominant for larger $\frac{\beta}{H}$.

#### Scenario 3: Runaway bubbles in vacuum

Bubble walls in a vacuum setting do necessarily, due to the lack of friction, become ultra-relativistic. A vacuum-dominated epoch can be realized by significant super-cooling $T_{\rm n} \ll T_{\rm c}$. In such a situation, there are either negligible or no plasma effects at all, implying

$$\Omega_{\rm GW} = \Omega_\phi$$

together with $\kappa_\phi = v_{\rm w} = 1$. Due to $\alpha \sim \rho_{\rm rel}^{-1}$ and vanishing relativistic energy density $\rho_{\rm rel}$ in vacuum, we have $\alpha \to \infty$ which lets the $\alpha$ dependence drop out of the spectrum (III.29). This scenario will not be relevant in any of the models we consider later.

### III.3.4 Parameter Dependence

Before starting to investigate specific models, it is useful to develop a feeling for the behavior of $\Omega_{\rm GW}(f)$ in dependence on the characteristic parameters. These will be varied in the following, where we highlight the changes in the GW spectrum.

PT strength $\alpha$ is the most intuitive paramter and it just impacts the height of the signal (see Fig. III.4, left). The scaling goes with the power of 2 (3/2) for the scalar and sound wave contribution (turbulent contribution) in case of small $\alpha$ but saturates for $\alpha > 1$. Varying $\frac{\beta}{H}$ impacts the spectrum as follows: Larger values represent slower PTs and, assuming $R \sim v_{\rm w}\beta^{-1}$ and constant $H$, smaller bubbles at the time of collision. Those in turn cause weaker anisotropies on smaller scales, compared to those created by larger bubbles. The characteristic wavelength depends on the size of a source, so overall the spectrum becomes weaker and moves to higher frequencies (see Fig. III.4, right). The scaling with nucleation temperature $T_{\rm n}$ is less obvious. Naively, one would expect a more redshifted peak frequency for earlier PTs, i.e. for larger $T_{\rm n}$.

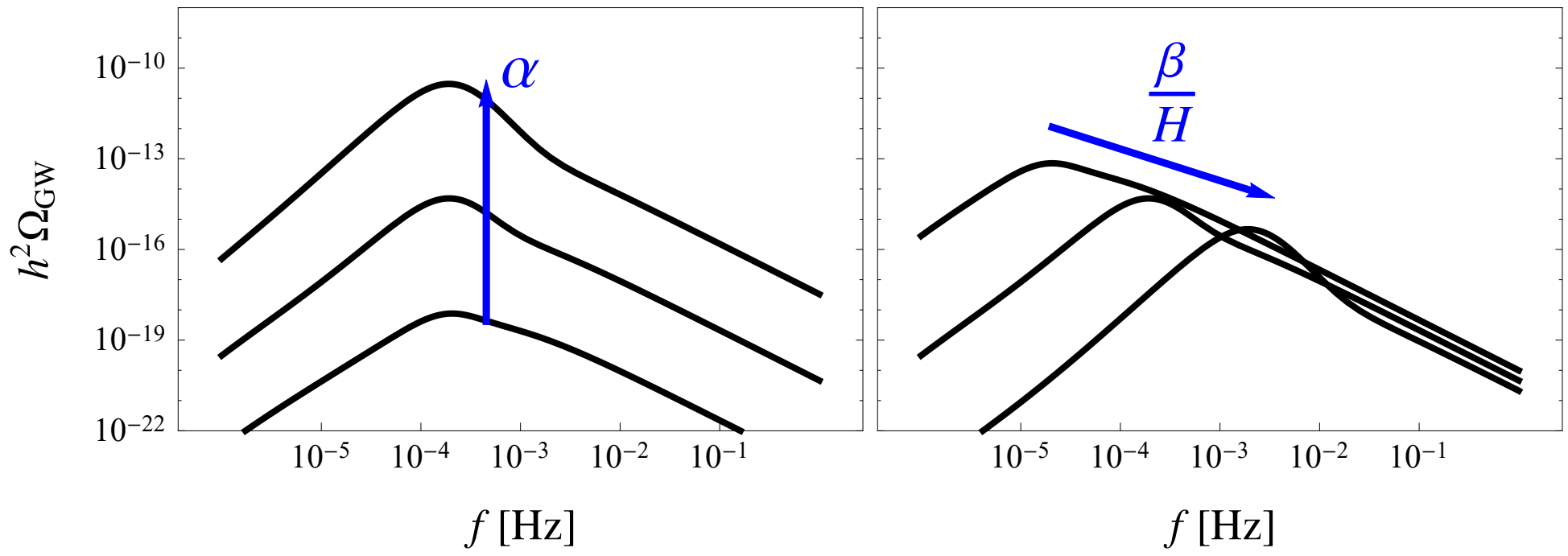

Figure III.4: Spectrum at $T_{\rm n} = 100\,\text{GeV}$ with $\frac{\beta}{H} = 10$, $\alpha = \{10^{-3}, 10^{-2}, 10^{-1}\}$ (left) and $\frac{\beta}{H} = \{1, 10, 100\}$, $\alpha = 10^{-2}$ (right).

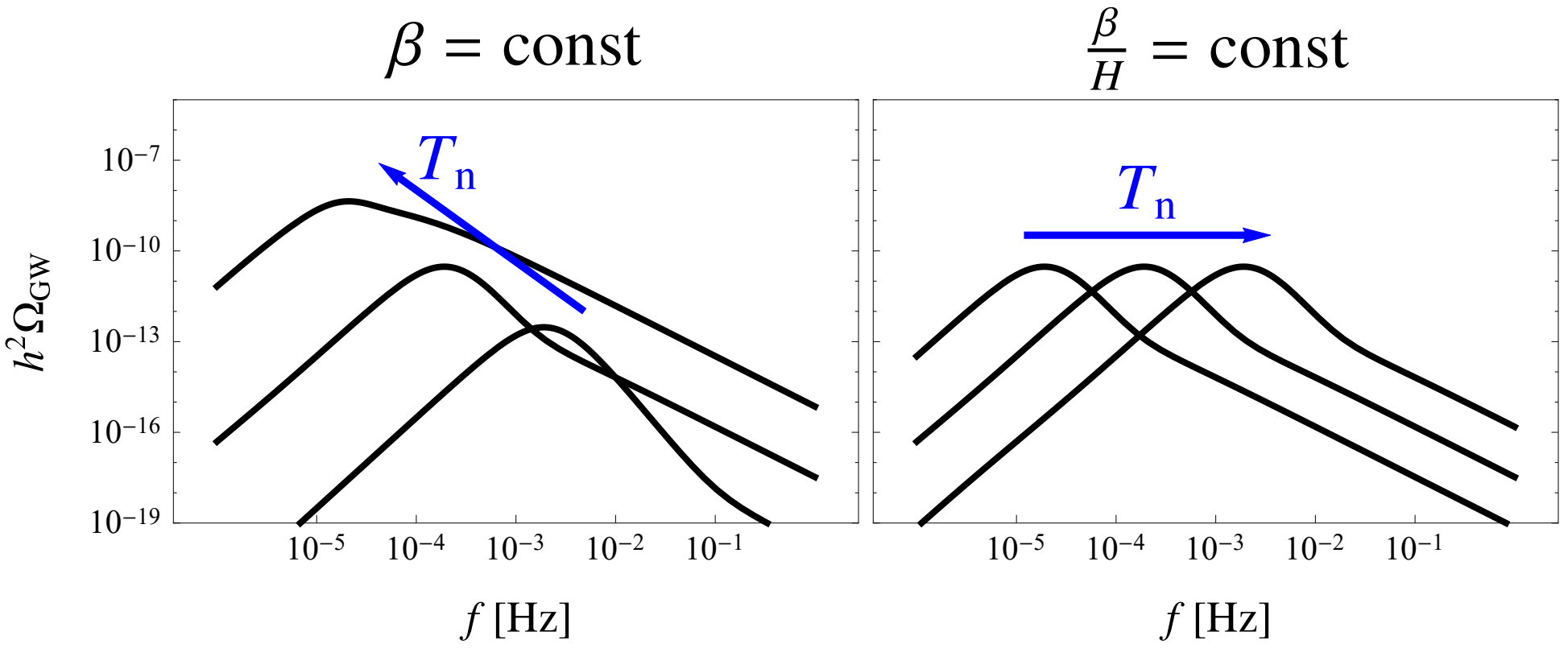

Figure III.5: Spectrum with $\alpha = 0.1$, $T_{\rm n} = \{10^2, 10^3, 10^4\}\,\text{GeV}$, $\frac{\beta}{H} = \{10^3, 10^1, 10^{-1}\}$ (left) and $\frac{\beta}{H} = 10$ (right).

This is indeed what happens for fixed $\beta$, as shown in Fig.III.5 (left). Due to $H \sim T^2$, $\frac{\beta}{H}$ decreases quadratically with $T_{\rm n}$. This explains the spectrum being shifted upwards and to lower frequencies, in addition to the effect of frequency redshift. Usually however, $\frac{\beta}{H}$ instead of $\beta$ is the variable of interest. Keeping this ratio fixed, the effect of peak frequency redshift, which scales linearly with $T_{\rm n}$, becomes overcompensated by the quadratic scaling of $\beta$ with $T_{\rm n}$. In other words, $T_{\rm n}$ is now the only varied scale, so the peak frequency goes linear with it while the spectrum is fixed in height (see Fig.III.5, right).

Note that in the figures demonstrating this scaling behavior, a non-runaway scenario with $\alpha_\infty = \alpha$ is shown. This analysis aimed at showing only the qualitative behavior, neglecting the slight differences between the different contributions. These differences are however still visible in the changing shapes of the spectra.

## III.4 Detectability

### III.4.1 Observatories

Aiming at the detection of gravitational radiation, detectors of different kinds have been built and improved upon in the past. After half a century of fruitless attempts, the first direct detection of a binary black hole merger was announced by LIGO in 2016. In the following, a brief overview over the most relevant GW observatories in the context of this thesis will be provided.

#### Interferometers

One possible approach of GW detection is the concept of laser interferometry, as sketched in Fig. II.7, where the interferometer's arm length determines the sensitive frequency band. The most sensitive operational *ground-based* experiments are LIGO [63] and Virgo [64]. They both have arm lengths of $\mathcal{O}(\text{km})$ leading to sensitivities in the range 10 to 1000 Hz. This frequency domain is interesting for the observation of pulsars, supernovae and mergers of black holes and neutron stars. As we will see, first-order PTs around the electroweak scale and below cause GW spectra at much lower frequencies. In the context of this work, the planned *space-based* observatories will be more interesting. In these experiments, the laser source and the test masses are placed in separate satellites with huge distances in between, together sitting in an earth-like orbit around the sun. The European project LISA [65], scheduled to launch in 2034, will feature three spacecraft forming a triangle of 2.5 Gm side length connected by six laser links leading to a sensitivity in the mHz region. BBO [66] is the proposed successor of LISA with an arm length of 50 Mm and sensitivity around 10 Hz. The Japanese project DECIGO [67] is scheduled to launch after 2030 and will be sensitive in the same region as BBO. A precursor mission with shorter arm length is called B-DECIGO [68]. Ultimate DECIGO is a so far purely hypothetical observatory whose sensitivity is limited only by quantum mechanics. The sensitivity of interferometers is in general limited by the quantum noise and radiation pressure on the test masses, both due to the laser, and by instrumental and thermal noise. Ground-based experiments are further prone to seismic and environmental effects.

#### Pulsar timing arrays

*Millisecond pulsars* are magnetized rotating neutron stars that emit beams of electromagnetic radiation in the direction of their rotational axes. The pulsar's beams hit Earth periodically with an extremely stable frequency, thereby providing precise astrophysical clocks. Tiny variations in the arrival time of the pulses can reveal the presence of metric perturbations. This allows us to search for GWs by simply timing pulsars. To cancel atmospheric effects, many pulsars are considered at the same time, forming a *pulsar timing array* (PTA). This kind of GW search is suited for probing supermassive black hole binaries, cosmic strings and stochastic backgrounds in the frequency range from $10^{-9}$ to $10^{-6}$ Hz. The projects NANOGrav [69] and EPTA [70] are now running for several years but were only able to provide limits so far. The SKA [71] will be sensitive to significantly weaker GWs and aims to be operational in 2020.

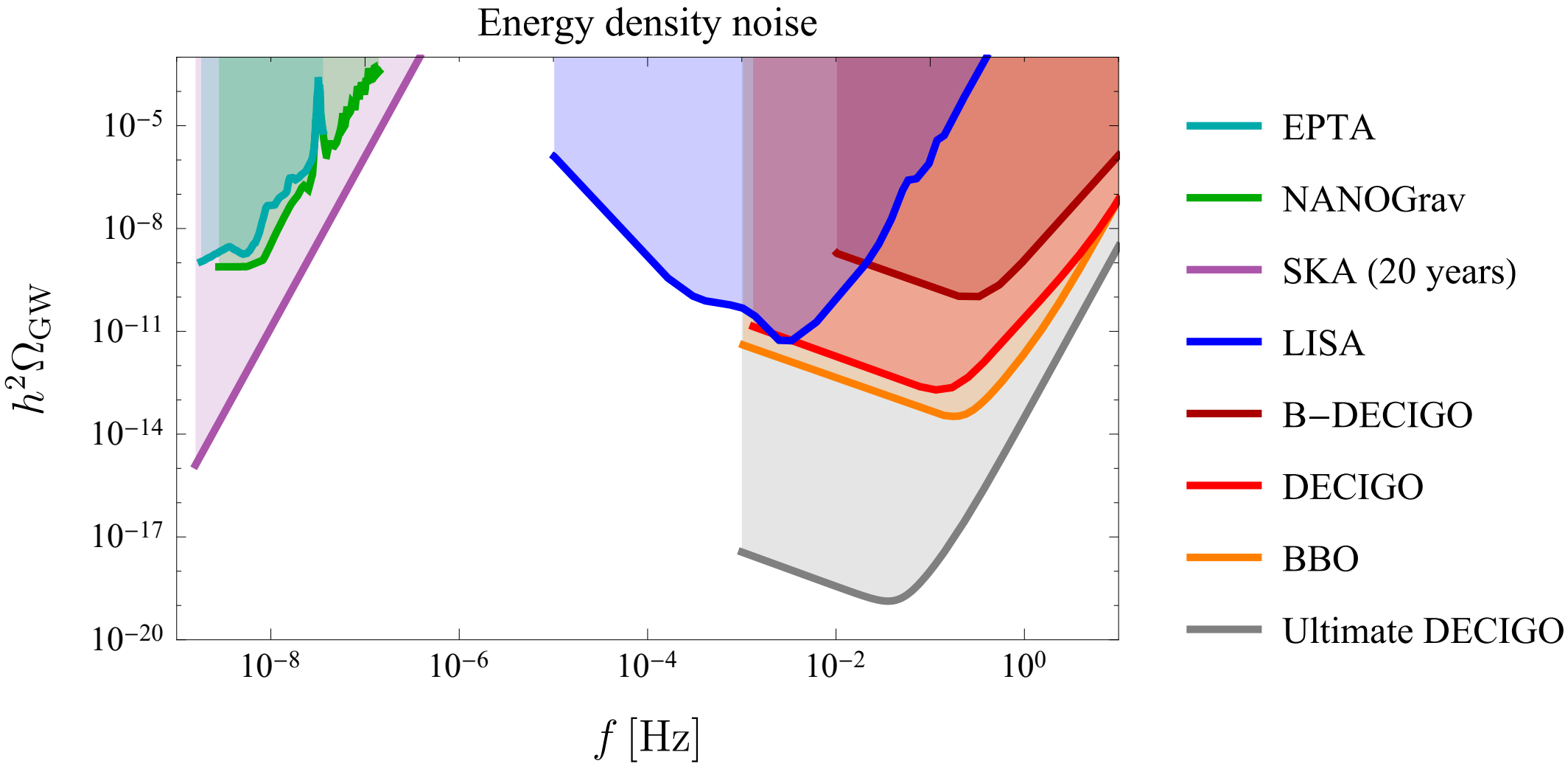

Figure III.6: Energy density noise curves for (planned) space-based experiments and PTAs. The sources are [72] (EPTA), [73] (NANOGrav), [74, 75, 76] (SKA), [77] (LISA), [78] (B-DECIGO), [79] (DECIGO, BBO) and [80, 81] (Ultimate DECIGO). Note that we use the sky-averaged noise curves. The noise for DECIGO and BBO corresponds to only a single L-shaped interferometer of which there will be many in the final configurations.

* * *

Detector noise curves are not always presented in the same way, but they can be converted as follows [82]: Given the strain amplitude $h(f)$ of a signal or the background, one can calculate the *power spectral density* by

$$S(f) = \frac{h^2(f)}{f} \tag{III.32}$$

or the *energy density spectrum* with

$$\Omega_{\rm GW}(f) = \frac{2\pi^2}{3H_0^2} f^3 S(f). \tag{III.33}$$

A collection of noise curves for the considered detectors is given in Fig. III.6. The sensitivity drop at $f = 1\,\text{yr}^{-1} \approx 3 \times 10^{-8}\,\text{Hz}$ in the PTA curves is attributed to Earth's rotation around the Sun. LISA's sensitivity has a bump around $f \sim 10^{-3}\,\text{Hz}$ due to galactic confusion noise in the form of unresolved binary systems. Note that the inconsistent appearance of the curves, some smooth and others more detailed, is because they partly originate from approximate formulae but are otherwise exactly copied from plots.

| Detector | $T$ in years | $\mathrm{SNR}_{\mathrm{thr}}$ | |
|---|---|---|---|
| EPTA | 18 | 1.19 | [72] |
| NANOGrav | 11 | 0.697 | [73] |
| SKA | 5, 10, 20 | 4 | [76] |
| LISA | 4 | 10 | [46] |
| B-DECIGO | 4 | 8 | [78] |
| DECIGO | 4 | 25 | [79] |
| BBO | 4 | 25 | [79] |
| Ultimate DECIGO | 4 | 25 | [79] |

Table III.2: Assumed observation periods and SNR thresholds. The citations indicate where $\mathrm{SNR}_{\mathrm{thr}}$ has been extracted from.

### III.4.2 Signal-to-Noise Ratio

If a predicted signal intersects a detector noise curve, this does not automatically imply that it is detectable. The signal must lie above the noise curve in a sufficiently large frequency band and for a long enough time, in order to become visible to the detector. Integrating over frequency and time yields the *signal-to-noise ratio*

$$\mathrm{SNR} \equiv \sqrt{T \int_{-\infty}^{\infty} \mathrm{d}f \left( \frac{\Omega_{\mathrm{signal}}(f)}{\Omega_{\mathrm{noise}}(f)} \right)^2} \tag{III.34}$$

which is a measure for the detectability of a signal. $T$ is the duration of the respective detector mission, which can simply be factored out as we are dealing with spectra that are constant in time. A signal is detectable if the SNR is above a certain threshold $\mathrm{SNR}_{\mathrm{thr}}$, a number determined individually for the different detectors. Tab. III.2 summarizes the values used in this work, together with the assumed observation periods.

With this at hand, meaningful sensitivity curves can be constructed. Here we follow the procedure used in [74]. A single power-law curve $\Omega_\gamma(f) = \Omega_\gamma \cdot (f/\mathrm{Hz})^\gamma$ is detectable if its amplitude is at least

$$\Omega_\gamma = \mathrm{SNR}_{\mathrm{thr}} \left[ T \int_{-\infty}^{\infty} \mathrm{d}f \left( \frac{(f/\mathrm{Hz})^\gamma}{\Omega_{\mathrm{noise}}(f)} \right)^2 \right]^{-1/2} . \tag{III.35}$$

In order to generalize for any combination of power-laws, $\Omega_\gamma$ has to be calculated for several different exponents $\gamma$. The envelope $\max_\gamma[\Omega_\gamma(f)]$ embodies the *power-law integrated sensitivity curve* of a detector (see Fig. III.7). Detectability is now implied whenever a power-law spectrum touches the derived envelope. Fig. III.8 shows the resulting curves for the detectors under consideration.

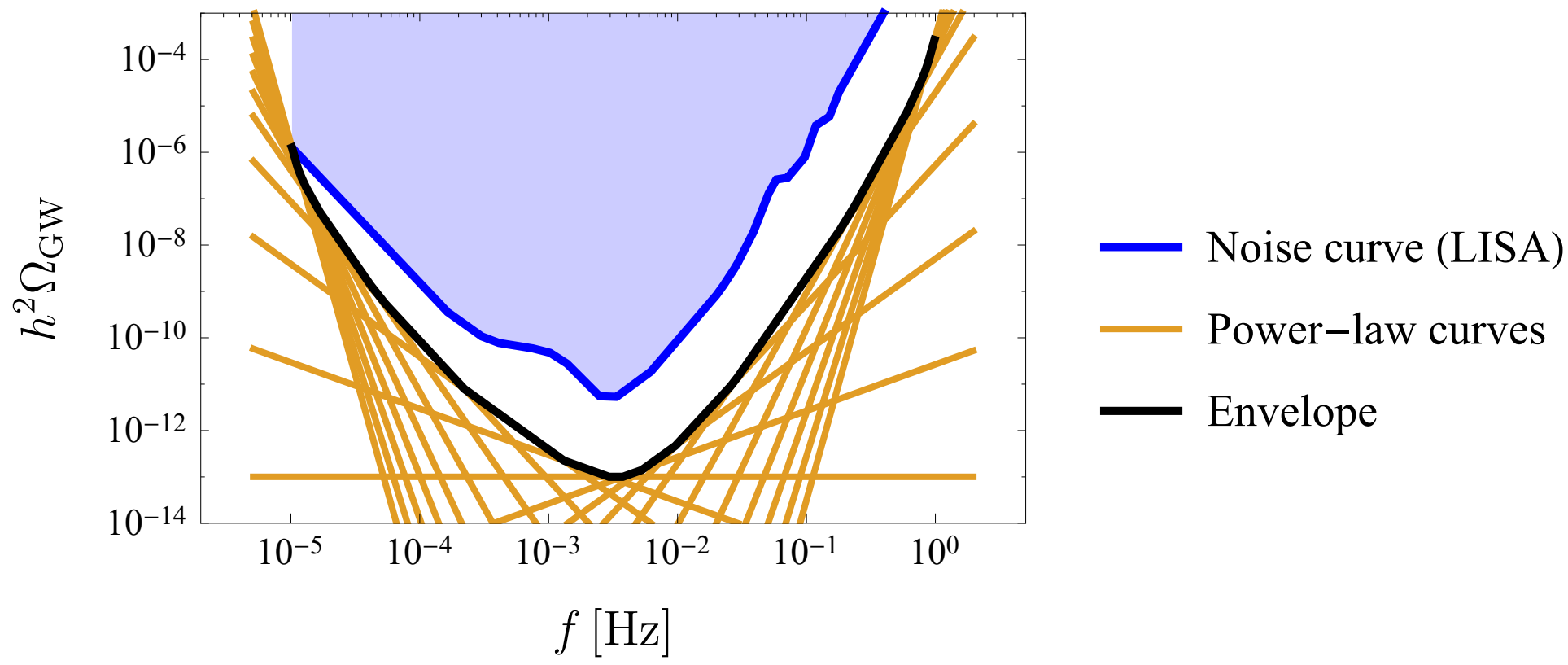

Figure III.7: Construction of the power-law integrated sensitivity curves for LISA with $\gamma = \{-10, -9, \ldots, 9, 10\}$.

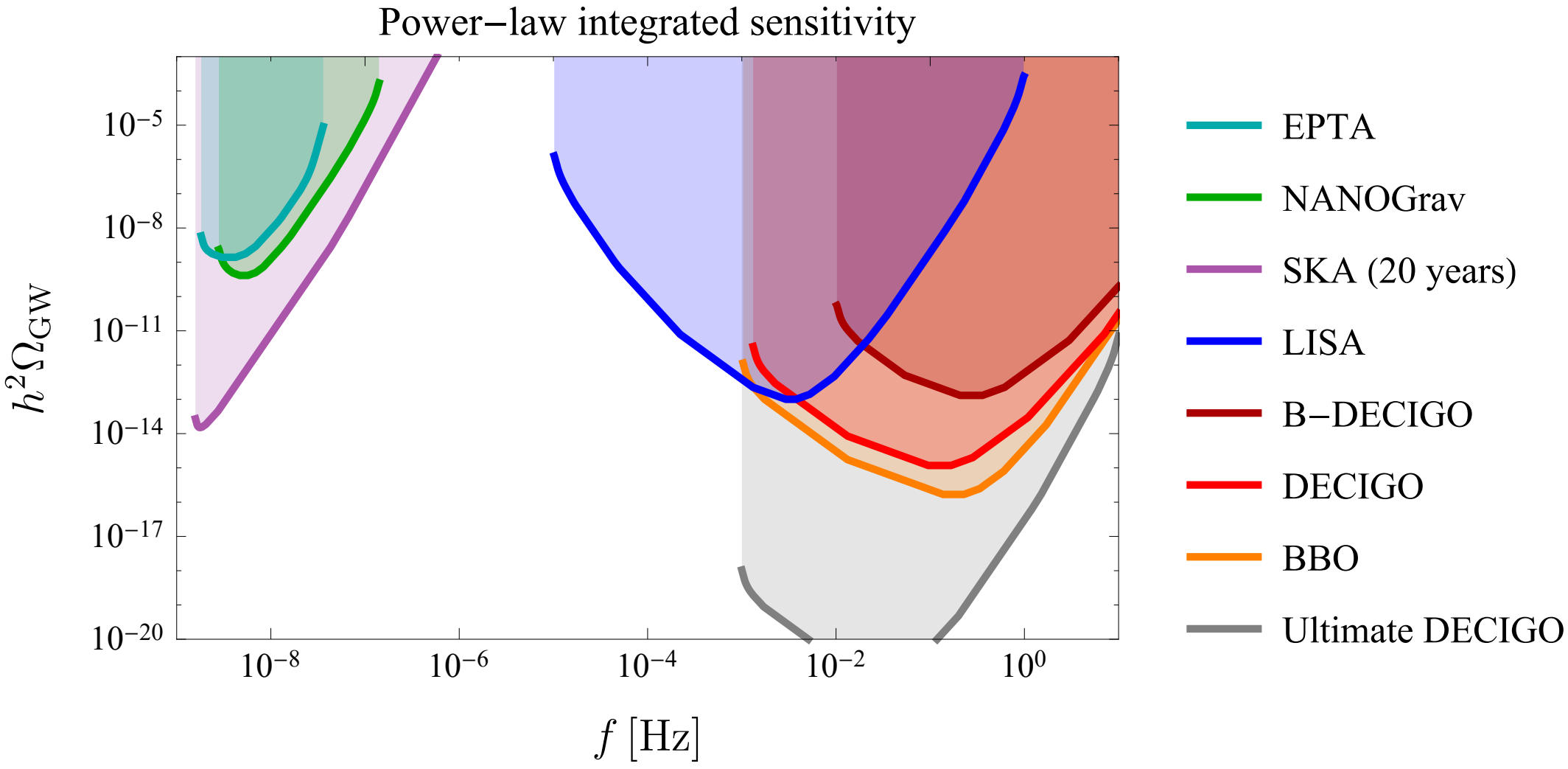

Figure III.8: Power-law integrated sensitivity curves, assuming the thresholds and observation periods in Tab. III.2.

### III.4.3 Detectable Regions

Now that it is clear how to assess the detectability of a given spectrum, we are able to identify detectable regions in the space of the characteristic GW parameters $\alpha$, $\frac{\beta}{H}$ and $T_*$, where the latter is equal to $T_{\rm n}$ in the scenarios we consider.

Fig. III.9 shows the sensitive regions for a non-runaway spectrum, i.e. with only the sound wave and turbulence contribution included. The characteristic shapes in the $\frac{\beta}{H}$-$T_*$-plane are a result of the signal's composite nature: For small $\frac{\beta}{H}$, the flat high-frequency tail of $\Omega_{\rm turb}(f)$ is detectable. This contribution becomes suppressed for increasing $\frac{\beta}{H}$ while at the same time the spectrum moves towards higher frequencies (see Fig.III.4, right). For certain values of $T_*$ it can occur that the spectrum is then invisible for an intermediate $\frac{\beta}{H}$ range, until it becomes visible again at some larger $\frac{\beta}{H}$ due to $\Omega_{\text{sw}}(f)$ with its steeper high-frequency tail. This behavior is not present in Fig. III.10, showing the sensitivity for runaway bubbles with the scalar field contribution as the dominating component, i.e. $\alpha_\infty \ll \alpha$. Runaway bubbles are detectable towards much lower $T_*$. This can be attributed to the high-frequency tail of $\Omega_\phi(f)$ which is even flatter compared to the tails of the non-runaway contributions (see Fig. III.3). The visible kinks at fixed temperatures in the $\frac{\beta}{H}$-$T_*$-plane of Fig. III.9 and III.10, e.g. around $T_* \sim 10^{-0.5}$, are due to the stepwise nature of $g_{\text{rel}}(T)$ which enters through the redshift calculation.

Overall, the plots allow to assign PTs at a given temperature to the corresponding class of detectors that is suited best for their detection. PTs in the range from 100 keV to 100 MeV are interesting for PTAs while future space-based experiments are sensitive to temperatures from 10 GeV to tens of TeV and beyond. These ranges are shifted towards lower temperatures for larger values of $\frac{\beta}{H}$. This behavior can be understood by the fact that the redshifted peak frequency increases with $\frac{\beta}{H}$ such that $T_{\text{n}}$ needs to be decreased to stay in the detectable range.

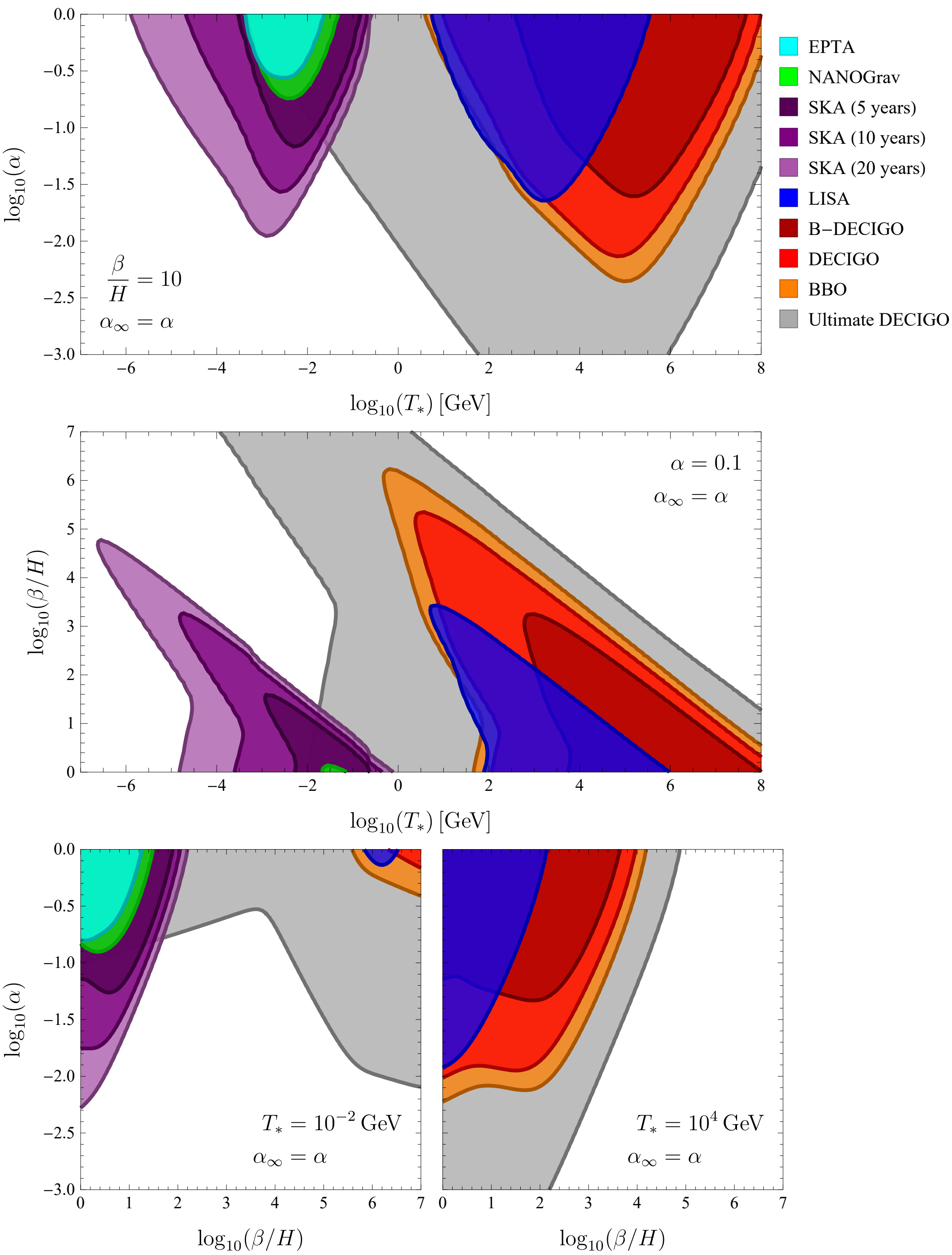

Figure III.9: Sensitive regions assuming non-runaway bubbles.

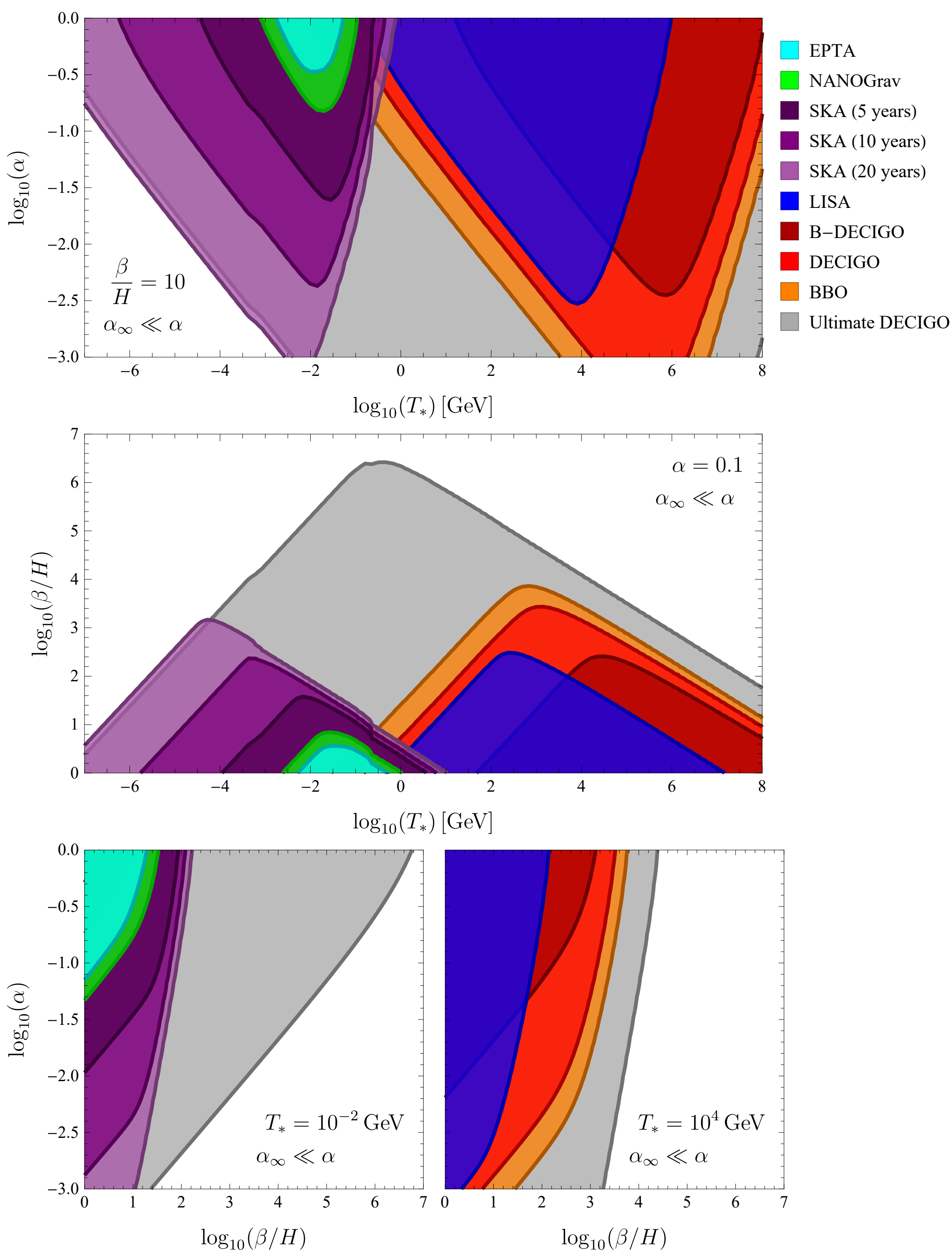

Figure III.10: Sensitive regions assuming runaway bubbles.

# IV Investigating Models

## IV.1 The Standard Model

Although the SM does not feature a first-order PT, we will in the following give details of the electroweak sector as it forms the basis for all possible extensions. We will further apply the machinery developed in Chapters II and III to the plain SM. This can be seen as a consistency check of the theoretical considerations that were made, but also as a cross-check of `CosmoTransitions`' results.

The SM is a self-consistent theoretical structure of particle physics, which was developed in the 20th century by Glashow, Weinberg, Salam [83, 84, 85] and many others and describes nature stunningly well up to energies of at least $\mathcal{O}(\text{TeV})$, with some exceptions. It incorporates three of the four known fundamental forces together with all known particles. The SM is a gauge theory where the underlying symmetries are $SU(3)_c$, providing the *strong force*, and $SU(2)_L \times U(1)_Y$, giving rise to *electroweak* interactions.

The matter content of the SM consists of three fermion generations in two sets, *leptons* and *quarks*, where only the latter are $SU(3)_c$ triplets and carry color charge. There are furthermore 12 spin-1 gauge bosons, corresponding to the three local gauge symmetries: 8 gluons, 3 $W$-bosons and a single $B$-boson. On top of that, there is a spin-1 Higgs boson, which is a $SU(2)_L$ doublet carrying $U(1)_Y$ hypercharge and induces *electroweak symmetry breaking* (EWSB). Below a certain temperature, the Higgs scalar field acquires a VEV and thereby breaks electroweak gauge symmetry down to electromagnetism:

$$SU(2)_L \times U(1)_Y \to U(1)_{\text{em}}$$

In the broken symmetry phase, gauge bosons mix and become

$$\begin{aligned} W_\mu^\pm &= \frac{1}{\sqrt{2}}(W_\mu^1 \mp W_\mu^2), \\ Z_\mu &= \frac{1}{\sqrt{g_2^2 + g_1^2}}(g_2 W_\mu^3 - g_1 B_\mu), \\ A_\mu &= \frac{1}{\sqrt{g_2^2 + g_1^2}}(g_2 W_\mu^3 + g_1 B_\mu) \qquad \text{(massless photon)} \end{aligned} \tag{IV.1}$$

with gauge couplings $g_2 \approx 0.65$ and $g_1 \approx 0.35$ for $SU(2)_L$ and $U(1)_Y$ respectively. The amount of mixing between $W^3$ and $B$ is parametrized by the *weak mixing angle*

| | Chirality | Field | $SU(3)_c$ | $SU(2)_L$ | $U(1)_Y$ |
|---|---|---|---|---|---|
| Higgs doublet | - | $H$ | 1 | 2 | $+\frac{1}{2}$ |
| 3 lepton doublets | (LH) | $L_L^i$ | 1 | 2 | $-\frac{1}{2}$ |
| 3 quark doublets | (LH) | $Q_L^i$ | 3 | 2 | $+\frac{1}{6}$ |
| 3 charged leptons | (RH) | $e_R^i$ | 1 | 1 | $-1$ |
| 3 up-type quarks | (RH) | $u_R^i$ | 3 | 1 | $+\frac{2}{3}$ |
| 3 down-type quarks | (RH) | $d_R^i$ | 3 | 1 | $-\frac{1}{3}$ |

Table IV.1: SM electroweak gauge charges in the unbroken phase.

| | Chirality | Field | $U(1)_{\text{em}}$ | Mass at $T=0$ |
|---|---|---|---|---|
| Goldstones & massive Higgs | - | $\begin{pmatrix} G^+ \\ h + iG^0 \end{pmatrix}$ | $0$<br>$-1$ | $m_h = 125\,\text{GeV}$ |
| 6 leptons | (LH) | $\begin{pmatrix} \nu \\ e \end{pmatrix}_L^i$ | $0$<br>$-1$ | $0$<br>$511\,\text{keV}, 106\,\text{MeV}, 1.8\,\text{GeV}$ |
| 6 quarks | (LH) | $\begin{pmatrix} u \\ d \end{pmatrix}_L^i$ | $+\frac{2}{3}$<br>$-\frac{1}{3}$ | $2.4\,\text{MeV}, 1.3\,\text{GeV}, 172\,\text{GeV}$<br>$4.8\,\text{MeV}, 95\,\text{MeV}, 4.2\,\text{GeV}$ |
| 3 charged leptons | (RH) | $e_R^i$ | $-1$ | as LH |
| 3 up-type quarks | (RH) | $u_R^i$ | $+\frac{2}{3}$ | as LH |
| 3 down-type quarks | (RH) | $d_R^i$ | $-\frac{1}{3}$ | as LH |

Table IV.2: SM electromagnetic charges and tree-level masses in the broken phase.

defined as

$$\tan\theta_w \equiv \frac{g_1}{g_2} \approx 0.53.$$

According to the *Goldstone theorem*, the three broken generators correspond to massless Goldstone bosons which are formally equivalent to longitudinal gauge boson polarization modes, leading to the masses

$$m_W \approx 80.4\,\text{GeV},$$
$$m_Z \approx 91.2\,\text{GeV}$$

at $T = 0$.[1] During EWSB also all fermions except for neutrinos obtain masses proportional to their Yukawa couplings. The gauge charges and tree-level masses of the SM matter content are listed in Tab. IV.1 and IV.2.

[1]The actual temperature today of $T = 240\,\mu\text{eV}$ is not zero, but totally negligible compared to the weak scale. $T = 0$ will therefore mark today's values.

### IV.1.1 Potential and Masses

The relevant Higgs Lagrangian in the SM is given by

$$\mathcal{L} \supset (D_\mu H)^\dagger (D^\mu H) - V_{\text{tree}}(H) \tag{IV.2}$$

with

$$V_{\text{tree}}(H) = -\mu^2 H^\dagger H + \lambda (H^\dagger H)^2. \tag{IV.3}$$

The part involving $\mu^2$ carries a negative sign and thereby allows for spontaneous symmetry breaking, as explained in Section II.2.4. The Higgs doublet written in components is

$$H = \begin{pmatrix} G^+ \\ \frac{1}{\sqrt{2}}(h + iG^0) \end{pmatrix} \tag{IV.4}$$

and the tree-level potential in terms of the neutral CP-even component reads

$$V_{\text{tree}}(h) = -\frac{1}{2}\mu^2 h^2 + \frac{1}{4}\lambda h^4. \tag{IV.5}$$

In the above expression, all terms including $G^+$ or $G^0$ have been dropped because they will not obtain a VEV and are therefore irrelevant in light of the constant background field method which was used to derive the effective potential. In order to determine the field dependent masses of the different components, one has to go a step back, write out the VEV explicitly by $h \to h_\text{c} + h$ and determine the second derivatives with respect to fields $G^+$, $G^0$ and $h$. Keeping this in mind, we will from now on and for the sake of simplicity again stick with the somewhat sloppy notation where $h = h_\text{c}$.

The above tree-level potential has a minimum at $v \equiv \frac{\mu}{\sqrt{\lambda}} \approx 246\,\text{GeV}$ with $\mu \approx 88.4\,\text{GeV}$ and $\lambda \approx 0.13$. To fix the Higgs VEV and mass after adding the Coleman-Weinberg contribution at $T = 0$, finite counterterms are needed. Imposing the renormalization conditions in (II.85) yields

$$V_{\text{ct}}(h) = -\frac{1}{2}\delta\mu^2 h^2 + \frac{1}{4}\delta\lambda h^4 \tag{IV.6}$$

with

$$\begin{aligned} \delta\mu^2 &= \frac{3V'_{\text{CW}}(v)}{2v} - \frac{V''_{\text{CW}}(v)}{2}, \\ \delta\lambda &= \frac{V'_{\text{CW}}(v)}{2v^3} - \frac{V''_{\text{CW}}(v)}{2v^2}. \end{aligned}$$

Light particles lead to negligibly small contributions to the effective potential. We therefore only include the third up-type quark (top), the massive gauge bosons ($W^\pm$, $Z$) and the components of $H$. The masses can be derived from the full Lagrangian

and amount to

$$
\begin{aligned}
m_h^2(h) &= -\mu^2 + 3\lambda h^2, \\
m_{G^+,G^0}^2(h) &= -\mu^2 + \lambda h^2, \\
m_t(h) &= y_t h \equiv m_t \frac{h}{v}, \\
m_{W^{1,2}}^2(h) &= m_{W^\pm}^2(h) = \frac{1}{4} g_2^2 h^2 \equiv m_W \frac{h^2}{v^2}, \\
m_{(W^3,B)}^2(h) &= \frac{h^2}{4} \begin{pmatrix} g_2^2 & -g_2 g_1 \\ -g_2 g_1 & g_1^2 \end{pmatrix}
\end{aligned}
\tag{IV.7}
$$

with eigenvalues

$$
\begin{aligned}
m_Z^2(h) &= \frac{1}{4} \sqrt{g_1^2 + g_2^2} h^2 \equiv m_Z \frac{h^2}{v^2}, \\
m_A^2(h) &= 0.
\end{aligned}
\tag{IV.8}
$$

The Debye masses are given by

$$
\begin{aligned}
\Pi_H(T) &= \left( \frac{1}{2}\lambda + \frac{1}{4} y_t^2 + \frac{1}{16} g_1^2 + \frac{3}{16} g_2^2 \right) T^2, \\
\Pi_{W^i}^{\mathrm{L}}(T) &= \frac{11}{6} g_2^2 T^2, \\
\Pi_B^{\mathrm{L}}(T) &= \frac{11}{6} g_1^2 T^2,
\end{aligned}
\tag{IV.9}
$$

where the superscript 'L' marks longitudinal modes. Note that the Goldstone bosons and the photon are massless at $T = 0$ and $h = v$, but in general receive thermal or field dependent masses. In order to evaluate the daisy potential, the longitudinal masses must be added in the unbroken phase, for example

$$
\left( m_{(W^3,B)}^{\mathrm{L}}(T) \right)^2 \equiv m_{(W^3,B)}^2(h) + \begin{pmatrix} \Pi_{W^i}^{\mathrm{L}}(T) & 0 \\ 0 & \Pi_B^{\mathrm{L}}(T) \end{pmatrix},
\tag{IV.10}
$$

and are then diagonalized. The procedure works analogous for any mass matrix and will not be shown explicitly again in the subsequent sections.

## IV.1.2 Critical Temperature

Before evaluating the SM one-loop effective potential with the assistance of `CosmoTransitions`, an estimate for the critical temperature shall be made. To do so, consider only tree-level and finite temperature potential, the latter only up to cubic order in the high-temperature expansion. The resulting potential can be parametrized

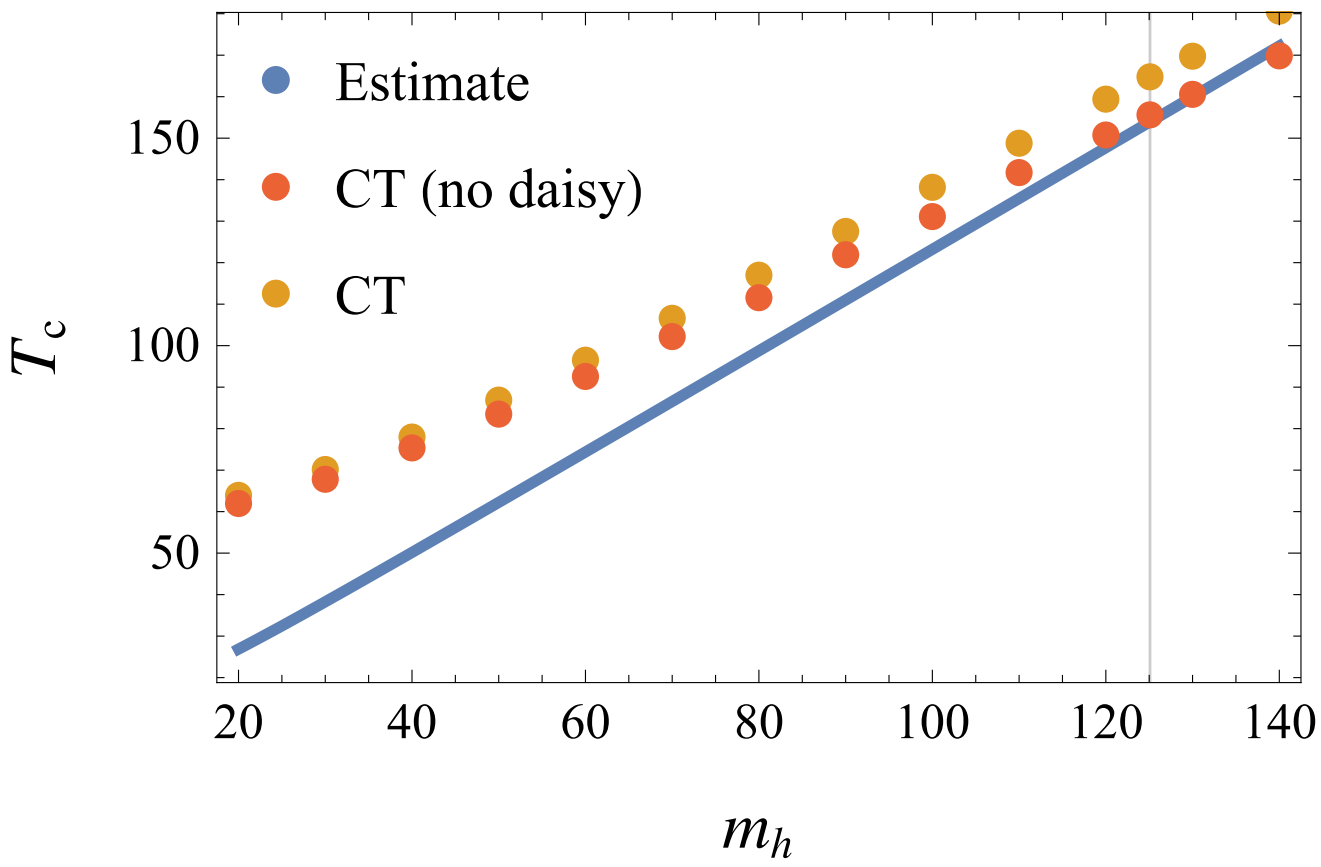

Figure IV.1: Estimate and `CosmoTransitions` result of the EWPT critical temperature for different Higgs masses in the SM.

similarly to Section II.2.4

$$V_{\text{eff}}(h,T) \approx \left(AT^2 - \frac{1}{2}\mu^2\right) h^2 - BT|\phi|^3 + \frac{1}{4}\lambda^4 h^4 \tag{IV.11}$$

with

$$\begin{aligned} A &= \frac{1}{24v^2}(6m_W^2 + 3m_Z^2) + \frac{1}{48v^2}(12m_t^2), \\ B &= \frac{1}{12\pi v^3}(6m_W^3 + 3m_Z^3). \end{aligned} \tag{IV.12}$$

After plugging the tree-level masses in, the estimated critical temperature for the EWPT evaluates to

$$T_{\text{c}} = \frac{\mu}{\sqrt{2(A - B^2/\lambda)}} \approx 154\,\text{GeV}$$

which is very close to the `CosmoTransitions` result $T_{\text{c}} = 155.3\,\text{GeV}$ (with the daisy contribution turned off) and also not too far from the lattice simulation result [86].

The critical temperature for varying Higgs mass is shown Fig. IV.1. For small $m_h$, the estimate seems to differ from the one-loop result obtained by `CosmoTransitions`. This can be attributed to approximations that went into the estimate, such as neglecting $V_{\text{CW}}$ and using the high-temperature expansion of $V_T$. For larger $m_h$, tree-level parameter $\mu \sim m_h$ also becomes large and dominates the effective potential compared to the approximated one-loop parts. Hence in this regime the estimate of $T_{\text{c}}$ deviates less from the `CosmoTransitions` result.

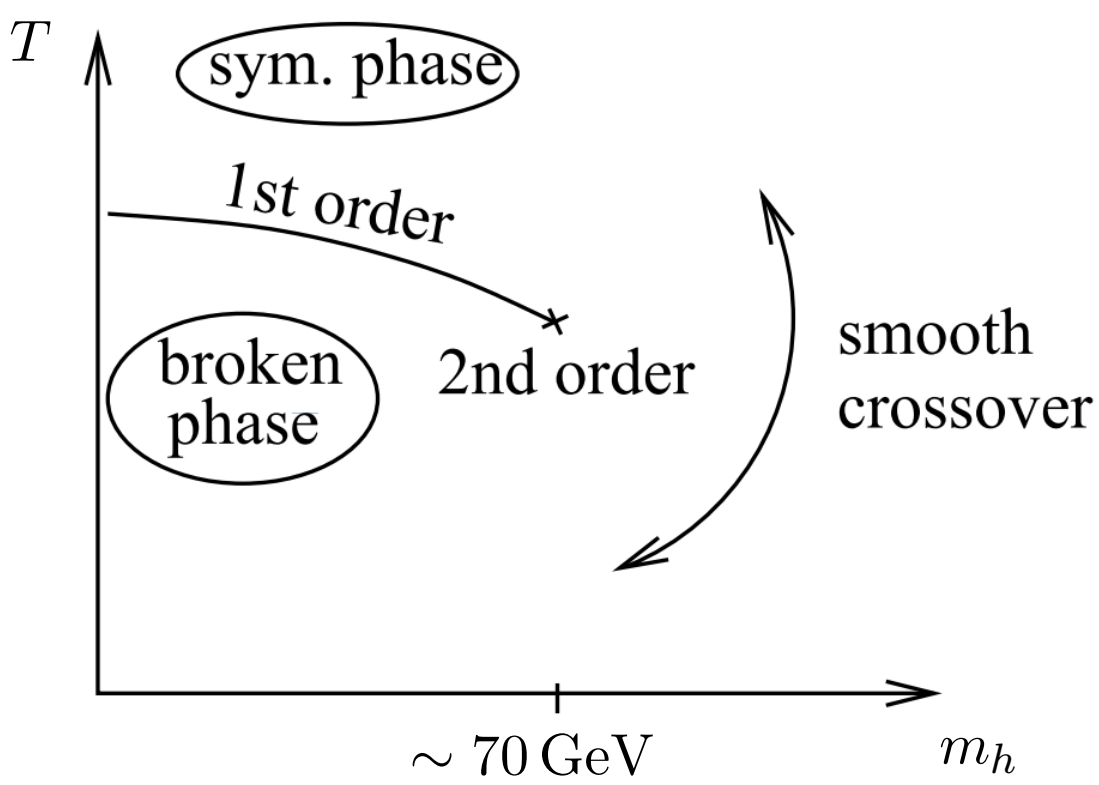

Figure IV.2: Phase diagram for the EWPT in the SM. Figure taken from [90].

### IV.1.3 Validity of the Perturbative Expansion

For the whole examined range of Higgs masses, `CosmoTransitions` classifies the EWPT as a first-order transition with GW parameters $\alpha = 10^{-4} \sim 10^{-3}$ and $\frac{\beta}{H} = 10^5 \sim 10^6$. This is expected at the level of a one-loop analysis, because the Higgs field carries gauge charges. At this point, we should check the validity of the analysis. A quick glimpse at the PT order parameter

$$\begin{aligned} \frac{\phi_{\min}(T_c)}{T_c} &= \frac{\frac{1}{2\lambda}\left[3BT_c + \sqrt{T_c^2(9B^2 - 8A\lambda) + 4\lambda\mu^2}\right]}{T_c} \\ &= \frac{2B}{\lambda} \sim \frac{m_W^3}{\lambda v^3} \sim \sqrt{\lambda}\frac{m_W^3}{m_h^3} \end{aligned} \tag{IV.13}$$

reveals that for large Higgs masses $m_h \gtrsim m_W$ the perturbative expansion is not appropriate [37]. In other words, the result of our one-loop analysis, suggesting a first-order PT at $m_h = 125\,\text{GeV}$, cannot be trusted. And indeed, corrections of higher order seem to overpower the gauge boson induced potential barrier and render the EWPT a *smooth crossover*.[2] This is a result of various lattice simulations, in which the EWPT turned out to be first-order only for Higgs masses $\lesssim 70\,\text{GeV}$ [87, 88, 89]. From the perspective of baryogenesis or GW production, the SM is therefore not very promising and we will now move on to extensions of it.

## IV.2 The Vev Flip-Flop

Extensive searches for many kinds of WIMP DM have thus far been unsuccessful. This provides strong motivation to consider classes of DM beyond the classical WIMP paradigm. One interesting new mechanism is given by the so-called *vev flip-flop* [91, 92]. This model framework introduces a new scalar field which couples to the SM via a Higgs portal term. The presence of this additional field promotes the simple 1-dimensional effective potential to a more complex 2-dimensional one, allowing for a richer phase structure. By choosing the Lagrangian parameters in the right way, one can induce a multi-step PT where the new scalar acquires a VEV for a certain period. It must then be ensured that the universe evolves into the correct electroweak minimum today at $T = 0$.

[2]The denotation ‘smooth crossover’ can be understood in the mathematical sense: The order parameter has an infinite number of derivatives with respect to temperature, as opposed to a second-order PT, where it only needs to be continuous.

To put the flip-flop to a good use, a DM fermion that couples to the new scalar can be introduced. Depending on the desired scenario, either freeze-out or freeze-in, one has to introduce DM depleting or producing processes. Those are switched on and off by the new scalar's VEV. What is called 'freeze-out' here is not the classical decoupling process that sets in when $\Gamma \sim H$. The decoupling will instead be artificially induced and controlled by the dynamics of the flip-flop. That is why the DM candidate is called 'WIMP-like' in this class of models, as opposed to an actual vanilla WIMP which freezes out conventionally.

Depending on the exact model properties, the vev flip-flop features one or even two first-order PTs. This opens up the possibility of probing the novel framework by the search for stochastic GW signals. In the following, two models featuring the vev flip-flop will be presented while focusing on the aspect of possible GW signatures and their detectability.

### IV.2.1 Model A: Real Scalar Singlet

In its simplest form, the additional scalar $S$ is a real gauge singlet which is not charged under any symmetry. The allowed terms in the tree-level potential are

$$\begin{aligned} V_{\text{tree}}(H,S) = &- \mu_H^2 H^\dagger H + \lambda_H (H^\dagger H)^2 \\ &- \frac{1}{2}\mu_S^2 S^2 + \frac{\lambda_{S3}}{3!}\mu_S S^3 + \frac{\lambda_{S4}}{4!} S^4 \\ &+ \lambda_{\text{p3}}\mu_S S (H^\dagger H) + \lambda_{\text{p4}} S^2 (H^\dagger H) \end{aligned} \tag{IV.14}$$

with *portal couplings* $\lambda_{\text{p3}}$ and $\lambda_{\text{p4}}$, linking the hidden and visible sectors. In the broken electroweak phase, the doublet structure (IV.4) of the Higgs field $H$ becomes evident. The potential in terms of the neutral CP-even component reads

$$\begin{aligned} V_{\text{tree}}(h,S) = &- \frac{\mu_H^2}{2} h^2 + \frac{\lambda_H}{4} h^4 \\ &- \frac{1}{2}\mu_S^2 S^2 + \frac{\lambda_{S3}}{3!}\mu_S S^3 + \frac{\lambda_{S4}}{4!} S^4 \\ &+ \frac{\lambda_{\text{p3}}}{2}\mu_S S h^2 + \frac{\lambda_{\text{p4}}}{4} S^2 h^2. \end{aligned} \tag{IV.15}$$

| Field | Type | DOF | $\mathbb{Z}_2$ | $SU(3)_c$ | $SU(2)_L$ | $U(1)_Y$ | Mass at $T=0$ |
|---|---|---|---|---|---|---|---|
| $S$ | real scalar | 1 | 0° | 1 | 1 | 0 | $\mathcal{O}(100\,\text{GeV})$ |
| $\chi$ | Dirac fermion | 4 | 180° | 1 | 1 | 0 | $\mathcal{O}(100\,\text{GeV})$ |

Table IV.3: Additional fields with charges and masses in the real scalar singlet version of the vev flip-flop.

As DM, a Dirac fermion $\chi$ with $\mathbb{Z}_2$ charge is considered.[3] This discrete symmetry is required to prevent $\chi$ from acting like a right-handed neutrino with terms like $\bar{L}_L H \chi$ that would make it unstable. It comes with a bare mass, which is allowed as it is a gauge singlet, and couples to $S$ via a Yukawa term:

$$\mathcal{L} \supset \bar{\chi}(i\partial\!\!\!/ - m_\chi + y_\chi S)\chi \tag{IV.16}$$

We will chose a small $y_\chi$ in this model to allow for a freeze-in scenario where $\chi$ must not be thermalized after reheating. A Yukawa term $\bar{\chi}H\chi$ is not allowed, because $\chi$ carries neither $SU(2)_L$ nor $U(1)_Y$ charge. The whole model will live at the electroweak mass scale (see Tab. IV.3). Applying the earlier developed formulas yields the field dependent masses

$$\begin{aligned}
&m^2_{(h,S)}(h,S) = \\
&\qquad\begin{pmatrix} -\mu_H^2 + 3\lambda_H h^2 + \lambda_{\text{p}3}\mu_S S + \frac{1}{2}\lambda_{\text{p}4} S^2 & (\lambda_{\text{p}3}\mu_S + \lambda_{\text{p}4} S)h \\ (\lambda_{\text{p}3}\mu_S + \lambda_{\text{p}4} S)h & -\mu_S^2 + \lambda_{S3}\mu_S S + \frac{1}{2}\lambda_{S4} S^2 + \frac{1}{2}\lambda_{\text{p}4} h^2 \end{pmatrix}, \\
&m^2_{G^+,G^0}(h,S) = -\mu_H^2 + \lambda_H h^2 + \lambda_{\text{p}3}\mu_S S + \frac{1}{2}\lambda_{\text{p}4} S^2
\end{aligned} \tag{IV.17}$$

and Debye masses

$$\begin{aligned}
\Pi_H(T) &= \left(\frac{1}{2}\lambda_H + \frac{1}{24}\lambda_{\text{p}4} + \frac{1}{4}y_t^2 + \frac{1}{16}g_1^2 + \frac{3}{16}g_2^2\right)T^2, \\
\Pi_S(T) &= \left(\frac{1}{24}\lambda_{S4} + \frac{1}{6}\lambda_{\text{p}4}\right)T^2
\end{aligned} \tag{IV.18}$$

where $y_\chi$ has been neglected. The $W^i$ and $B$ (i.e. $W^\pm$, $Z$ and $A$ in the broken electroweak phase) as well as the top quark are untouched by the new scalar field. The respective field dependent and Debye masses can therefore be taken from (IV.7) and (IV.9). For the calculation of the one-loop effective potential, the mass matrix $m^2_{(h,S)}$ has to be diagonalized. Note that a non-diagonal mass matrix calls for a distinction between mass and flavor states and, if being rigorous, one would have to invent separate labels. We will however still call the masses $m_h(T)$, $m_S(T)$ and assign them to the eigenvalues in a meaningful way, such that they assume their respective

[3] A $\mathbb{Z}_2$ symmetry transformation is a 180° rotation in the complex plane, i.e. a sign flip.

physical values at $T = 0$, where the mixing ceases.

To allow for a two-step PT, multiple non-zero minima are required at tree-level which is achieved by choosing both $\mu_H^2$ and $\mu_S^2$ to be positive. Independent of the temperature driven dynamics, one should arrive at $h = v = 246\,\text{GeV}$ for $T = 0$ to keep the electroweak sector intact, so $\mu_H$ and $\lambda_H$ will be exactly the SM couplings. In order to not overcomplicate the model, we avoid mass mixing between $H$ and $S$ at $T = 0$ by choosing $\lambda_{\text{p}3} \ll 1$. To make the flip-flop work, we want $S = 0$ at $T = 0$. Considering only the tree-level part, this is realized by $\frac{\lambda_{\text{p}4}}{2}v^2 > \mu_S^2$ which makes

$$m_S^2 \equiv m_S^2(h = v, S = 0) = -\mu_S^2 + \frac{\lambda_{\text{p}4}}{2}v^2 \tag{IV.19}$$

positive. This in turn indicates that the extremum at $(h, S) = (v, 0)$ is a minimum. (IV.19) makes it possible to treat $m_S$ as an input parameter and to calculate $\mu_S$ as a function of it. To ensure a symmetric potential with $S \approx 0$ at $T \to \infty$, $\lambda_{S3}$ is also kept small for now. The smallness of $\lambda_{\text{p}3}$, $\lambda_{S3}$ and $y_\chi$ does not violate t'Hooft's definition of naturalness [93], since a $\mathbb{Z}_2$ symmetry would be restored if all three couplings vanish.

Finally, the Higgs VEV and the masses of $h$ and $S$ have to be fixed at $T = 0$. This is done by adding the finite counterterms

$$V_{\text{ct}}(h, S) = -\frac{1}{2}\delta\mu_H^2 h^2 + \frac{1}{4}\delta\lambda_H h^4 - \frac{1}{2}\delta\mu_S^2 S^2 \tag{IV.20}$$

which are determined by

$$\begin{aligned}
\delta\mu_H^2 &= \frac{3}{2v}\,\frac{\partial V_{\text{CW}}}{\partial h}\bigg|_{\substack{h=v\\ S=0}} - \frac{1}{2}\,\frac{\partial^2 V_{\text{CW}}}{\partial h^2}\bigg|_{\substack{h=v\\ S=0}}\,,\\
\delta\lambda_H &= \frac{1}{2v^3}\,\frac{\partial V_{\text{CW}}}{\partial h}\bigg|_{\substack{h=v\\ S=0}} - \frac{1}{2v^2}\,\frac{\partial^2 V_{\text{CW}}}{\partial h^2}\bigg|_{\substack{h=v\\ S=0}}\,,\\
\delta\mu_S &= \frac{\partial^2 V_{\text{CW}}}{\partial S^2}\bigg|_{\substack{h=v\\ S=0}}\,.
\end{aligned}$$

All required ingredients are now gathered and the full one-loop potential together with the daisy corrections, as summarized in (II.123), can be investigated.

#### IV.2.1.1 Dynamics of the Flip-Flop

To study the dynamics of the vev flip-flop, we will pick the parameter point $m_S = 140\,\text{GeV}$, $\lambda_{S3} = -0.1$, $\lambda_{S4} = 6$, $\lambda_{\text{p}4} = 1$. The choice of $\lambda_{S4}$ seems arbitrary, but is based on the fact that one could also normalize the quartic term with $\frac{1}{4}$ instead of $\frac{1}{4!}$ which would correspond to a coupling of 1. In order to track the VEV behavior including all occurring PTs, `CosmoTransitions` is employed. The starting temperature is set to $T_{\text{R}} = 500\,\text{GeV}$.

At very high temperatures, $V_T$ dominates the potential in both field directions and makes the 2-dimensional potential parabolic (Fig. IV.3, top left). The Higgs VEV is zero at this time, while the $S$ VEV is small but non-zero due to the asymmetry induced by $\lambda_{S3}$. With decreasing temperature, the tree-level minimum in the $S$-direction becomes more and more dominant, causing a continuously increasing VEV (Fig. IV.3, top right). This transition, which marks the 'flip' in 'flip-flop', is smooth since there are no barriers which could cause a first-order PT. The absence of a barrier is due to the absence of gauge bosons coupling to $S$. Fig. IV.4 demonstrates the dynamics up to this point in the $h = 0$ slice.

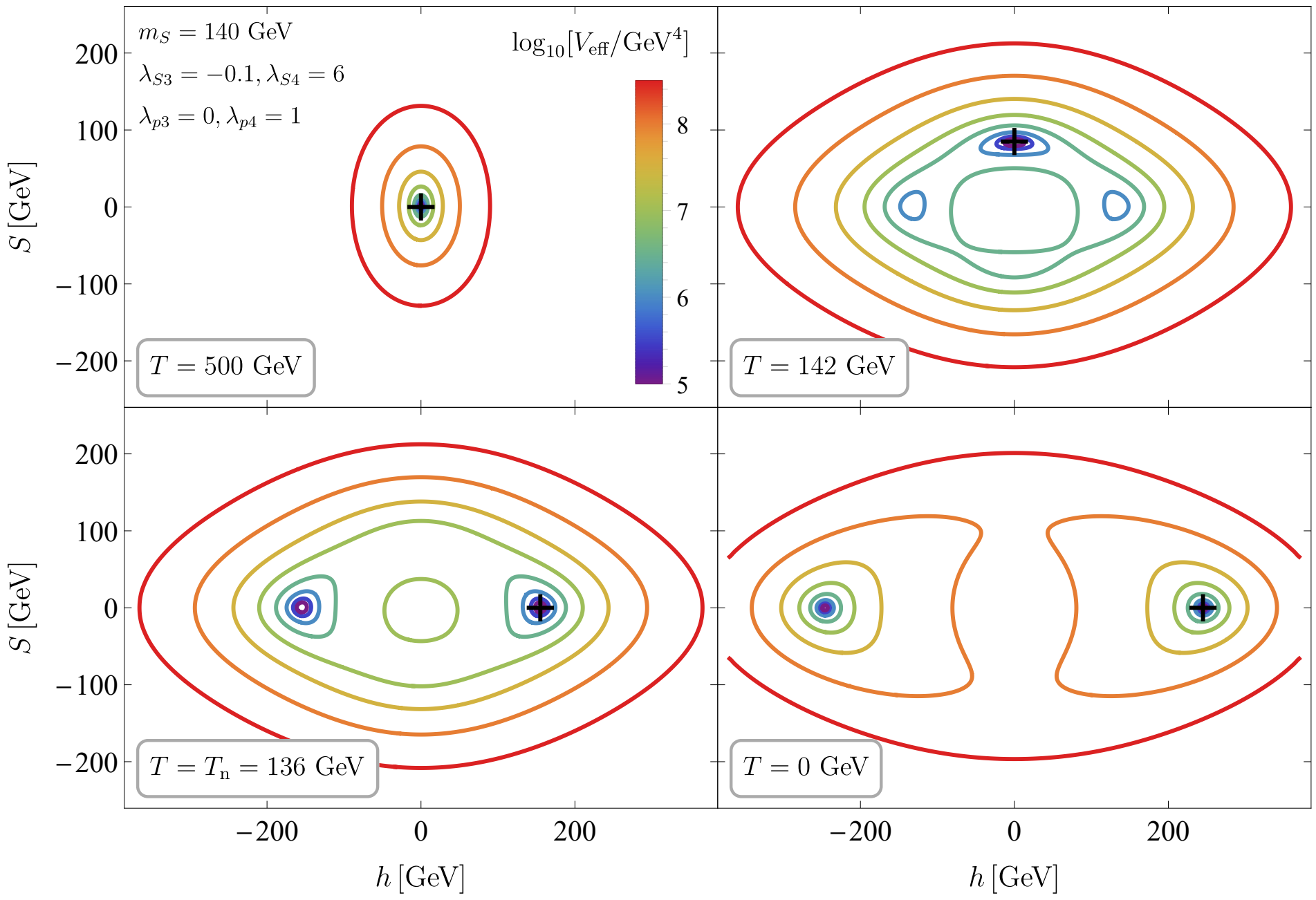

Figure IV.3: Evolution of $V_{\text{eff}}$ with temperature. The black cross marks the current vacuum state.

At the critical temperature $T_{\text{c}} \approx 142\,\text{GeV}$, the tree-level induced minimum at $h > 0$ becomes the global minimum. Instead of occupying it immediately, the universe enters a period of supercooling due to the tree-level barrier between the minima. From this point on, it is just a 'matter of time' until the nucleation condition (III.2) is met and the universe tunnels to the energetically favorable state at $S = 0$, $h > 0$ (Fig. IV.3, bottom left). This transition, the 'flop', happens at $T_{\text{n}} \approx 136\,\text{GeV}$ and takes the role of the EWPT in this model, which is now first-order in contrast to the one in the vanilla SM. Further decreasing the temperature, the VEV stays in the electroweak minimum and gradually moves towards higher field values until $h = v$ at $T = 0$ (Fig. IV.3, bottom right).

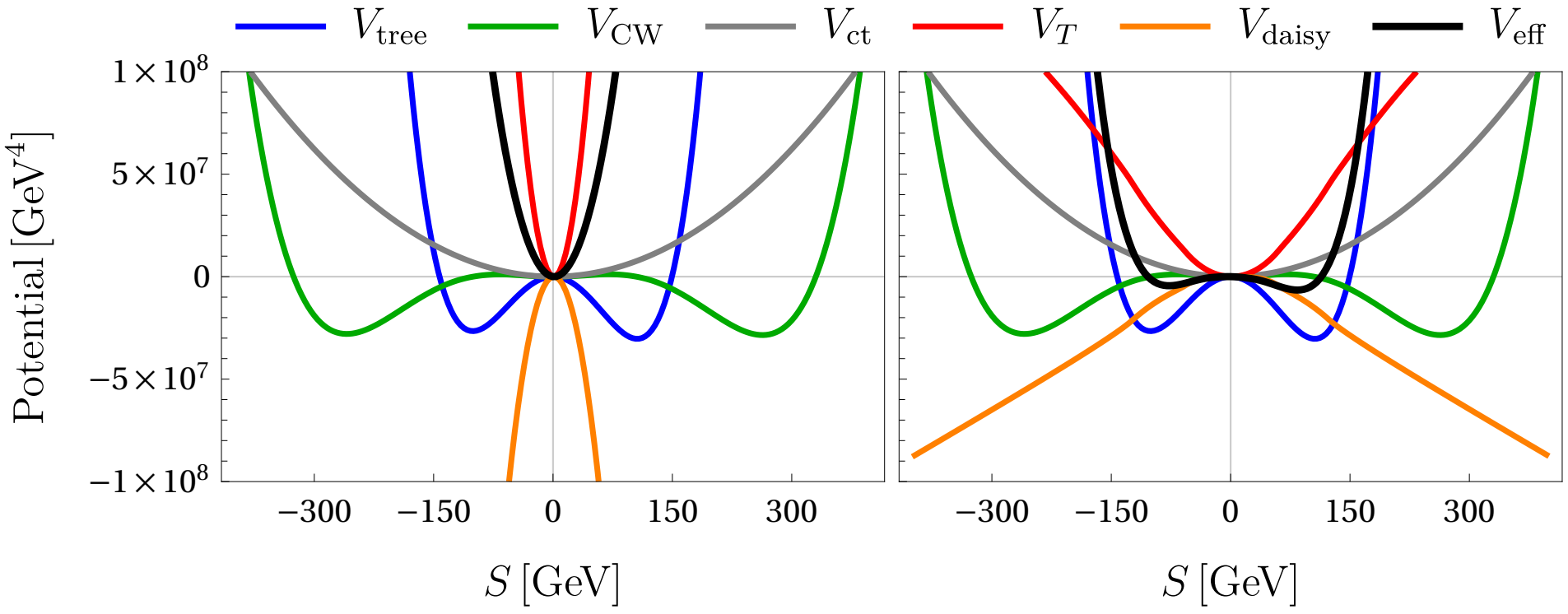

Figure IV.4: $V_{\text{eff}}$ and its components at $h = 0$ for $T = 500\,\text{GeV}$ (left) and $T = 142\,\text{GeV}$ (right). Note that in all potential plots of this thesis, constants have been subtracted in order to align all contributions at the origin. Constant terms in the potential do not influence any dynamics.

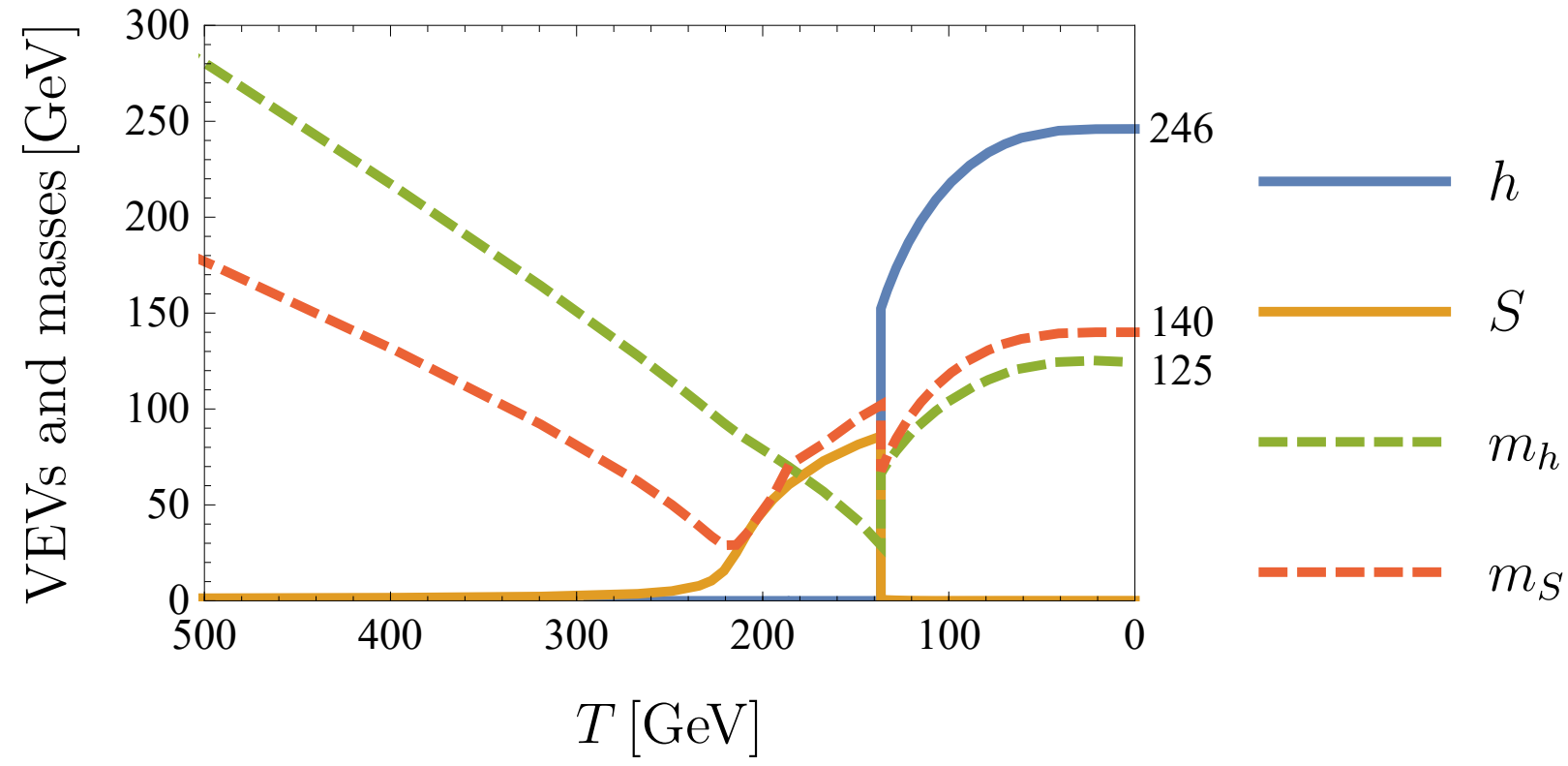

Figure IV.5: Evolution of the scalar VEVs and masses with decreasing temperature, as obtained by using `CosmoTransitions`.

#### IV.2.1.2 Dark Matter Production Mechanisms

The DM aspect of this model is thoroughly discussed in our publication [92] where three different freeze-in scenarios are considered. The most instructive one in the context of the vev flip-flop will be briefly reviewed in the following.

After reheating, we want the abundance of $\chi$ to be negligible, which is realized by the smallness of $y_\chi$. In contrast, $S$ is thermalized due to the $\mathcal{O}(1)$ portal coupling $\lambda_{\text{p4}}$. The possible DM production channels with their respective widths and cross sections

are given by

$$\sim \frac{y_\chi^2}{8\pi m_S^2(T)}(m_S^2(T) - 4m_\chi^2)^{3/2}, \tag{IV.21}$$

$$\sim y_\chi^2(\lambda_{S3}\mu_S + \lambda_{S4}S)^2 \frac{(E^2 - 4m_\chi^2)^{3/2}}{8\pi E^2(m_S^2(T) - E^2)^2\sqrt{E^2 - 4m_S^2(T)}}, \tag{IV.22}$$

$$\sim y_\chi^2(\lambda_{\mathrm{p}3}\mu_S + \lambda_{\mathrm{p}4}S)^2 \frac{(E^2 - 4m_\chi^2)^{3/2}}{8\pi E^2(m_S^2(T) - E^2)^2\sqrt{E^2 - 4m_h^2(T)}}, \tag{IV.23}$$

where $E$ is the center-of-mass energy. The scalar masses are understood to include the one-loop and daisy corrections. In practice, they are obtained by calculating the Hessian matrix of $V_{\mathrm{eff}}$ with respect to $(h, S)$ and diagonalizing it. The decay process contributes only at early times where $m_S(T) > 2m_\chi$ (see Fig. IV.5) and is therefore only of minor importance if we choose a reheating temperature $T_{\mathrm{R}} \lesssim 500\,\mathrm{GeV}$. Both annihilation processes are either realized through the couplings $\lambda_{S3}$ and $\lambda_{\mathrm{p}3}$ or through the $S$ VEV together with the $\mathcal{O}(1)$ couplings $\lambda_{S4}$ and $\lambda_{\mathrm{p}4}$. As already motivated, we consider both trilinear couplings to be small. As a consequence, both diagrams give their main contribution during the intermediate $S \neq 0$ phase of the vev flip-flop. After the EWPT, the DM abundance is almost fixed. The remaining production processes, which can still be active due to non-zero trilinear couplings, will completely come to a halt after Boltzmann suppression sets in at $T \sim 100\,\mathrm{GeV}$. The tiny relic abundance of $S$ can decay through processes like $S \to H \to b\bar{b}$ which requires $\lambda_{\mathrm{p}3}$ to be not exactly zero.

Summarizing, we see that the DM relic abundance is determined by the dynamics of the flip-flop and can be further tuned by changing $y_\chi$ in order to reach exactly the DM abundance observed today.

#### IV.2.1.3 Parameter Space Estimates

Before scanning over the parameter space with `CosmoTransitions`, one can determine approximate bounds of interesting regions. The derived limits will subsequently be displayed as curves in the 2D parameter plots.

#### Ensure minimum in $S$-direction

In order to allow for the desired two-step PT, a tree-level minimum in the $S$-direction is required. This is realized by

$$0 < \mu_S^2 = \frac{\lambda_{\mathrm{p4}}}{2} v^2 - m_S^2 \tag{IV.24}$$

which gives a lower (upper) limit on $\lambda_{\mathrm{p4}}$ ($m_S$), indicated by an orange curve in the subsequent plots.

#### Ensure electroweak vacuum at $T = 0$

At zero temperature, i.e. today, we want the universe to be in the electroweak minimum. There are tree-level minima in both the $h$ and the $S$ direction, so it has to be guaranteed that the global minimum is the one at $(h, S) = (v, 0)$ rather than $(h, S) = (0, S_{\mathrm{min}})$.[4] The condition reads

$$V_{\mathrm{tree}}(v, 0) < V_{\mathrm{tree}}(0, S_{\mathrm{min}}) \tag{IV.25}$$

and evaluates to a quite lengthy expression. This tree-level bound will be drawn as a blue curve in the 2D parameter plots and the excluded region will be labeled by '$S \neq 0$ at $T = 0$'.

#### Ensure symmetry restoration for $T \to \infty$

At high temperatures, the dominant contributions to the effective potential are $V_T$ and $V_{\mathrm{daisy}}$. Due to its negative sign (see Fig. IV.4), the daisy contribution makes the first transition happen earlier, i.e. it increases the critical temperature $T_{\mathrm{c}}$. Depending on the model parameters, this effect can be so strong that a minimum at $S \neq 0$ persists even for $T \to \infty$. To find the corresponding model parameter constraint, $V_{\mathrm{daisy}}$ needs to be compared to $V_T$ in the high-temperature limit. Setting $h = 0$, the daisy contribution can be rewritten as

$$\begin{aligned}
V_{\mathrm{daisy}} &= -\frac{T}{12\pi} \sum_b n_b^{\mathrm{L}} \left[ \left( m^2(S) + \Pi \right)_b^{3/2} - \left( m^2(S) \right)_b^{3/2} \right] \\
&\approx -\frac{T^4}{12\pi} \sum_b n_b^{\mathrm{L}} \left[ \left( \frac{\Pi_b}{T^2} \right)^{3/2} + \frac{3}{2} \sqrt{\frac{\Pi_b}{T^2}} \frac{m_b^2(S)}{T^2} - \left( \frac{m_b^2(S)}{T^2} \right)^{3/2} \right] \\
&\supset -\frac{T^4}{12\pi} \sum_b n_b^{\mathrm{L}} \frac{3}{2} \sqrt{\frac{\Pi_b}{T^2}} \frac{m_b^2(S)}{T^2} \sim T^2 S^2
\end{aligned} \tag{IV.26}$$

[4] There can be two non-degenerate minima in the $S$-direction if $\lambda_{S3} \neq 0$, but $S_{\mathrm{min}}$ refers to the deeper one. The Higgs field is in contrast $\mathbb{Z}_2$ symmetric and we can restrict the analysis to $h \geq 0$ without loss of generality.

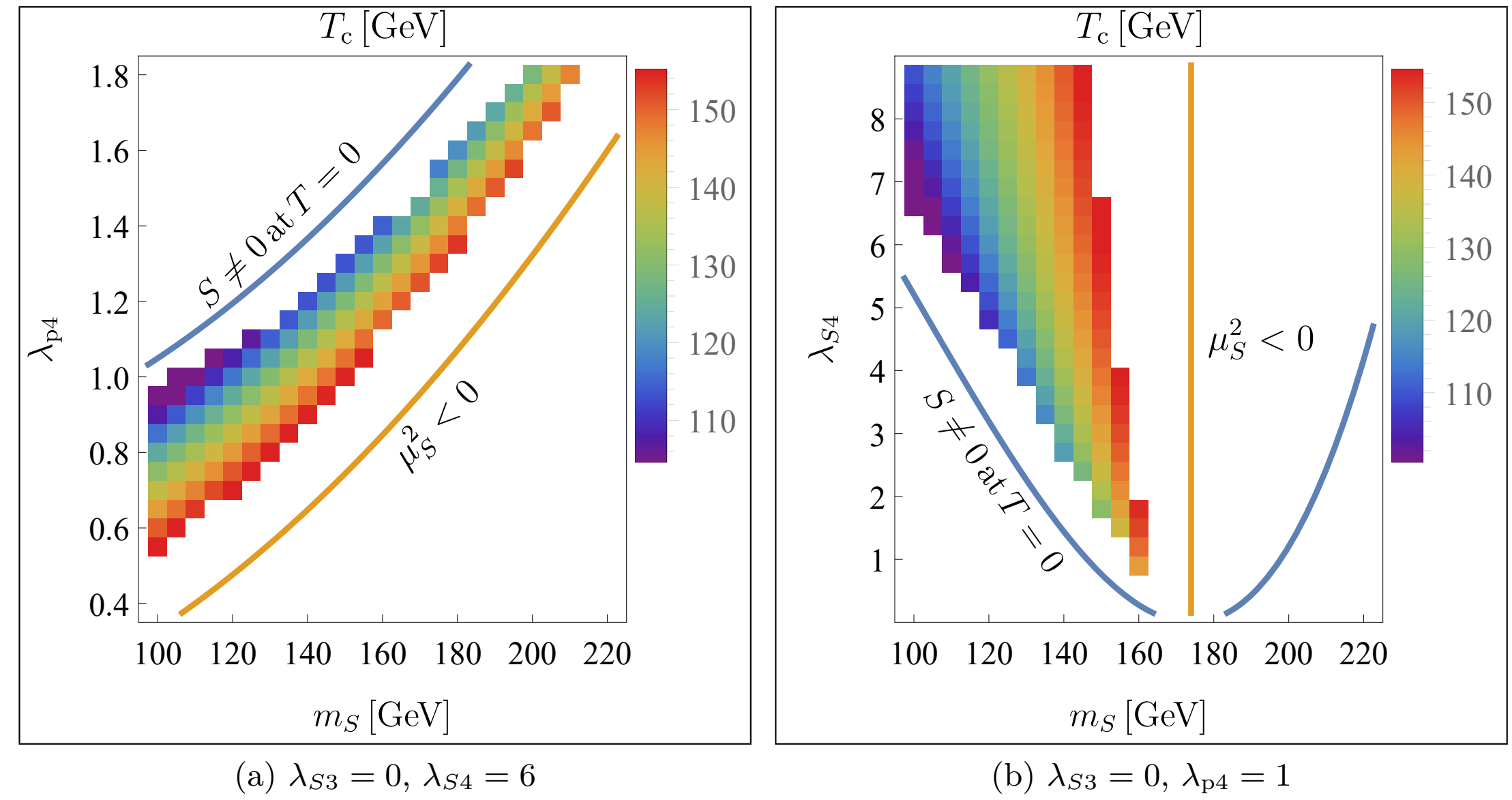

(a) $\lambda_{S3} = 0$, $\lambda_{S4} = 6$

(b) $\lambda_{S3} = 0$, $\lambda_{p4} = 1$

Figure IV.6: Critical temperature of the EWPT in the real singlet vev flip-flop, together with the derived approximate bounds.

In the second step, $(\bullet)^{3/2}$ was expanded around $\frac{m^2}{T^2} = 0$ and in the third one, only the terms $\sim S^2$ have been kept. If the coefficients of $S^2$ in $V_T$ and $V_{\text{daisy}}$ add up to an overall positive value, $S = 0$ is ensured for $T \to \infty$. The condition reads

$$0 < \sum_b \frac{n_b}{24} \frac{\partial^2 m_b^2(S)}{\partial S^2} + \sum_f \frac{n_f}{48} \frac{\partial^2 m_f^2(S)}{\partial S^2} - \sum_b \frac{3 n_b^{\mathrm{L}}}{24\pi} \sqrt{\frac{\Pi_b}{T^2}} \frac{\partial^2 m_b^2(S)}{\partial S^2}. \tag{IV.27}$$

Note that when inserting the Debye masses only up to order $\sim T^2$, the r.h.s. becomes independent of $T$ and $S$ and contains only model parameters. After plugging in the numeric values of all SM couplings, a relation between $\lambda_{S4}$ and $\lambda_{p4}$ is found which can be solved (numerically) for one of the two variables. It turns out that for this specific model, the condition is fulfilled for any combination of the two couplings, assuming they are not both larger than $\mathcal{O}(10)$ at the same time. This limit will therefore not bother us for now, but the calculation will be useful for the next considered model.

* * *

The `CosmoTransitions` results approximately confirm the derived tree-level limits (see Fig. IV.6). The deviations originate from the thermal and non-thermal one-loop contributions. The plots further show a decreasing critical temperature for larger $\lambda_{p4}$ or smaller $m_S$ which results both in a larger $\mu_S^2$, deepening the tree-level $S$ minimum and thereby delaying the second transition of the flip-flop, i.e. the EWPT. Lowering $\lambda_{S4}$ also deepens the $S$ minimum and therefore has the same effect.

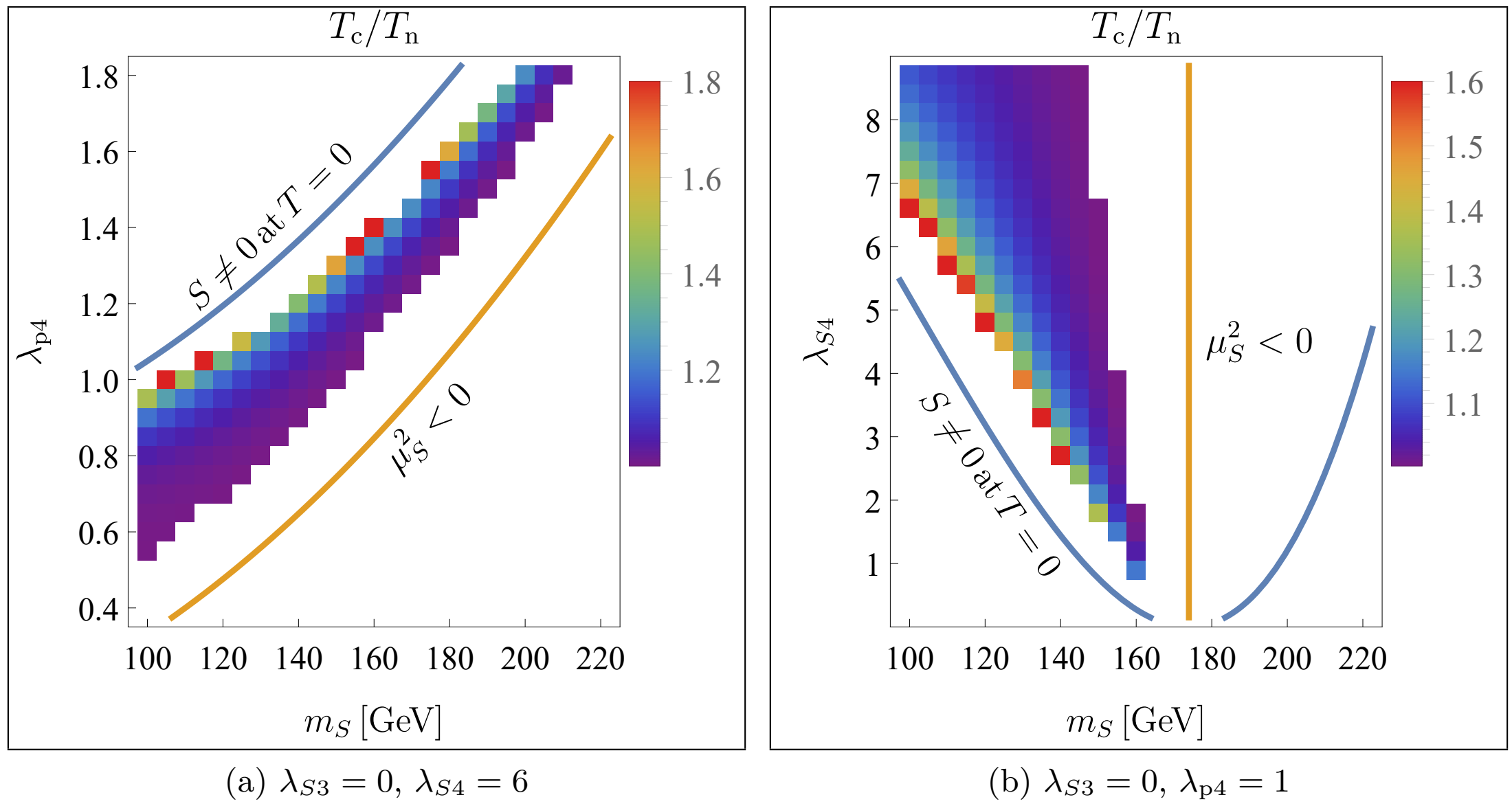

(a) $\lambda_{S3} = 0$, $\lambda_{S4} = 6$

(b) $\lambda_{S3} = 0$, $\lambda_{p4} = 1$

Figure IV.7: Extent of supercooling of the EWPT in the real singlet vev flip-flop, measured by $\frac{T_c}{T_n}$.

#### IV.2.1.4 Gravitational Waves

In the following, the model parameter space will be explored with the objective to find regions in which the EWPT features a sizable GW spectrum. The more supercooling is occurring before nucleation, the stronger the expected PT and the resulting GW spectrum. According to Fig. IV.7, the strongest supercooling occurs for large $\lambda_{p4}$ or small $\lambda_{S4}$ and $m_S$. As explained above, this is the region with a deep $S$ minimum and small $T_c$. At lower temperatures, the tree-level barrier is more dominant compared to the temperature dependent features of the potential. A more dominant barrier in turn delays the nucleation and increases the ratio $\frac{T_c}{T_n}$. More supercooling has two consequences: Firstly, the potential difference $|\Delta V_{\text{eff}}|$ increases and with it the latent heat $\epsilon$ and thus $\alpha$. Secondly, it takes more time to fulfill the nucleation condition (III.2), hinting at a slowly varying $\frac{S_3}{T}$. The parameter $\beta$ is the time derivative of the action and thus also becomes small. Both, large $\alpha$ and small $\frac{\beta}{H}$, lead to strong GW signals.

The most optimistic points in the explored parameter space, as can be seen in Fig. IV.8 and IV.9, are around $\lambda_{p4} \sim 1$, $\lambda_{S4} \sim 6$, $m_S \sim 100\,\text{GeV}$ and yield $\alpha \sim 0.1$, $\frac{\beta}{H} \sim 300$. The ratio $\frac{\alpha}{\alpha_\infty}$ ranges from $10^2$ to $10^3$ (not shown as figure), which clearly indicates the runaway regime.

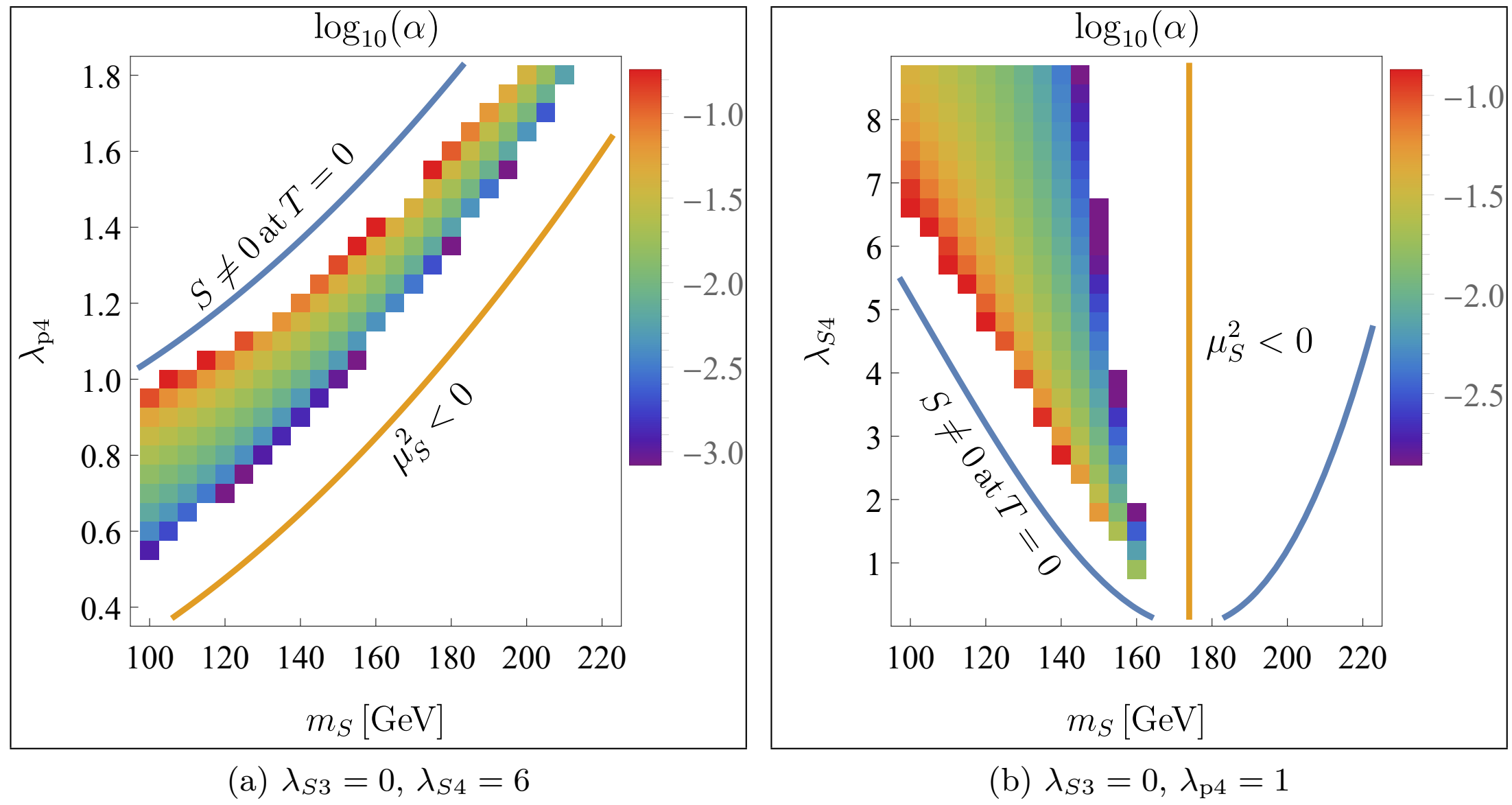

(a) $\lambda_{S3} = 0,\ \lambda_{S4} = 6$

(b) $\lambda_{S3} = 0,\ \lambda_{\text{p}4} = 1$

Figure IV.8: Strength $\alpha$ of the EWPT in the real singlet vev flip-flop.

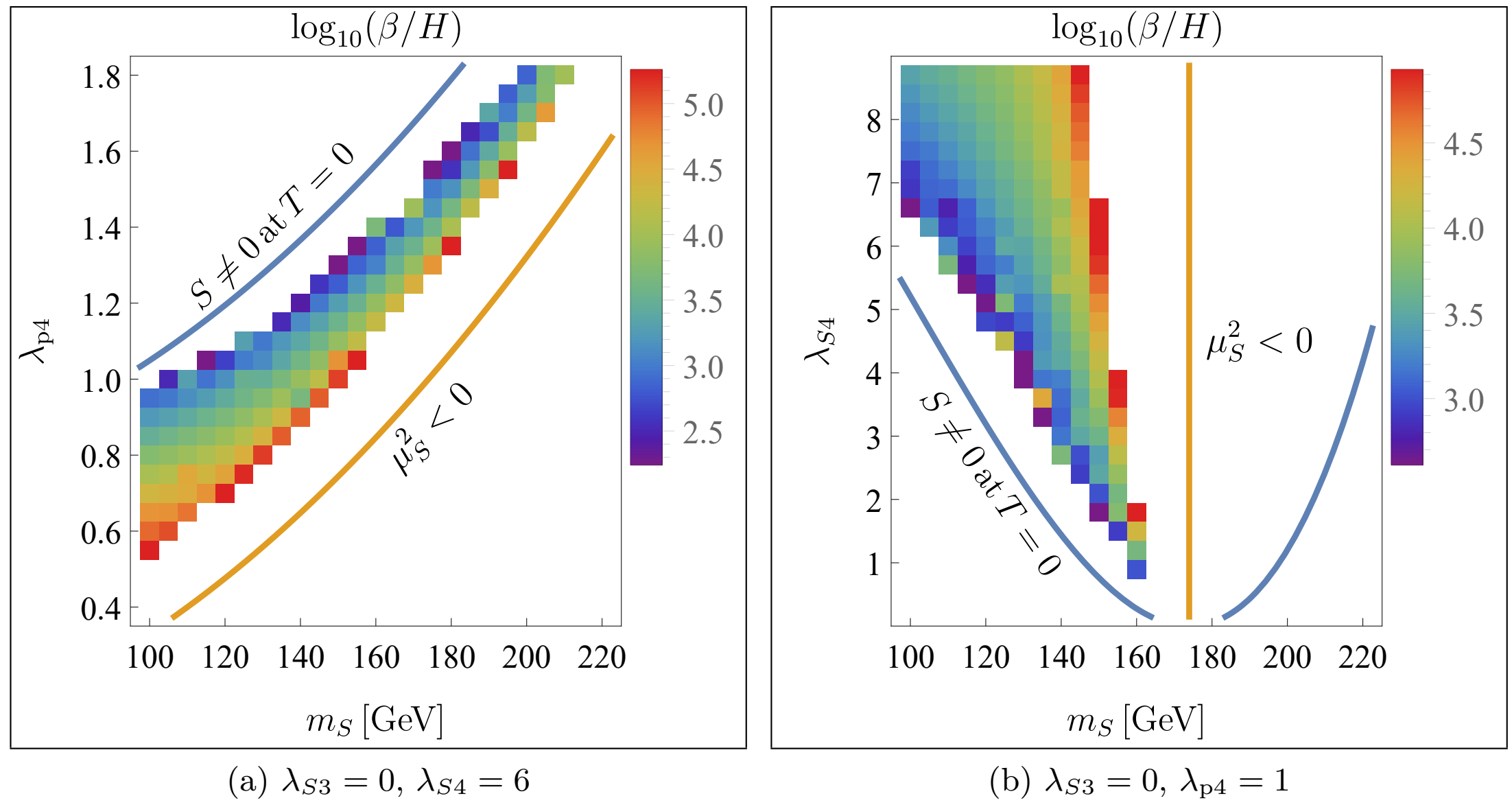

(a) $\lambda_{S3} = 0,\ \lambda_{S4} = 6$

(b) $\lambda_{S3} = 0,\ \lambda_{\text{p}4} = 1$

Figure IV.9: Inverse time scale $\frac{\beta}{H}$ of the EWPT in the real singlet vev flip-flop. The instabilities are caused by `CosmoTransitions` are not physical.

### Detectability

Due to $\frac{\alpha}{\alpha_\infty} \gg 1$, the scalar contribution is the dominant GW production process. Applying the formulae of Section III.3 yields a redshifted peak frequency of $\mathcal{O}(\text{mHz})$ for the promising parameter region. This is in the range of planned space-based experiments. An explicit calculation of the SNRs reveals however, that only Ultimate DECIGO would be capable of probing most of the parameter space (see Fig. IV.10). BBO and DECIGO cover only the most optimistic regions, while the signal would be completely out of reach for LISA or B-DECIGO.

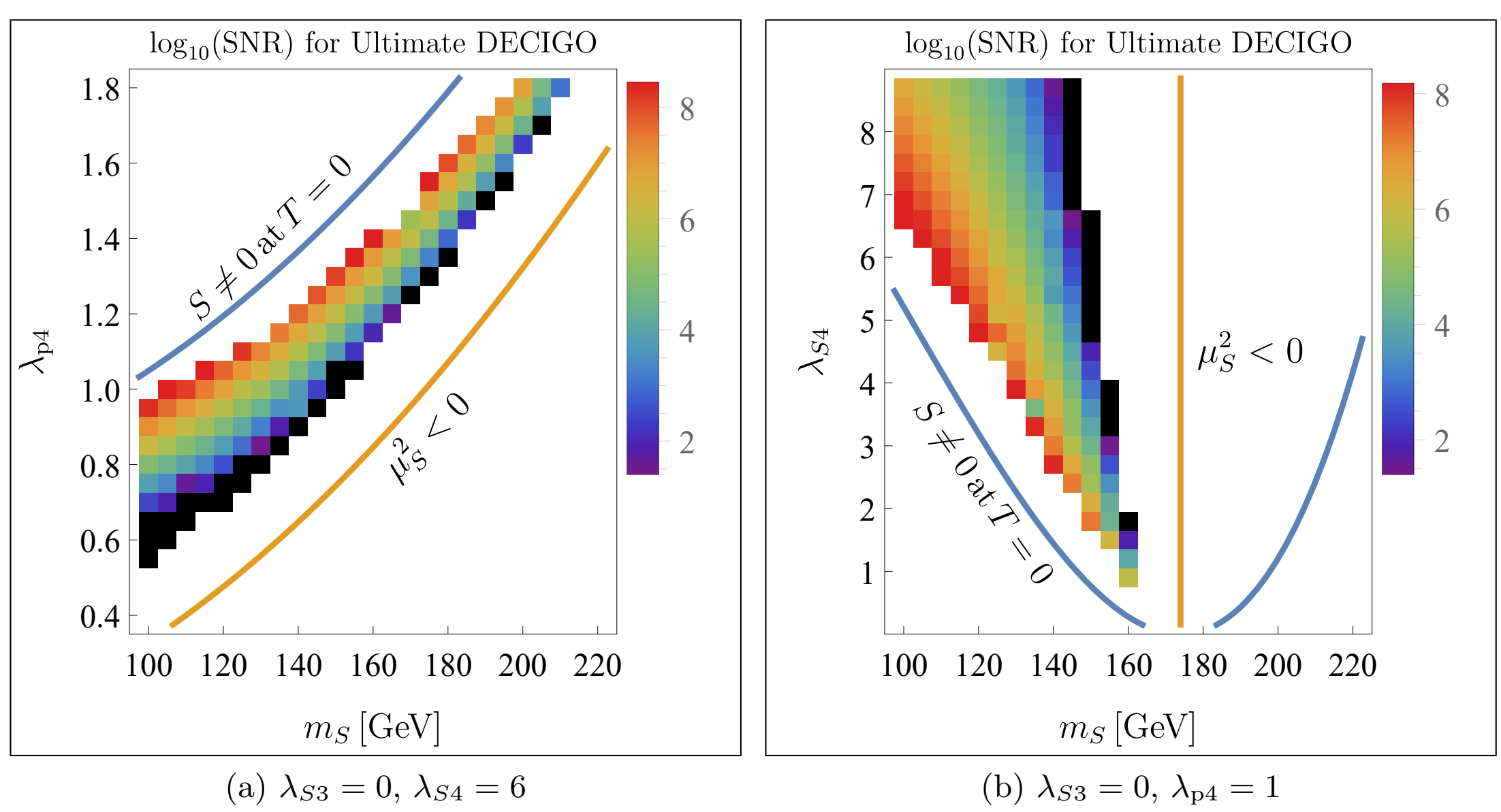

(a) $\lambda_{S3} = 0$, $\lambda_{S4} = 6$ (b) $\lambda_{S3} = 0$, $\lambda_{\mathrm{p}4} = 1$

Figure IV.10: SNR of the EWPT in the real singlet vev flip-flop for Ultimate DECIGO. Black regions are below the detectable threshold.

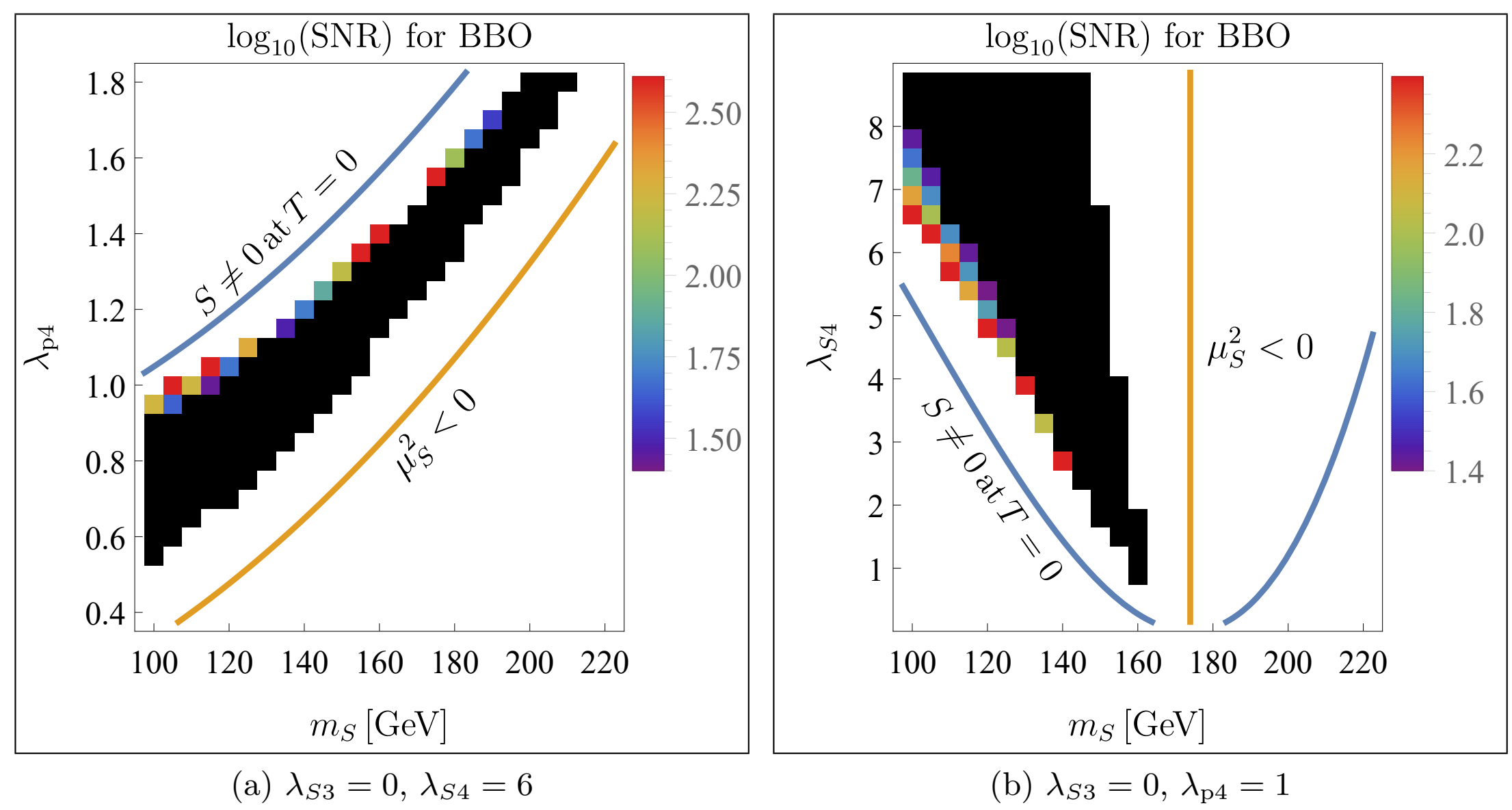

(a) $\lambda_{S3} = 0$, $\lambda_{S4} = 6$ (b) $\lambda_{S3} = 0$, $\lambda_{\mathrm{p}4} = 1$

Figure IV.11: SNR of the EWPT in the real singlet vev flip-flop for BBO. The plot for DECIGO is almost identical and therefore not shown.

#### IV.2.1.5 Cubic Tree-Level Term

Since the scalar field $S$ is not coupled to any gauge bosons, no thermally induced barrier exists that could render the first PT first-order. However, there is still the tree-level term $\sim \lambda_{S3}$. In the examined region, it turns out that for any $|\lambda_{S3}|$ the barrier in $V_{\text{eff}}$ forms too late, when the VEV already occupies the deeper minimum (see Fig. IV.12, left). As pointed out in [60], the barrier appears early enough in a region of large $|\lambda_{S3}|$ and $\lambda_{S4}$ (see Fig. IV.12, right). Unfortunately, the $S$ minimum is too deep in this region and the electroweak minimum will not be occupied at $T = 0$ (see Fig. IV.13).

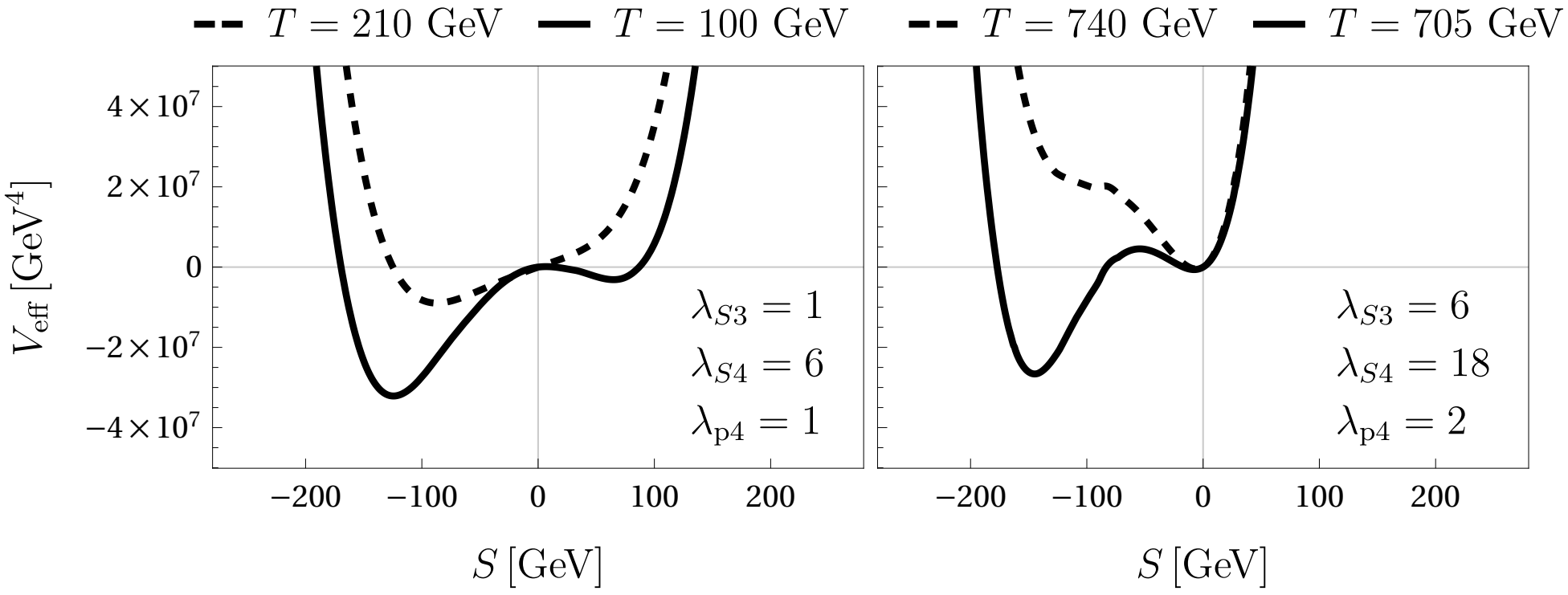

Figure IV.12: $V_{\text{eff}}$ before (dashed curve) and after (solid curve) the tree-level barrier starts to become visible. $m_S = 140\,\text{GeV}$ in both plots.

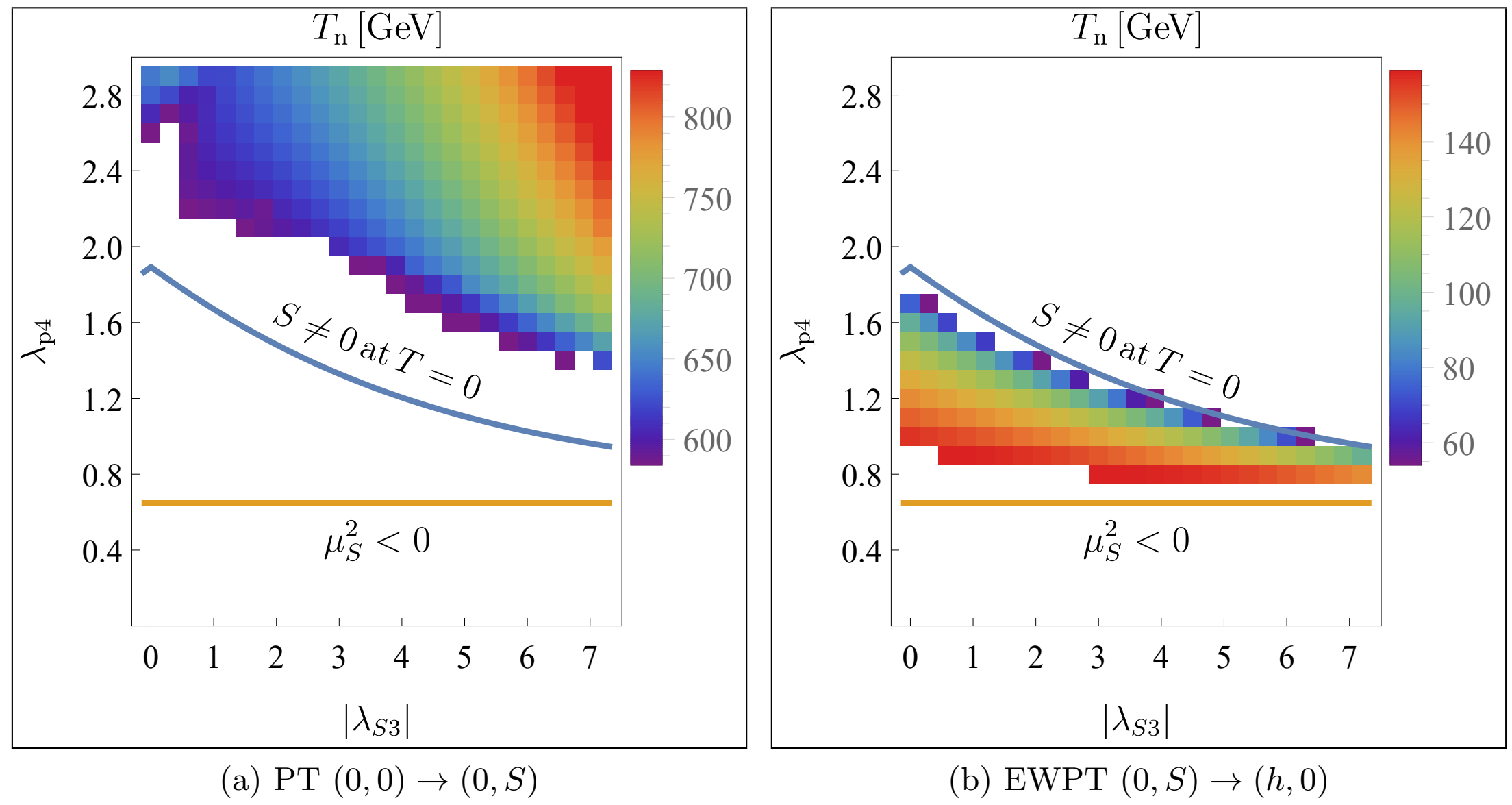

(a) PT $(0, 0) \to (0, S)$

(b) EWPT $(0, S) \to (h, 0)$

Figure IV.13: Nucleation temperature of the first (left) and second (right) PT with $m_S = 140\,\text{GeV}$, $\lambda_{S4} = 18$. The first transition is first-order only in a region where the electroweak minimum is not restored at $T = 0$.

| Field | Type | DOF | $\mathbb{Z}_3$ | $SU(3)_c$ | $SU(2)_L$ | $U(1)_Y$ | Mass at $T=0$ |
|---|---|---|---|---|---|---|---|
| $S$ | complex scalar | 6 | $+120°$ | 1 | 3 | 0 | $\mathcal{O}(100\,\mathrm{GeV})$ |
| $\chi$ | Dirac fermion | 4 | $+120°$ | 1 | 1 | 0 | $\mathcal{O}(\mathrm{TeV})$ |
| $\Psi$ | Dirac fermion | 12 | $-120°$ | 1 | 3 | 0 | $\mathcal{O}(\mathrm{TeV})$ |
| $\Psi'$ | Dirac fermion | 12 | $-120°$ | 1 | 3 | 0 | $\mathcal{O}(\mathrm{TeV})$ |

Table IV.4: Additional fields with charges and masses in the scalar triplet version of the vev flip-flop.

### IV.2.2 Model B: Scalar Triplet

Another variant of the vev flip-flop is realized by considering a scalar with SM gauge charge. In this specific model, the SM is augmented by a $SU(2)_L$ scalar triplet $S$, a fermion $\chi$ which serves as DM and two auxiliary fermions $\Psi^{(\prime)}$ (see Tab. IV.4). The additional particle content lives in the 100 GeV to TeV range and is charged under a $\mathbb{Z}_3$,[5] which is required to stabilize the DM. The allowed terms in the tree-level potential are

$$\begin{aligned} V_{\mathrm{tree}}(H,S) = &-\mu_H^2 H^\dagger H + \lambda_H (H^\dagger H)^2 \\ &-\mu_S^2 S^\dagger S + \lambda_S (S^\dagger S)^2 + \tilde{\lambda}_S (S^\dagger T_3^i S)(S^\dagger T_3^i S) \\ &+\lambda_{\mathrm{p}} (H^\dagger H)(S^\dagger S) + \tilde{\lambda}_{\mathrm{p}} (H^\dagger T_2^i H)(S^\dagger T_3^i S) \end{aligned} \tag{IV.28}$$

where a sum runs over $i = 1,2,3$ with $SU(2)$-generators $T_2^i$ ($T_3^i$) in the fundamental doublet (adjoint triplet) representation. The components of $S$ are given by

$$S = \begin{pmatrix} s^+ \\ \frac{1}{\sqrt{2}}(s+ia) \\ s^- \end{pmatrix}.$$

The neutral CP-even component $s$ will acquire a VEV and thereby break $SU(2)_L$, giving a mass to $W^i$. Note that this was not the case in the singlet version of the flip-flop model. The dark sector fermions carry bare masses and their Yukawa couplings are

$$\mathcal{L} \supset y_\chi S^\dagger \bar{\chi} \Psi + y'_\chi S^\dagger \bar{\chi} \Psi' + y_\Psi \epsilon^{ijk} S^i \bar{\Psi}^j (\Psi'^k)^c + \mathrm{h.c.} \tag{IV.29}$$

while the $\mathbb{Z}_3$ symmetry, a possible remnant of a $U(1)$ broken at higher energy, prevents direct couplings between $\chi$ and the SM. We consider the case of tiny $\tilde{\lambda}_{\mathrm{p}}$ and $y_\chi^{(\prime)}$, which will be motivated further below. These couplings will therefore not appear in any of the (thermal) masses.

The field dependent masses are provided by [91] and reveal a mixing in the basis $(h,s)$ and $(s^+, s^-, G^+)$ respectively. It turns out that for $s=0$, both mass matrices

[5] A $\mathbb{Z}_3$ symmetry transformation is a $\pm 120°$ rotation in the complex plane.

are diagonal, i.e. no mixing is present at $T = 0$ and the mass of $s$ is given by

$$m_s^2 \equiv m_s^2(h = v, s = 0) = -\mu_S^2 + \frac{\lambda_\mathrm{p}}{2} v^2. \tag{IV.30}$$

Interestingly,

$$m_{W^\pm}^2(h, s) = \frac{1}{4} g_2^2 (h^2 + 4s^2) \tag{IV.31}$$

is now also a function of the additional scalar VEV due to $S$ being a gauge triplet. The Debye masses amount to

$$\begin{aligned}
\Pi_H(T) &= \left( \frac{1}{2}\lambda_H + \frac{1}{4}\lambda_\mathrm{p} + \frac{1}{4} y_t^2 + \frac{1}{16} g_1^2 + \frac{3}{16} g_2^2 \right) T^2, \\
\Pi_S(T) &= \left( \frac{2}{3}\lambda_S + \frac{1}{3}\tilde{\lambda}_S + \frac{1}{6}\lambda_\mathrm{p} + \frac{1}{2} g_2^2 \right) T^2, \\
\Pi_{W^i}^\mathrm{L}(T) &= \frac{5}{2} g_2^2 T^2
\end{aligned} \tag{IV.32}$$

while $\Pi_B$ is unchanged by the new scalar field. For the calculation of $\Pi_{W^i}^\mathrm{L}$ we used the characteristic constants given by $\frac{1}{2}\delta^{ij} = \mathrm{Tr}[T_2^i T_2^j]$ and $2\delta^{ij} = \mathrm{Tr}[T_3^i T_3^j]$.

Lastly, finite counterterms have to be added in order to fix $v$, $m_h$ and $m_s$ at $T = 0$. They are identical to the singlet model counterterms (IV.20) after simply exchanging $S \leftrightarrow s$. Overall, apart from the different normalization $\frac{\lambda_{S4}}{6} \leftrightarrow \lambda_S$, there are many similarities between the singlet and the triplet model. The dynamics of the vev flip-flop are qualitatively the same, with a subtle difference: The $W$ boson now induces a cubic term in $V_T$ which gives rise to a barrier, making the first PT first-order.

#### IV.2.2.1 Parameter Space Estimates

The parameter limits for a working flip-flop can be derived in the same fashion as for the singlet model. The condition for ending up in the electroweak minimum now reads

$$\begin{aligned}
& V_\mathrm{tree}(v, 0) < V_\mathrm{tree}(0, s_\mathrm{min}) \\
\Leftrightarrow \qquad & \lambda_H \lambda_S v^4 < \left( \frac{\lambda_\mathrm{p}}{2} v^2 - m_s^2 \right)^2 .
\end{aligned} \tag{IV.33}$$

The requirement for $\mathbb{Z}_3$ symmetry restoration at $T \to \infty$ now constrains the parameter space of interest, which was not the case for the singlet model. The corresponding curve in the 2D parameter plots will be displayed in green and the excluded region marked by '$s \neq 0$ at $T \to \infty$'.

#### IV.2.2.2 Dark Matter Decay Mechanisms

In contrast to the singlet case, this version of the model utilizes DM freeze-out to generate the correct relic abundance [91]. In order to allow for a freeze-out in the first place, $\chi$ must be abundant after reheating. This is realized by interactions which froze out far above the electroweak scale. Since then, $\chi$ is not in thermal equilibrium with the SM due to the smallness of $y_\chi^{(\prime)}$, with an abundance far above the SM equilibrium abundance. $S$ and $\Psi^{(\prime)}$ are in contact with the thermal bath as we chose sizable $\lambda_\mathrm{p}$ and $y_\Psi$.

The first transition of the vev flip-flop, giving $s$ a VEV and thereby breaking $\mathbb{Z}_3$ symmetry, opens the decay channels $\chi \to \Psi^{(\prime)} \to \Psi^{(\prime)} W$ by allowing mixing between $\chi$ and $\Psi^{(\prime)}$. The inverse processes $\Psi^{(\prime)} W \to \chi$ are in principle also possible, but they play no important role as long as $\chi$'s abundance is far above equilibrium. The 'flop' restores $\mathbb{Z}_3$ and fixes the DM yield. As shown in [91], the correct relic abundance can be achieved at any parameter point that features the flip-flop by simply tuning $y_\chi^{(\prime)}$ and $m_\chi$.

This scenario is not identical to the conventional WIMP paradigm where DM is assumed to couple weakly but directly to the SM and to be in thermal equilibrium until it freezes-out when $\Gamma \sim H$. The DM abundance here is determined by the dynamics of the vev flip-flop. In order to make everything work, a few additional assumptions have to be made: The abundance of charged DM is highly constrained by experiments [94]. The neutral components of $\Psi^{(\prime)}$ and $S$, which are not completely depleted when the VEV turns off, thus need to be lighter than the charged ones. This is only the case for small $\tilde{\lambda}_\mathrm{p}$. We furthermore want $\chi$ and not $S$ or $\Psi^{(\prime)}$ to be DM, which imposes relative constraints on the masses.

#### IV.2.2.3 Gravitational Waves

As already mentioned, the gauge charge of $S$ renders the first of the two PTs first-order, making it interesting from the GW point of view. In contrast to the second PT, the barrier is here induced by thermal one-loop rather than tree-level effects. A strong GW spectrum is therefore expected in regions where the tree-level features are subdominant compared to the size of the one-loop barrier. This is achieved through small tree-level parameters $\lambda_S$, $\tilde{\lambda}_S$ and $\mu_S$, where the latter is small for either large $m_s$ or small $\lambda_\mathrm{p}$.

The `CosmoTransitions` results confirm these expectations (see Fig. IV.14 and IV.15). Unfortunately, even at the most optimistic points, the smallness of $\alpha \lesssim 10^{-3}$ in combination with the large $\frac{\beta}{H} \gtrsim 10^5$ makes the spectrum completely undetectable for any of the considered GW observatories. The weakness of the PT can be attributed to the fact that only 3 gauge bosons couple to $S$.[6]

[6] In the Economical 3-3-1 model [95], the SM $SU(2)_L$ symmetry is promoted to a $SU(3)_L$ and the EWPT proceeds in two steps. The total of now 8 gauge bosons induces a barrier which makes the transitions strongly first-order. Despite multiple attempts however, we were not able to reproduce the results of [96] where a detectable double bump spectrum is claimed.

The GW parameters of the second transition $(0, s) \to (h, 0)$ are almost unchanged compared to those of the singlet vev flip-flop. The SNRs and resulting detectable regions are very similar to the ones shown in Fig. IV.10 and IV.11. We therefore do not display them again.

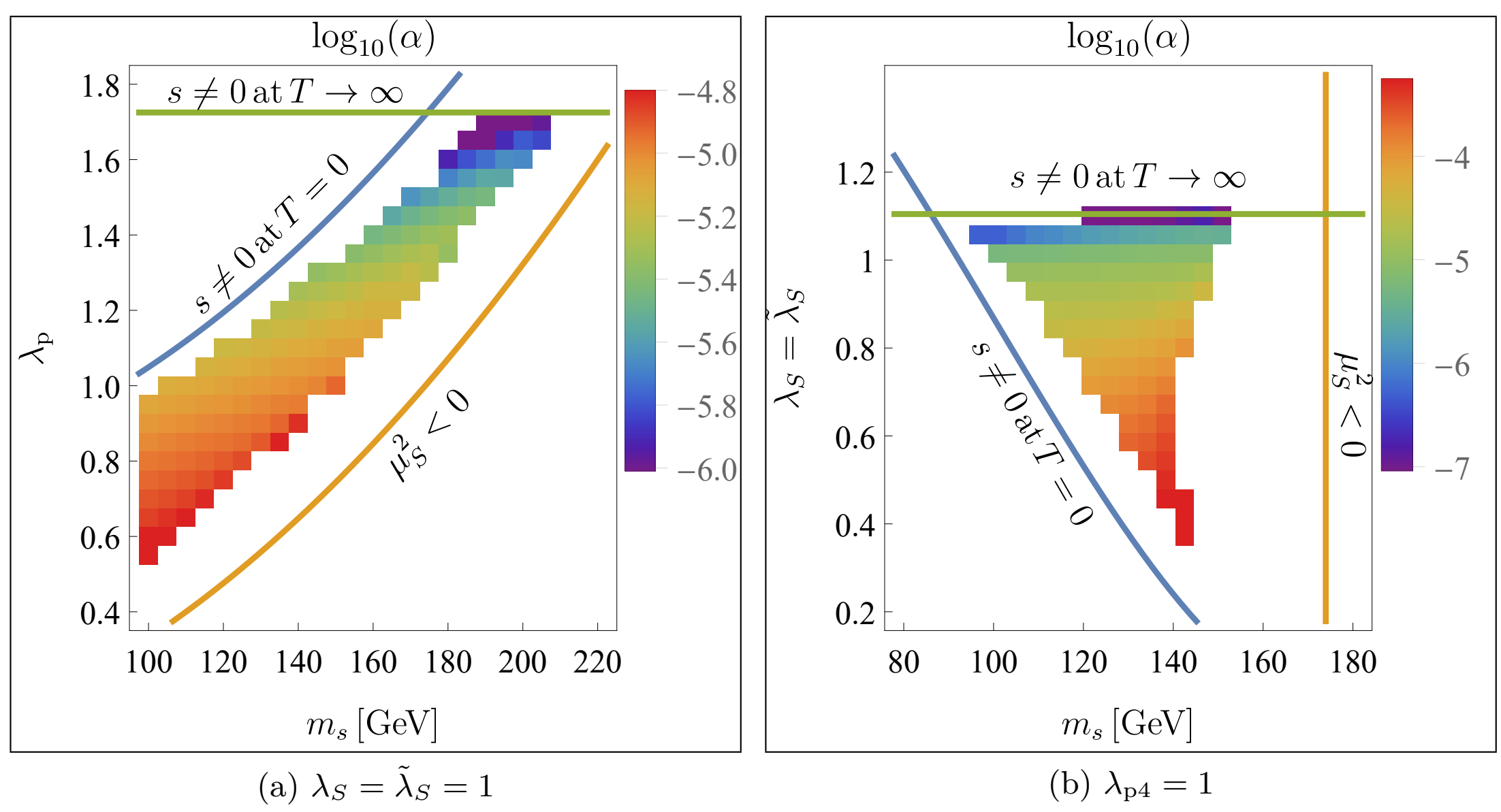

(a) $\lambda_S = \tilde{\lambda}_S = 1$ (b) $\lambda_{\mathrm{p4}} = 1$

Figure IV.14: Strength $\alpha$ of the first transition ('flip') in the scalar triplet vev flip-flop.

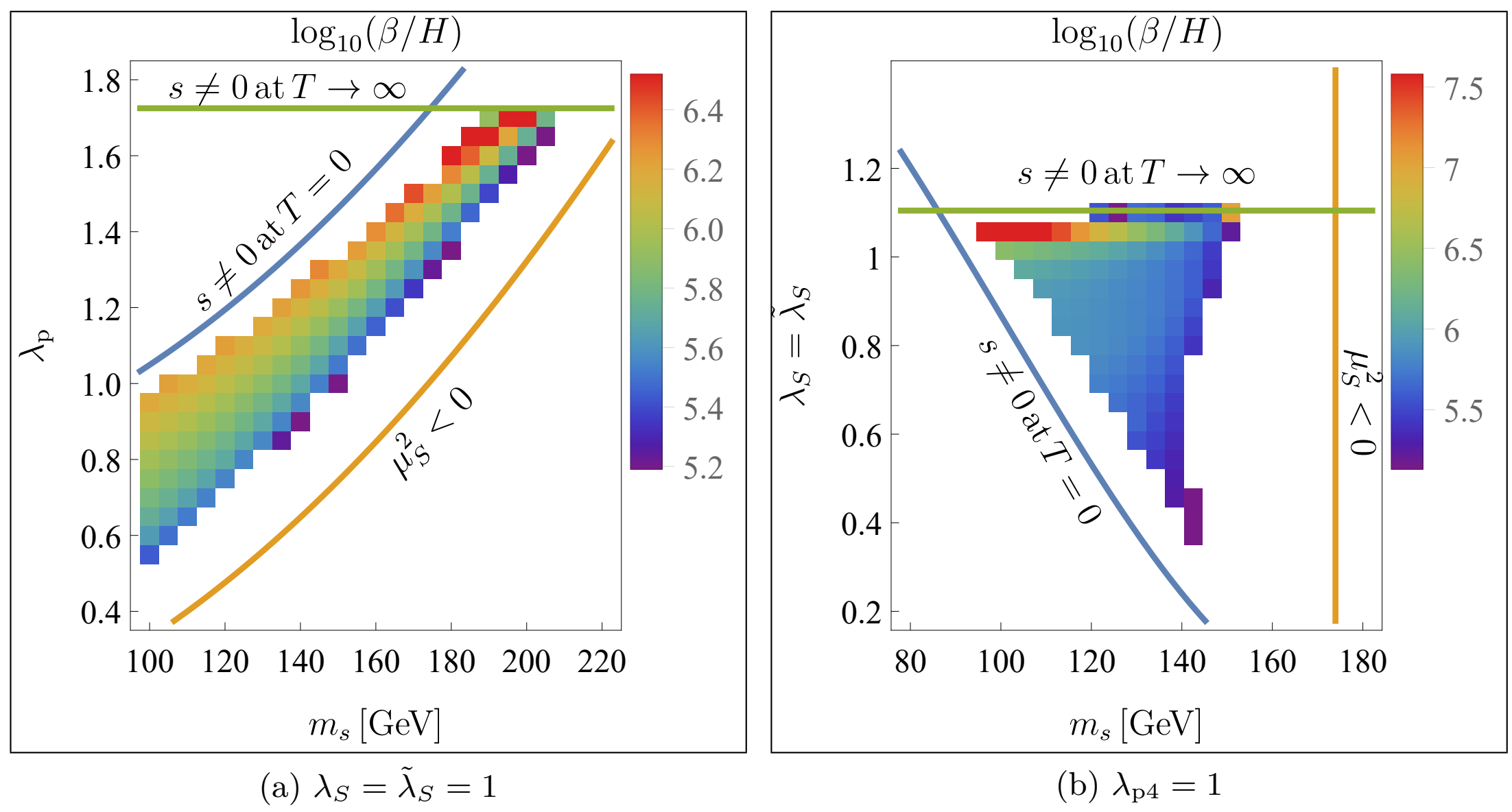

(a) $\lambda_S = \tilde{\lambda}_S = 1$ (b) $\lambda_{\mathrm{p4}} = 1$

Figure IV.15: Inverse time scale $\frac{\beta}{H}$ of the first transition ('flip') in the scalar triplet vev flip-flop.

## IV.3 The Dark Photon

So far, only models with discrete symmetries in the dark sector were considered. We will now move on to a very minimalistic extension of the SM featuring a dark abelian gauge symmetry. This gives rise to 'dark electromagnetism' mediated by the *dark photon*, an idea first proposed already 30 years ago [97]. We further introduce a DM fermion $\chi$ and a complex dark scalar $S$, both charged under the new gauge symmetry $U(1)_D$. In order to allow for a gauge invariant Lagrangian, the covariant derivative

$$D_\mu = \partial_\mu + \cdots + i g_D q_D B'_\mu \tag{IV.34}$$

is augmented by a term involving the hidden gauge field $B'_\mu$ with coupling strength $g_D$ and charge $q_D$ which we set to 1 for $S$ and $\chi$. The Lagrangian reads

$$\begin{aligned} \mathcal{L} \supset &(D_\mu S)^* (D^\mu S) + \bar{\chi}(i\not{D} - m_\chi)\chi \\ &- \frac{1}{4} F'_{\mu\nu} F'^{\mu\nu} - \frac{\epsilon}{2} F_{\mu\nu} F'^{\mu\nu} - V(H,S) \end{aligned} \tag{IV.35}$$

where the new kinetic terms are shown explicitly to illustrate how the new gauge boson enters. $F^{(\prime)}_{\mu\nu} = \partial_\mu B^{(\prime)}_\nu - \partial_\nu B^{(\prime)}_\mu$ are the $U(1)_Y$ ($U(1)_D$) field strength tensors. The term $\sim \epsilon$ induces an effective mixing between visible and dark photon, connecting the two sectors. Note that the gauge structure forbids a Yukawa term $\sim S\bar{\chi}\chi$ for vector-like DM.[7]

Measurements of positronium decays constrain $\epsilon$ for long-range dark forces [100]. A force with heavy mediator in turn has a limited range and is therefore not as constrained. This motivates a dark equivalent of the Higgs mechanism where $U(1)_D$ is spontaneously broken and the dark gauge boson receives a mass.[8] The breaking is induced by the scalar

$$S = \frac{1}{\sqrt{2}}(s + ia) \tag{IV.36}$$

where the CP-even component $s$ acquires a VEV. We assume that our DM candidate $\chi$ froze out far above the $U(1)_D$ symmetry breaking scale, providing the correct relic abundance. This gives us more freedom to explore the PT itself.

The given symmetries allow a potential of the form

$$\begin{aligned} V(H,S) = &- \mu_H^2 H^\dagger H + \lambda_H (H^\dagger H)^2 \\ &- \mu_S^2 S^* S + \lambda_S (S^* S)^2 + \lambda_\mathrm{p} (S^* S)(H^\dagger H) \end{aligned} \tag{IV.37}$$

where we require a tiny $\lambda_\mathrm{p}$ to keep the coupling between the sectors small. The field

[7] Chiral DM requires a set of additional fermions in order to cancel $U(1)_D$ gauge anomalies [98].
[8] Another way of making vector bosons massive is provided by the Stueckelberg mechanism [101].

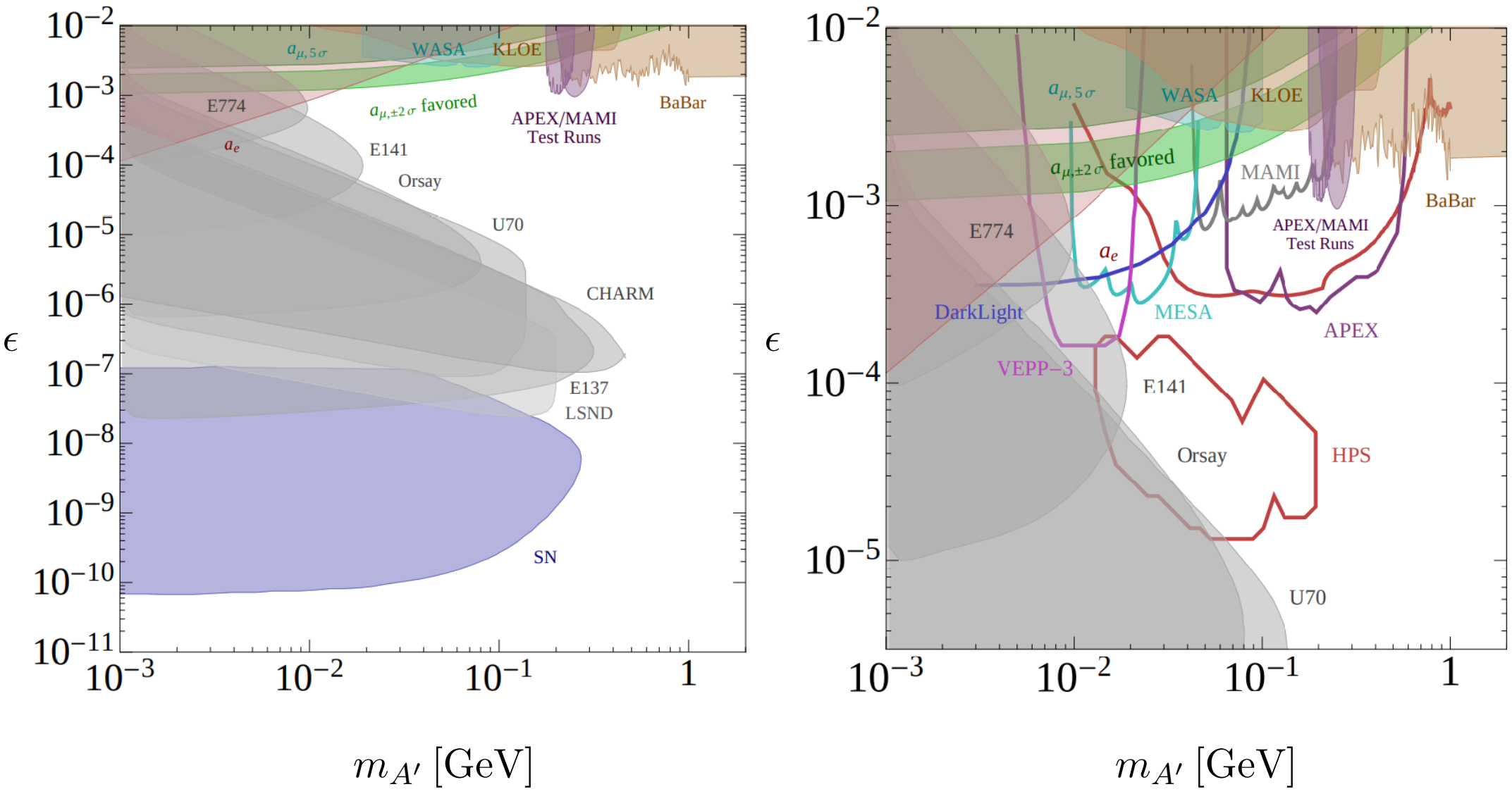

Figure IV.16: Dark photon parameter space with 90% exclusion regions of several measurements (shaded) and proposed projects (unshaded). Figure taken from [99].

dependent masses are given by

$$
\begin{aligned}
m^2_{(h,s)}(h,s) &= \begin{pmatrix} -\mu_H^2 + 3\lambda_H h^2 + \frac{1}{2}\lambda_{\mathrm{p}} s^2 & \lambda_{\mathrm{p}} hs \\ \lambda_{\mathrm{p}} hs & -\mu_S^2 + 3\lambda_S s^2 + \frac{1}{2}\lambda_{\mathrm{p}} h^2 \end{pmatrix}, \\
m^2_{G^+,G^0}(h,s) &= -\mu_H^2 + \lambda_H h^2 + \frac{1}{2}\lambda_{\mathrm{p}} s^2, \qquad \text{(IV.38)} \\
m^2_a(h,s) &= -\mu_S^2 + \lambda_S s^2 + \frac{1}{2}\lambda_{\mathrm{p}} h^2,
\end{aligned}
$$

while the remaining SM particle content is unchanged. The kinetic mixing induces a mass mixing between $W^3$, $B$ and $B'$ which become the $Z$, the massless photon $A$ and the *dark photon* $A'$ with mass

$$
m_{A'}(s) \approx g_D s \qquad \text{(IV.39)}
$$

for small $\epsilon$.[9] The parameter space spanned by $\epsilon$ and $m_{A'}$ is constrained by several experiments involving e.g. meson decays, anomalous magnetic moments and supernova cooling constraints (see Fig. IV.16). Note that, as opposed to the flip-flop models, we now work in the regime $\frac{\lambda_{\mathrm{p}}}{2}v^2 < \mu_S^2$ where $s = v_s > 0$ and $m_{A'} > 0$ at $T = 0$. This further implies a tree-level mixing between $h$ and $s$. It is thus not $h$ but the mixed eigenstate with mass 125 GeV that embodies the physical Higgs boson. In the limit

[9] A full derivation of the kinetically induced mass mixing can be found in [102].

of vanishing $\epsilon$, the relevant Debye masses amount to

$$\begin{aligned}
\Pi_H(T) &= \left(\frac{1}{2}\lambda_H + \frac{1}{12}\lambda_{\mathrm{p}} + \frac{1}{4}y_t^2 + \frac{1}{16}g_1^2 + \frac{3}{16}g_2^2\right)T^2, \\
\Pi_S(T) &= \left(\frac{1}{3}\lambda_S + \frac{1}{6}\lambda_{\mathrm{p}} + \frac{1}{4}g_D^2\right)T^2, \\
\Pi^{\mathrm{L}}_{W^i}(T) &= \frac{2}{3}g_D^2 T^2.
\end{aligned} \tag{IV.40}$$

If we further set the second portal $\lambda_{\mathrm{p}}$ to zero, the tree-level relations $\mu_S^2 = \lambda_S v_s^2$ and $m_s^2 = 2\lambda_S v_s^2$ can be used and the finite counterterms are completely analogous to those of the SM Higgs (IV.6) after replacing $h \to s$.

### IV.3.1 Gravitational Waves

We will now focus on the dynamics of the dark symmetry breaking and thus set $\lambda_{\mathrm{p}} = \epsilon = 0$ which separates the dark sector from the visible one. In the triplet scalar vev flip-flop model, 3 gauge bosons were available to induce a thermal barrier, where in the present case we only have 1. This however does not preclude the possibility of sizable GW signals as this PT occurs independently of EWSB. The GW signature of a $U(1)_D$ symmetry breaking in context with future space-based experiments such as LISA and DECIGO has recently been studied. Detectability for $m_{A'} > 10\,\mathrm{MeV}$ is claimed by [103], while [104] also considers collider constraints and states $m_{A'} = 25 \sim 100\,\mathrm{GeV}$ as the sensitive region. Our investigation will therefore focus on PTA detectability which lives around the MeV scale. The scale of our PT is set by $T_{\mathrm{n}} \sim v_s$.

Most promising in terms of the parameters $\alpha$ and $\frac{\beta}{H}$ is the region at large $g_D$ and small $\lambda_S$, as can be seen in Fig. IV.17. Large $g_D$ refers to a large thermally induced barrier. $\lambda_S$ in turn controls how dominant the tree-level features are in comparison to the thermal barrier. The ratio $\frac{\alpha}{\alpha_\infty}$ is close to 1 in the explored region (not shown as figure), which implies bubble walls in the runaway regime or very close to it. Note that the quantities $\frac{\beta}{H}$ and $\frac{\alpha}{\alpha_\infty}$ are completely scale independent, while $\alpha$ changes slightly with the relativistic DOF. The detectability, of course, is highly frequency and thus scale dependent.

As shown in Fig. IV.18, the PT would be visible to SKA for $v_s$ around the scale $10 \sim 100\,\mathrm{keV}$. This is a bit below the naively expected MeV range, which is a consequence of the large $\frac{\beta}{H} \gtrsim 1000$. In terms of $\lambda_S$, the detectable region is located at $\lambda_S \lesssim 0.1$. A quartic coupling of this size implies that $s$ will be the lightest hidden sector particle. We still want $\chi$ to be the DM candidate however, therefore a channel for further decay has to be introduced. For this, we will in the following consider two options:

(A) $s$ decays into SM photons through the portal couplings

(B) $s$ decays into some unspecified dark radiation with massless mediator which we do not detail here

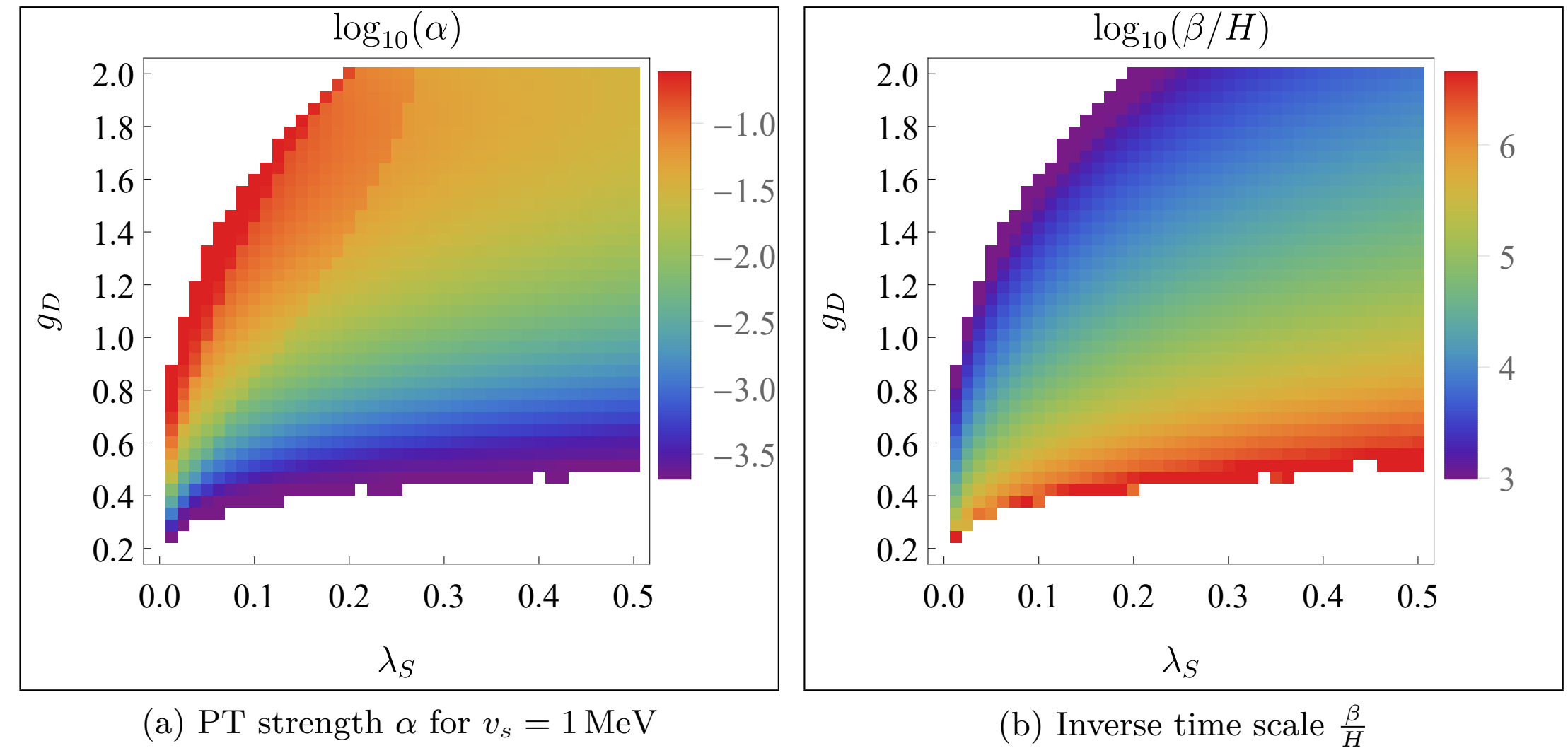

(a) PT strength $\alpha$ for $v_s = 1\,\mathrm{MeV}$

(b) Inverse time scale $\frac{\beta}{H}$

Figure IV.17: GW parameters of the $U(1)_D$ PT. The discontinuities in $\alpha$ are due to changing $g_{\mathrm{rel}}(T_{\mathrm{n}})$. The white regions do not feature a first-order PT: For too large $g_D$ and small $\lambda_S$, the combination of $V_{\mathrm{CW}}$ and $V_{\mathrm{ct}}$ produces a stable minimum at $s = 0$, $T = 0$. For small $g_D$, the thermally induced barrier is too weak to render the PT first-order.

If we consider the tail of the detectable region at the highest feasible scale, i.e. where $v_s \approx 1\,\mathrm{MeV}$ and $\lambda_S \approx 0.02$, the depletion of $s$ would start at $T \sim m_s = \sqrt{2\lambda_S} v_s \approx 200\,\mathrm{keV}$. This is clearly after the SM neutrino decoupling around $T \approx 1\,\mathrm{MeV}$ and also during BBN which falls in the range $T = 50\,\mathrm{keV} \sim 1\,\mathrm{MeV}$ [105]. In option (A), the photon thermal bath with temperature $T$ is reheated such that the temperature ratio after electron-positron annihilation and $s$ depletion (II.36) changes to

$$\frac{T_\nu}{T} = \left(\frac{2}{2 + \frac{7}{8} \cdot 4 + 1}\right)^{1/3} = \left(\frac{4}{13}\right)^{1/3} \tag{IV.41}$$

which changes the number of effective relativistic DOF. In option (B), the dark radiation directly adds 2 relativistic DOF. The effective relativistic DOF, which were in the SM case given by (II.37), now become

$$g_{\mathrm{rel}}(T < m_e, m_s) = \begin{cases} 2 + \frac{7}{8} \cdot 2 \cdot N_{\mathrm{eff}} \cdot \left(\frac{4}{13}\right)^{4/3} & \text{for option (A)} \\ 2 + \frac{7}{8} \cdot 2 \cdot N_{\mathrm{eff}} \cdot \left(\frac{4}{11}\right)^{4/3} + 2 & \text{for option (B)} \end{cases}. \tag{IV.42}$$

A change in $g_{\mathrm{rel}}$ can come into conflict with observations for two reasons. Firstly, it changes the expansion rate due to $H^2 \sim g_{\mathrm{rel}}$. If this is the case during BBN, the resulting light element yields will differ from the observed values. Secondly, the CMB angular power spectrum would be altered. A change in $g_{\mathrm{rel}}$ is not distinguishable from a change in $N_{\text{eff}}$. The mentioned constraints are therefore stated in terms of $N_{\text{eff}}$ and amount to $2.85 \pm 0.28$ (BBN, [106]) and $3.15 \pm 0.23$ (CMB, [9]) respectively, where the SM prediction is $N_{\text{eff}}^{\text{SM}} = 3.046$ [107]. Considering our two options again, the observed value $N_{\text{eff}}^{\text{obs}}$ would be

$$\begin{aligned} & N_{\text{eff}}^{\text{obs}} \cdot \left(\frac{4}{11}\right)^{4/3} = N_{\text{eff}}^{\text{SM}} \cdot \left(\frac{4}{13}\right)^{4/3} \\ \Rightarrow \quad & N_{\text{eff}}^{\text{obs}} \approx N_{\text{eff}}^{\text{SM}} - 0.6 \end{aligned} \tag{IV.43}$$

for option (A) and

$$\begin{aligned} & \frac{7}{8} \cdot 2 \cdot N_{\text{eff}}^{\text{obs}} \cdot \left(\frac{4}{11}\right)^{4/3} = \frac{7}{8} \cdot 2 \cdot N_{\text{eff}}^{\text{SM}} \cdot \left(\frac{4}{11}\right)^{4/3} + 2 \\ \Rightarrow \quad & N_{\text{eff}}^{\text{obs}} \approx N_{\text{eff}}^{\text{SM}} + 4.4 \end{aligned} \tag{IV.44}$$

for option (B). Both deviations are beyond the error bands of the experimentally determined values. Note that the calculation is built upon the assumption that the dark photon becomes non relativistic before neutrino decoupling, which is true for the parameter point we were looking at ($v_s \sim 1\,\text{MeV}$, $g_D = 1$).

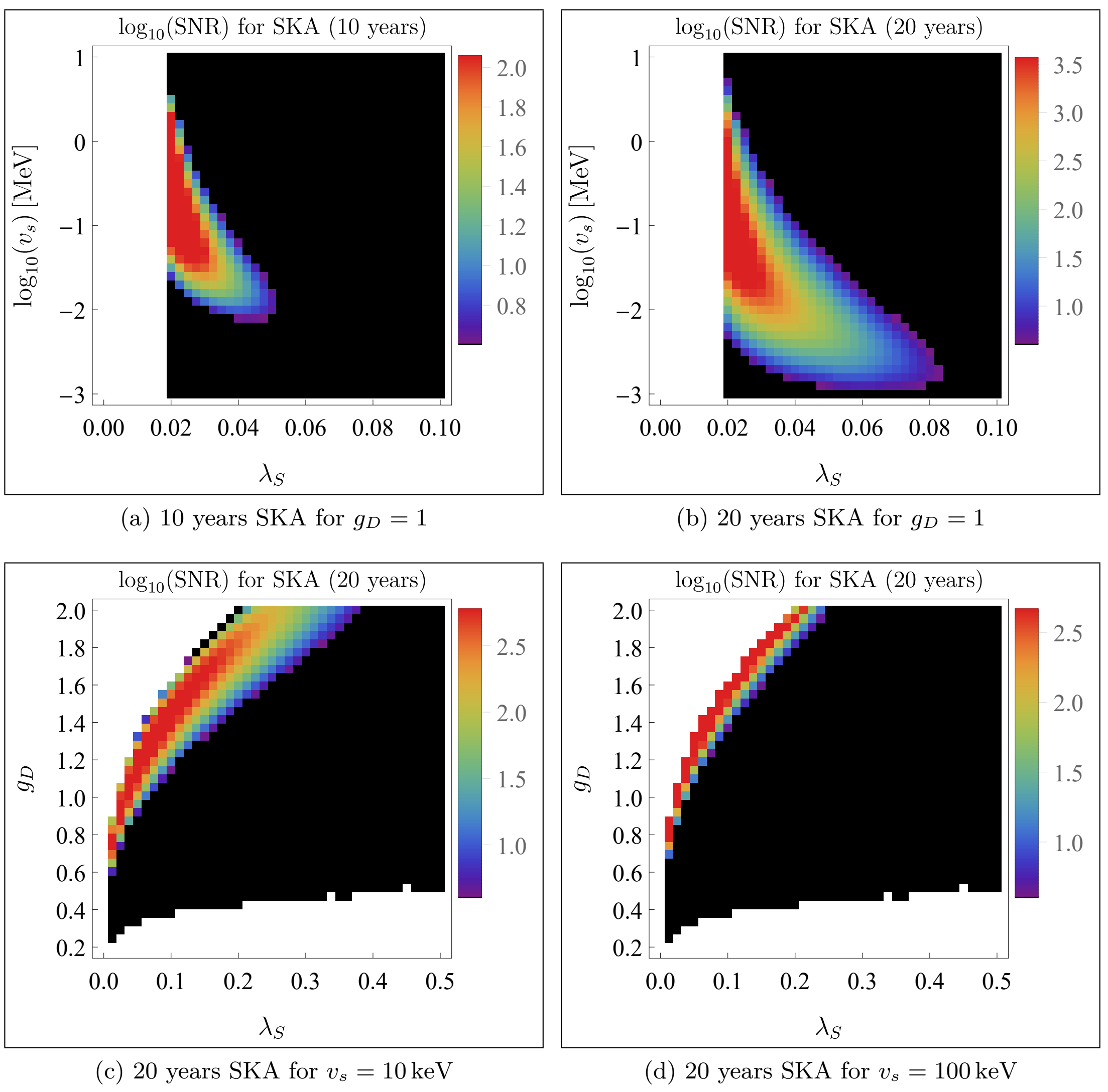

(a) 10 years SKA for $g_D = 1$

(b) 20 years SKA for $g_D = 1$

(c) 20 years SKA for $v_s = 10\,\mathrm{keV}$

(d) 20 years SKA for $v_s = 100\,\mathrm{keV}$

Figure IV.18: Detectability of the $U(1)_D$ PT with the planned PTA project SKA. Detectable regions are drawn in color. The sensitive regions with EPTA and NANOGrav are negligibly small and therefore not shown.

### IV.3.2 Decoupled Dark Temperature

A way out of the $N_{\text{eff}}$ problem can be a temperature offset between dark and visible sector which is possible due to the small portal couplings. A temperature offset in one or another direction is generated whenever a SM or a dark species drops out of equilibrium by annihilation, after the two sectors are already thermally decoupled. According to entropy conservation, the temperature ratio is determined by the ratio of thermalized relativistic entropy DOF between the two sectors, i.e.

$$r_T \equiv \frac{T_{\text{d}}}{T_\gamma} = \left( \frac{g_{s,\gamma}}{g_{s,\gamma}^{\text{dec}}} \frac{g_{s,\text{d}}^{\text{dec}}}{g_{s,\text{d}}} \right)^{1/3} \tag{IV.45}$$

with 'd' ('$\gamma$') labeling dark (visible) sector quantities. The GW parameters accordingly scale as

$$\begin{aligned}
\frac{T_{\text{n}}}{T_{\text{n}}'} &= r_T^{-1}, \\
\frac{\alpha}{\alpha'} &= \frac{\rho_{\text{rel}}'}{\rho_{\text{rel}}} = \frac{(g_\gamma + g_{\text{d}})T_{\text{d}}^4}{g_\gamma T_\gamma^4 + g_{\text{d}} T_{\text{d}}^4} = \frac{g_\gamma + g_{\text{d}}}{g_\gamma r_T^{-4} + g_{\text{d}}} \approx r_T^4, \\
\frac{\alpha_\infty}{\alpha_\infty'} &= \frac{T_{\text{n}}^2}{T_{\text{n}}'^2} \frac{\rho_{\text{rel}}'}{\rho_{\text{rel}}} \approx r_T^2, \\
\frac{(\beta/H)}{(\beta/H)'} &= \frac{H'}{H} = \frac{T_{\text{n}}'^2}{T_{\text{n}}^2} = r_T^2
\end{aligned} \tag{IV.46}$$

where the primed quantities are the ones naively obtained by `CosmoTransitions`, i.e. without considering the temperature offset. We see that in both directions, $r_T < 1$ and $r_T > 1$, either $\alpha$ or $\frac{\beta}{H}$ becomes better while the other becomes worse in view of detectability. Note that for large $r_T$, the amplifying effect of increasing $\alpha$ saturates, as can be seen by (III.29).

Fig. IV.19 pictures the impact of $r_T$ on the 20 year SKA detectability for different $\lambda_S$, where the smallest $\lambda_S = 0.02$ yielded the most promising GW spectrum. We would expect that a higher (lower) temperature of the visible sector, i.e. $r_T < 1$ ($r_T > 1$), also shifts the detectable region towards higher (lower) $v_s$. This behavior is in principle apparent in Fig. IV.19, but the effect of decreasing $\alpha$ (increasing $\beta$) for $r_T < 1$ ($r_T > 1$) leads to a rapid shrinking of the detectable area and the best region appears to be around $r_T = 1 \sim 2$.

In order to circumvent the $N_{\text{eff}}$ constraints for option (A), we can simply choose a small $r_T$ such that the $s$ annihilation, occurring at $T_{\text{d}} \lesssim 200\,\text{keV}$ in the detectable $v_s$ region, takes place before the neutrino decoupling. In order to push $T_\gamma$ above the required $1\,\text{MeV}$, a ratio of $r_T \lesssim 0.2$ is required. Unfortunately, this is outside the detectable region as can be seen in Fig. IV.19. In case of option (B), $r_T$ can be used to redshift the dark relativistic DOF such that they contribute less to $g_{\text{rel}}$.

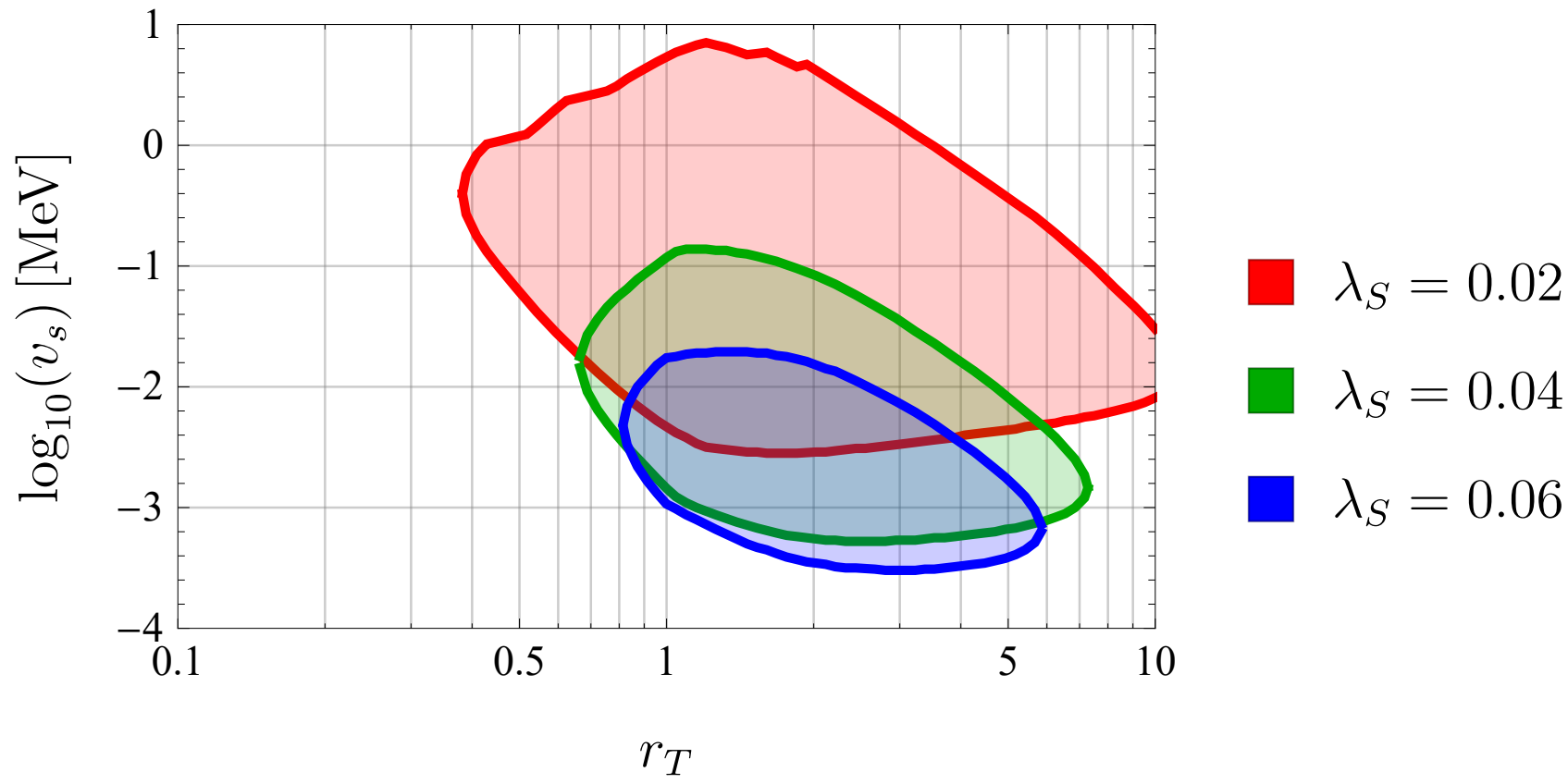

Figure IV.19: Deformation of the detectable region with changing temperature ratio $r_T = \frac{T_\mathrm{d}}{T_\gamma}$ for $g_D = 1$.

Quantitatively we have

$$
\begin{aligned}
g_\mathrm{rel}(T_\gamma < m_e, m_s) &= 2 + \frac{7}{8} \cdot 2 \cdot N_\mathrm{eff} \cdot \left(\frac{T_\nu}{T_\gamma}\right)^4 + 2 \cdot \left(\frac{T_\mathrm{d}}{T_\gamma}\right)^4 \\
&= 2 + \frac{7}{8} \cdot 2 \cdot N_\mathrm{eff} \cdot \left(\frac{4}{11}\right)^{4/3} + 2 \cdot r_T^4 .
\end{aligned}
\tag{IV.47}
$$

Being generous, we take the upper bound from the CMB constraint which gives us $N_\mathrm{eff}^\mathrm{obs} < 3.38$ as the maximally allowed value. This translates into the requirement

$$
\begin{aligned}
& \frac{7}{8} \cdot 2 \cdot 3.38 \cdot \left(\frac{4}{11}\right)^{4/3} > \frac{7}{8} \cdot 2 \cdot N_\mathrm{eff}^\mathrm{SM} \cdot \left(\frac{4}{11}\right)^{4/3} + 2 \cdot r_T^4 \\
\Rightarrow \quad & r_T \lesssim 0.52
\end{aligned}
\tag{IV.48}
$$

which, together with $v_s \sim 1\,\mathrm{MeV}$, lies still within the detectable region.

Another possibility to face the $N_\mathrm{eff}$ problem is provided by [105]. In this recent publication, a scenario is presented in which the dark and visible sector come into thermal contact after neutrino decoupling for the first time. The cooling effect due to the equilibration then compensates the heating effect due to the later annihilation of the DM candidate. For this to work, a light hidden particle with even lighter mediator is required, which could be provided by $A'$ and $s$ in our model. The applicability and details of a possible adaptation to our model remain to be investigated.

# V Summary and Outlook

The goal of this thesis was to acquire a profound understanding of the topic of GWs caused by cosmological PTs and to apply the accumulated knowledge by investigating different extensions of the SM with respect to the featured GW spectra and their detectability.

Chapter II aimed at building a solid foundation of the related theoretical background and methodology. This includes the basics of modern cosmology and gravitational radiation, aspects that are of great importance for the understanding of the topic. A quite technical section about the effective potential and its temperature dependent behavior followed, showing the derivation of a collection of handy and useful formulae, e.g. for the potential contributions at one-loop level or for thermal masses, that were later used in the model analyses. Furthermore, the mechanisms behind false vacuum decay and bubble formation were explained and the chapter closed with a brief description of the software packages `CosmoTransitions` and `SARAH` that were employed for this work.

The focus completely moved to GWs in the context of PTs in Chapter III. First, the canonical parameters $T_\mathrm{n}$, $\alpha$, $\beta$ and $v_\mathrm{w}$ were explained, based on the theoretic considerations of the previous chapter. Concerning the bubble wall velocity $v_\mathrm{w}$, it turned out that an exact determination for the considered models is beyond the scope of this thesis. Furthermore, the physics community seems to not have reached consensus on the topic of runaway bubbles yet. The idea behind this work is to demonstrate what might be possible to probe in the future, wherefore the most optimistic scenario of luminal bubble walls and an allowed runaway regime was chosen for the later analyses. The last two sections of the chapter contain the latest state-of-the-art formulae and parameters for the different contributions to the GW spectrum as well as recent sensitivity curves. The sensitive parameter regions of operational and planned PTAs and space-based interferometer missions have been computed and were displayed.

Chapter IV presented the results of model analyses which are based on the background developed in the preceding chapters. The first considered model framework was the vev flip-flop, which was initially introduced to provide a novel DM production mechanism. The flip-flop features a two-step PT at the electroweak scale, where first an additional scalar field and then the SM Higgs obtains a VEV. The required DM abundance is generated in the intermediate phase. The first transition ('flip') turned out to be first-order if the additional scalar field carries a gauge charge, giving rise to a thermally induced barrier in the effective potential at one-loop level. The corresponding GW spectrum is however below any of the considered sensitivities. The second transition ('flop') in turn features a sizable tree-level barrier inducing a strong first-order transition. This results in a GW spectrum that would be visible to fu-ture space-based observatories like BBO or DECIGO. Introducing more gauge bosons could push both transitions into the detectable region and cause a characteristic double bump spectrum. The Economical 3-3-1 model features a $SU(3)_L$ symmetry and thus additional gauge bosons, but a detectable double bump spectrum as claimed by [96] could not be reproduced. Finally, the dark photon model was considered. It adds a $U(1)_D$ gauge symmetry and a scalar that breaks it, making the dark photon massive. The corresponding PT lives at the sub-MeV scale and turned out to be detectable by the planned PTA project SKA. A transition at this scale however is in conflict with observations of the CMB and with the observed abundances of light elements. This issue has been solved by introducing a temperature ratio $r_T$ between the dark and visible sector. The derived $r_T$-scaling behavior of the GW parameters will also be useful for the analysis of other models in the future.

* * *

Modern-day physics builds on the great efforts and successes that were made in the last century, yet there are numerous unresolved phenomena such as dark matter, dark energy, inflation, baryogenesis and the hierarchy problem, just to name a few. It is these open questions that fascinate and motivate me and other physicists all over the world in their everyday efforts in pushing the boundaries of humankind's knowledge. Gravitational-wave physics is definitely a major bearer of hope for further progress in the 21st century and this thesis tried to make a few steps into this auspicious direction.

# Appendix

## A.1 Computation of Matsubara Sums

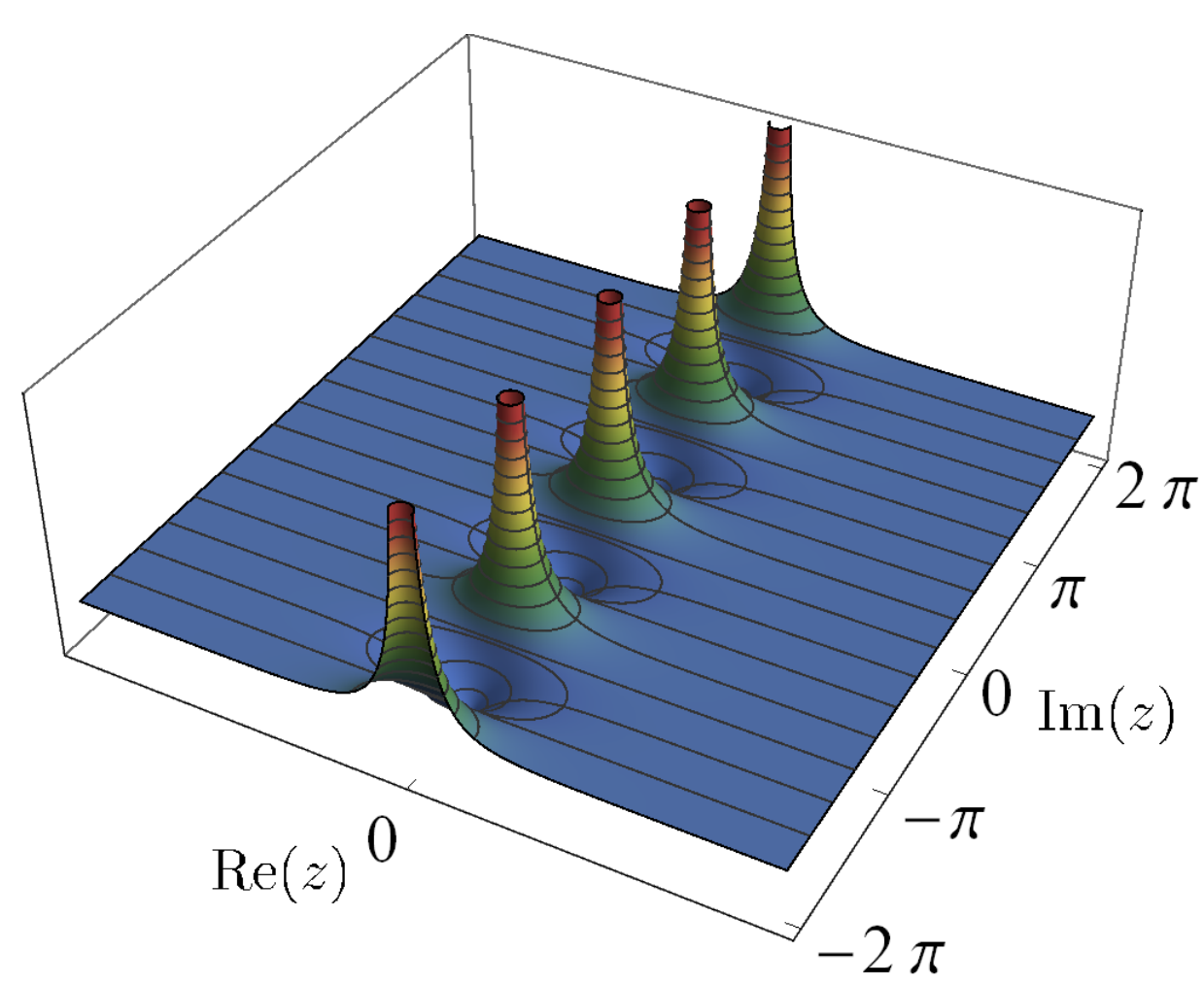

Figure A.1: Singularities of the hyperbolic cotangent $|\coth(z)|$.

To compute a Matsubara sum, one can make use of an elegant trick [108]. According to the *residue theorem*, when integrating in closed contours around the poles of a complex valued function, the result will be the sum of the function's residues at the poles. The functions

$$\xi_{b/f}(z) = \begin{cases} \coth\left(\frac{z}{2T}\right) = 1 + \frac{2}{e^{z/T}-1} & \text{(bosons)} \\ \tanh\left(\frac{z}{2T}\right) = 1 - \frac{2}{e^{z/T}+1} & \text{(fermions)} \end{cases} \tag{A.1}$$

have singularities at $z = i\omega_n$ along the imaginary axis, with residues[1]

$$\mathrm{Res}_{z=i\omega_n}\left[\xi_{b/f}\left(z\right)\right] = 2T. \tag{A.2}$$

Furthermore, if a function $f(z)$ has no poles that coincide with $i\omega_n$, it can simply be multiplied to the integrand. The residues thereby change only linearly, since the poles of $\xi_{b/f}(z)$ are of order one:

$$\mathrm{Res}_{z=i\omega_n}\left[f(z)\xi_{b/f}(z)\right] = 2Tf(i\omega_n) \tag{A.3}$$

[1] The residue for poles of order $n$ is $\mathrm{Res}_{z=z_0}[f(z)] = \frac{1}{(n-1)!}\lim\limits_{z\to z_0}\frac{\partial^{(n-1)}}{\partial z^{(n-1)}}[(z-z_0)^n f(z)]$.

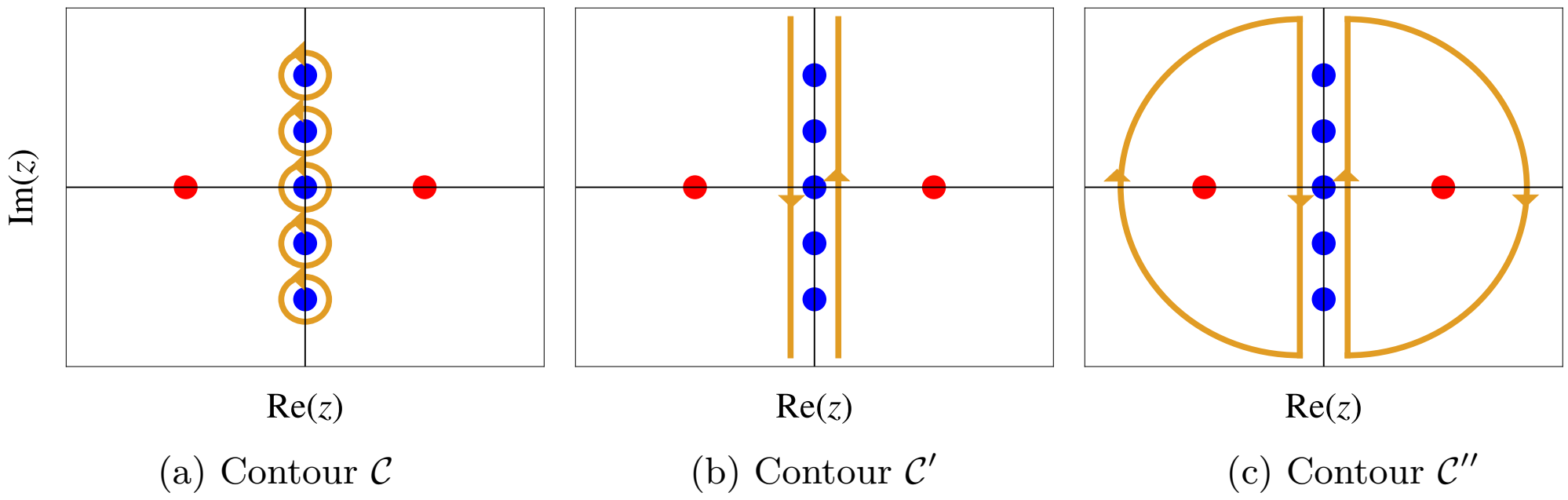

(a) Contour $\mathcal{C}$ (b) Contour $\mathcal{C}'$ (c) Contour $\mathcal{C}''$

Figure A.2: Poles of $\coth(\frac{z}{2T})$ (blue), poles of $f(z)$ (red) and integration contours (orange).

The infinite Matsubara sum can thus be rewritten as

$$T \sum_{n=-\infty}^{\infty} f(z = i\omega_n) = \frac{1}{2} \frac{1}{2\pi i} \int_{\mathcal{C}} \mathrm{d}z \, f(z) \xi_{b/f}(z) \tag{A.4}$$

where the integration contour $\mathcal{C}$ encircles each pole separately and counter-clockwise with infinitesimal radius. In the next step, the integration contour is deformed to $\mathcal{C}'$ and now runs from $-i\infty + \epsilon$ to $+i\infty + \epsilon$ and back from $+i\infty - \epsilon$ to $-i\infty - \epsilon$. This deformation requires $f(z)$ to have no poles along the imaginary axis at all. The two integration lines connect at $\pm i\infty$, implying that they can equivalently be closed by semicircles through $\pm\infty$ where $f(z)\xi_{b/f}(z)$ should vanish (contour $\mathcal{C}''$). Instead of the Matsubara poles $i\omega_n$, the two closed semicircles now contain all singularities $z_i$ of $f(z)$. They can be picked up by making use of the residue theorem again, i.e.

$$\frac{1}{2} \frac{1}{2\pi i} \int_{\mathcal{C}''} \mathrm{d}z \, f(z) \xi_{b/f}(z) = -\frac{1}{2} \sum_i \mathrm{Res}_{z=z_i} \left[ f(z) \xi_{b/f}(z) \right] . \tag{A.5}$$

The poles are now encircled clockwise, thus the minus sign. Putting everything together, we arrive at the identity

$$T \sum_{n=-\infty}^{\infty} f(z = i\omega_n) = -\frac{1}{2} \sum_i \mathrm{Res}_{z=z_i} \left[ f(z) \xi_{b/f}(z) \right] . \tag{A.6}$$

## A.2 Computation of Debye Masses

In the following, the thermal Debye masses for the toy Lagrangian (II.76) are derived to leading order. We consider the $T \gg m$ regime where masses are negligible and we further set all external momenta to zero. Before deriving the different thermal mass corrections, it is useful to pre-evaluate some typically arising integrals:

$$
\begin{aligned}
T\sum_n \int \frac{\mathrm{d}^3k}{(2\pi)^3}\frac{1}{\omega_n^2+|\boldsymbol{k}|^2} &= \int_0^\infty \frac{\mathrm{d}kk^2}{2\pi^2}\left[\frac{\xi_{b/f}(k)}{2k}\right] \\
&= \int_0^\infty \frac{\mathrm{d}kk^2}{2\pi^2}\left[\cancel{\frac{1}{2k}}+\frac{1}{k}\frac{\eta}{e^{k/T}-\eta}\right] \\
&= \begin{cases} \frac{T^2}{12} & \text{(bosons)} \\ -\frac{T^2}{24} & \text{(fermions)} \end{cases}
\end{aligned} \tag{A.7}
$$

$$
\begin{aligned}
T\sum_n \int \frac{\mathrm{d}^3k}{(2\pi)^3}\frac{\omega_n^2}{(\omega_n^2+|\boldsymbol{k}|^2)^2} &= \int_0^\infty \frac{\mathrm{d}kk^2}{2\pi^2}\left[\frac{k+2T\xi_{b/f}(k)-k\xi_{b/f}^2(k)}{8kT}\right] \\
&= \int_0^\infty \frac{\mathrm{d}kk^2}{2\pi^2}\left[\cancel{\frac{1}{4k}}+\frac{\eta}{2}\left(\frac{1}{k}-\frac{1}{T}\right)\frac{1}{e^{k/T}-\eta}-\frac{1}{2T}\frac{1}{(e^{k/T}-\eta)^2}\right] \\
&= \begin{cases} -\frac{T^2}{24} & \text{(bosons)} \\ \frac{T^2}{48} & \text{(fermions)} \end{cases}
\end{aligned} \tag{A.8}
$$

$$
\begin{aligned}
T\sum_n \int \frac{\mathrm{d}^3k}{(2\pi)^3}\frac{|\boldsymbol{k}|^2}{(\omega_n^2+|\boldsymbol{k}|^2)^2} &= \int_0^\infty \frac{\mathrm{d}kk^2}{2\pi^2}\left[\frac{-k+2T\xi_{b/f}(k)+k\xi_{b/f}^2(k)}{8kT}\right] \\
&= \int_0^\infty \frac{\mathrm{d}kk^2}{2\pi^2}\left[\cancel{\frac{1}{4k}}+\frac{\eta}{2}\left(\frac{1}{k}+\frac{1}{T}\right)\frac{1}{e^{k/T}-\eta}+\frac{1}{2T}\frac{1}{(e^{k/T}-\eta)^2}\right] \\
&= \begin{cases} \frac{T^2}{8} & \text{(bosons)} \\ -\frac{T^2}{16} & \text{(fermions)} \end{cases}
\end{aligned} \tag{A.9}
$$

The identity (A.6) came in handy for the Matsubara sums evaluation. Each integral splits up into a non-thermal and a temperature dependent part. The non-thermal one is UV divergent. The infinities in the scalar Debye masses can be canceled by the counterterms we already introduced, while the infinities arising in the gauge boson Debye masses will cancel amongst the three contributions. The divergent part has therefore already been dropped in the last steps of the above calculations.

### Scalar Debye masses

The leading diagrams are

$$
\begin{aligned}
&= 2 \cdot \frac{i}{2} \int \frac{\mathrm{d}^4 k}{(2\pi)^4} \frac{6\lambda}{k^2 + i\epsilon} \\
&\to T \sum_n \int \frac{\mathrm{d}^3 k}{(2\pi)^3} \frac{6\lambda}{\omega_n^2 + |\boldsymbol{k}|^2} \\
&\supset 2 \cdot \lambda \frac{T^2}{4} = 2 \cdot \frac{\partial^2 m_\phi^2(\phi)}{\partial \phi^2} \frac{T^2}{24},
\end{aligned}
\tag{A.10}
$$

$$
\begin{aligned}
&= -i \int \frac{\mathrm{d}^4 k}{(2\pi)^4} \frac{(y/\sqrt{2})^2 \mathrm{Tr}\,[\not{k}\not{k}]}{(k^2 + i\epsilon)^2} \\
&= -4 \frac{i}{2} \int \frac{\mathrm{d}^4 k}{(2\pi)^4} \frac{y^2}{(k^2 + i\epsilon)^2} \\
&\to -4 \frac{T}{2} \sum_n \int \frac{\mathrm{d}^3 k}{(2\pi)^3} \frac{y^2}{\omega_n^2 + |\boldsymbol{k}|^2} \\
&\supset 4 \cdot y^2 \frac{T^2}{48} = 4 \cdot \frac{\partial^2 m_\psi^2(\phi)}{\partial \phi^2} \frac{T^2}{48},
\end{aligned}
\tag{A.11}
$$

$$
\begin{aligned}
&= \frac{i}{2} \int \frac{\mathrm{d}^4 k}{(2\pi)^4} (2ig)\, \mathrm{Tr}\,[\Delta^{\mu\nu}(k)] \\
&\to 3 \frac{T}{2} \sum_n \int \frac{\mathrm{d}^3 k}{(2\pi)^3} \frac{2g}{\omega_n^2 + |\boldsymbol{k}|^2} \\
&\supset 3 \cdot g \frac{T^2}{12} = 3 \cdot \frac{\partial^2 m_A^2(\phi)}{\partial \phi^2} \frac{T^2}{24}
\end{aligned}
\tag{A.12}
$$

where $\lambda$, $y$ and $g$ are the couplings of our toy model Lagrangian (II.76). The prefactor 2 in the first diagram accounts for the loop scalar being complex.

### Gauge boson Debye masses

Polarization tensors of vector bosons can be split up into components of longitudinal (L) and transverse (T) polarization

$$
\Pi^{\mu\nu} = \Pi^{\mathrm{T}} T^{\mu\nu} + \Pi^{\mathrm{L}} L^{\mu\nu}
\tag{A.13}
$$

with projection operators $T^{\mu\nu} = \mathrm{diag}(0, 2, 2, 2)$ and $L^{\mu\nu} = \mathrm{diag}(-1, 0, 0, 0)$ in the IR limit [36]. For the considered abelian toy model, the leading thermal mass corrections are

$$+ \;=\; 2 \cdot \frac{i}{2} \int \frac{\mathrm{d}^4 k}{(2\pi)^4} \left[ \frac{-2g^2 g^{\mu\nu}}{k^2 + i\epsilon} - \frac{(2gk^\mu)(-2gk^\nu)}{(k^2 + i\epsilon)^2} \right] \tag{A.14}$$

with

$$\begin{aligned} \Pi^{\mathrm{L}} = -\Pi^{00} &= 2ig^2 \int \frac{\mathrm{d}^4 k}{(2\pi)^4} \left[ \frac{1}{k^2 + i\epsilon} - \frac{2k_0^2}{(k^2 + i\epsilon)^2} \right] \\ &\to 2g^2 T \sum_n \int \frac{\mathrm{d}^3 k}{(2\pi)^3} \left[ \frac{1}{\omega_n^2 + |\boldsymbol{k}|^2} - \frac{2\omega_n^2}{(\omega_n^2 + |\boldsymbol{k}|^2)^2} \right] \\ &= 2 \cdot g^2 \frac{T^2}{6}, \end{aligned} \tag{A.15}$$

$$\begin{aligned} \Pi^{\mathrm{T}} = \frac{1}{2}\Pi^{ii} &= ig^2 \int \frac{\mathrm{d}^4 k}{(2\pi)^4} \left[ \frac{1}{k^2 + i\epsilon} + \frac{2k_i^2}{(k^2 + i\epsilon)^2} \right] \\ &\to g^2 T \sum_n \int \frac{\mathrm{d}^3 k}{(2\pi)^3} \left[ \frac{1}{\omega_n^2 + |\boldsymbol{k}|^2} - \frac{\frac{2}{3}|\boldsymbol{k}|^2}{(\omega_n^2 + |\boldsymbol{k}|^2)^2} \right] \\ &= 0 \end{aligned} \tag{A.16}$$

and

$$\begin{aligned} &= -i \int \frac{\mathrm{d}^4 k}{(2\pi)^4} \frac{g^2 \mathrm{Tr}\left[\gamma^\mu \not{k} \gamma^\nu \not{k}\right]}{(k^2 + i\epsilon)^2} \\ &= -4ig^2 \int \frac{\mathrm{d}^4 k}{(2\pi)^4} \frac{k_\alpha k_\beta \left[ g^{\alpha\mu} g^{\beta\nu} - g^{\alpha\beta} g^{\mu\nu} + g^{\alpha\nu} g^{\beta\mu} \right]}{(k^2 + i\epsilon)^2} \\ &= -4ig^2 \int \frac{\mathrm{d}^4 k}{(2\pi)^4} \frac{2k^\mu k^\nu - k^2 g^{\mu\nu}}{(k^2 + i\epsilon)^2} \end{aligned} \tag{A.17}$$

with

$$\begin{aligned} \Pi^{\mathrm{L}} = -\Pi^{00} &= 4ig^2 \int \frac{\mathrm{d}^4 k}{(2\pi)^4} \frac{k_0^2 + |\boldsymbol{k}|^2}{(k^2 + i\epsilon)^2} \\ &\to 4g^2 T \sum_n \int \frac{\mathrm{d}^3 k}{(2\pi)^3} \frac{\omega_n^2 - |\boldsymbol{k}|^2}{(\omega_n^2 + |\boldsymbol{k}|^2)^2} \\ &= 4 \cdot g^2 \frac{T^2}{12}, \end{aligned} \tag{A.18}$$

$$\begin{aligned}\Pi^{\mathrm{T}} = \frac{1}{2}\Pi^{ii} &= -2ig^2 \int \frac{\mathrm{d}^4k}{(2\pi)^4} \frac{2k_i^2 + k_0^2 - |\boldsymbol{k}|^2}{(k^2 + i\epsilon)^2} \\ &\to 2g^2T \sum_n \int \frac{\mathrm{d}^3k}{(2\pi)^3} \frac{\frac{2}{3}|\boldsymbol{k}|^2 - \omega_n^2 - |\boldsymbol{k}|^2}{(\omega_n^2 + |\boldsymbol{k}|^2)^2} \\ &= 0.\end{aligned} \tag{A.19}$$

Note that all off-diagonal components of $\Pi^{\mu\nu}$ vanish as they would yield antisymmetric integrands. As the calculation shows, transverse bosonic modes do not receive thermal mass corrections due to protection by gauge symmetry [109].

## A.3 Higher Order Mass Corrections

Besides the daisy diagrams that were incorporated in the analyses of this thesis, there are other processes of higher loop order that can have sizable effects. Before writing down the scaling of some of these diagrams, note that we will now assume that daisy contributions are included. As a consequence, the mass scale used to fix dimensionality is now temperature dependent and denoted by $\mu_{\text{eff}}$. This accounts for the fact that propagators are now dressed, i.e. that they receive thermal corrections.

An example for a two-loop process is is the *sunset diagram* with superficial degree of divergence $D = 2$, scaling as

$$\Pi_{\text{sunset}} \equiv \text{[diagram]} \sim \lambda^2 T^2, \tag{A.20}$$

which is negligible compared to the hard thermal loop $\sim \lambda T^2$ if $\lambda$ is not too large. This is our general requirement for perturbativity. Further contributions are the *cactus diagram* with $D = 0$ for each of the lower loops and $D = 2$ for the upper one

$$\Pi_{\text{cactus}} \equiv \text{[diagram]} \sim \frac{1}{\mu_{\text{eff}}^2}(\lambda T)^2 \lambda T^2 = \beta^2 \lambda T^2 \tag{A.21}$$

and the $N$-loop *superdaisy* diagram with $D = 2$ for the two main loops as well as for each of the $N - 2$ petals

$$\begin{aligned}\Pi_{\text{superdaisy}} \equiv \text{[diagram]} &\sim \frac{1}{\mu_{\text{eff}}^{2N-4}} \lambda^2 T^2 (\lambda T^2)^{N-2} \\ &= \beta \alpha^{N-5/2} \lambda^{3/2} T^2 \sim \beta \lambda^{3/2} T^2\end{aligned} \tag{A.22}$$

where

$$\beta \equiv \lambda \frac{T}{\mu_{\text{eff}}} \tag{A.23}$$

was defined and $\alpha \sim 1$ around the critical temperature as demonstrated in Section II.2.6. Demanding $\Pi_{\text{cactus}}$ ($\Pi_{\text{superdaisy}}$) to be small compared to $\Pi_{\text{hard}}$ ($\Pi_{\text{daisy}}$) is tantamount to the requirement $\beta \lesssim 1$.

## A.4 Solving the Bounce Equation

The bounce equation, determining the bubble wall profile which minimizes the action, is usually solved numerically using the under-/overshoot method. For an attempt to apply this method, we will consider a potential of the generic form

$$V(\phi) = -a\phi^2 + b\phi^3 + c\phi^4. \tag{A.24}$$

The $O(3)$ symmetric bounce equation (II.134)

$$\phi''(r) = \left.\frac{\partial V}{\partial \phi}\right|_{\phi=\phi(r)} - \frac{2}{r}\phi'(r) \tag{A.25}$$

describes the corresponding tunneling process with boundary conditions $\phi'(0) = 0$ and $\phi(r \to \infty) = 0$. The above equation is mathematically equivalent to the differential equation of an oscillator with restoring force $F = \partial_\phi V(\phi)$,[2] damping factor $\frac{1}{r}$, displacement $\phi$ and time variable $r$. We can thus turn the *boundary value problem* into an *initial value problem* with $\phi(0) = \phi_0$ and $\phi'(0) = 0$, which can be solved easily by evaluating

$$\phi(r + \Delta r) = \phi(r) + \phi'(r)\Delta r, \tag{A.26}$$

$$\phi'(r + \Delta r) = \phi'(r) + \left( \left.\frac{\partial V}{\partial \phi}\right|_{\phi=\phi(r)} - \frac{2}{r}\phi'(r) \right) \Delta r \tag{A.27}$$

iteratively in small steps $\Delta r$, beginning with $r = 0$. The starting value $\phi_0$ has to be adjusted until the solution $\phi(r)$ satisfies the boundary condition $\phi(r \to \infty) = 0$ of the original problem, i.e. until $\phi(r)$ exactly comes to a halt at the local maximum of $-V(\phi)$ at $\phi = 0$. Before that is achieved, the solution will *undershoot* (*overshoot*) the true solution as long as $\phi_0 < \phi_{\text{sol}}$ ($\phi_0 > \phi_{\text{sol}}$), as depicted in Fig. A.3. The true solution represents the physical bubble profile (see Fig. A.4, b) and $\phi_{\text{sol}}$ refers to the endpoint of the corresponding tunneling process, which is typically close to the global minimum $\phi_{\text{min}}$ of potential $V(\phi)$.

[2] Note that usually $F = -\partial_\phi V(\phi)$, we thus have to flip our potential.

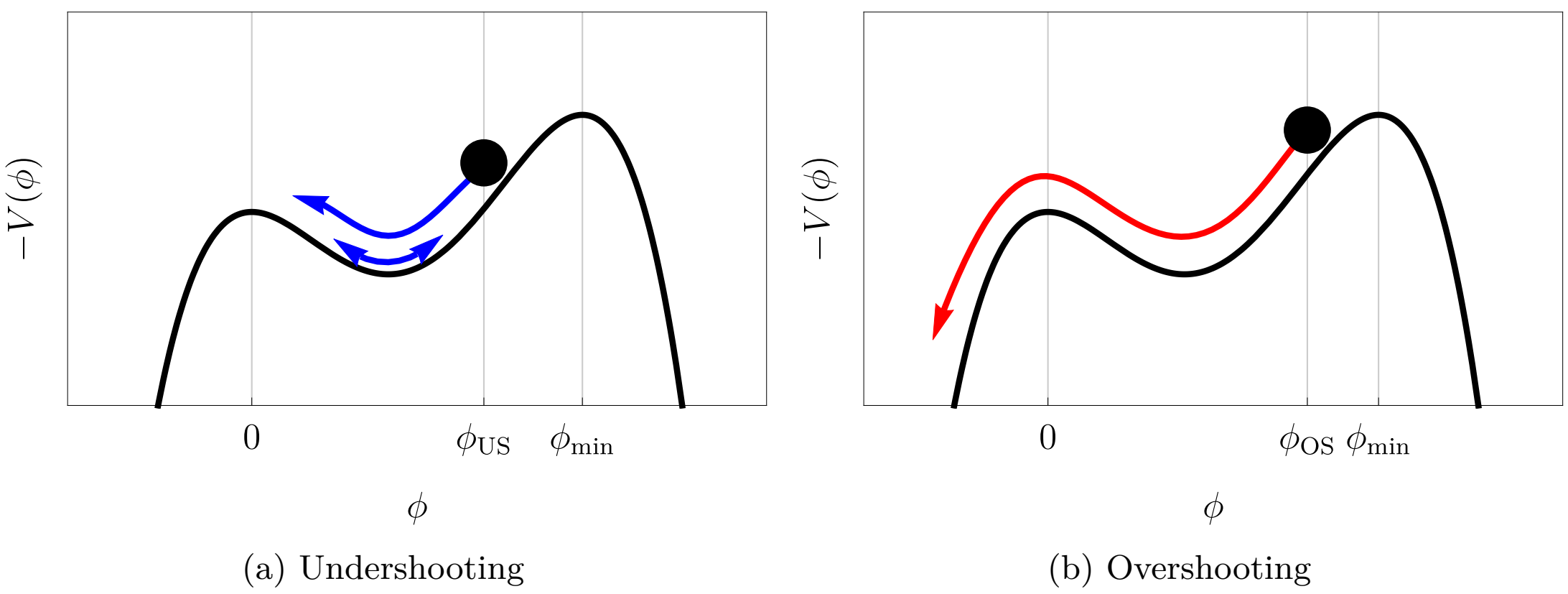

(a) Undershooting (b) Overshooting

Figure A.3: Schematic behavior of different solutions to the initial value problem.

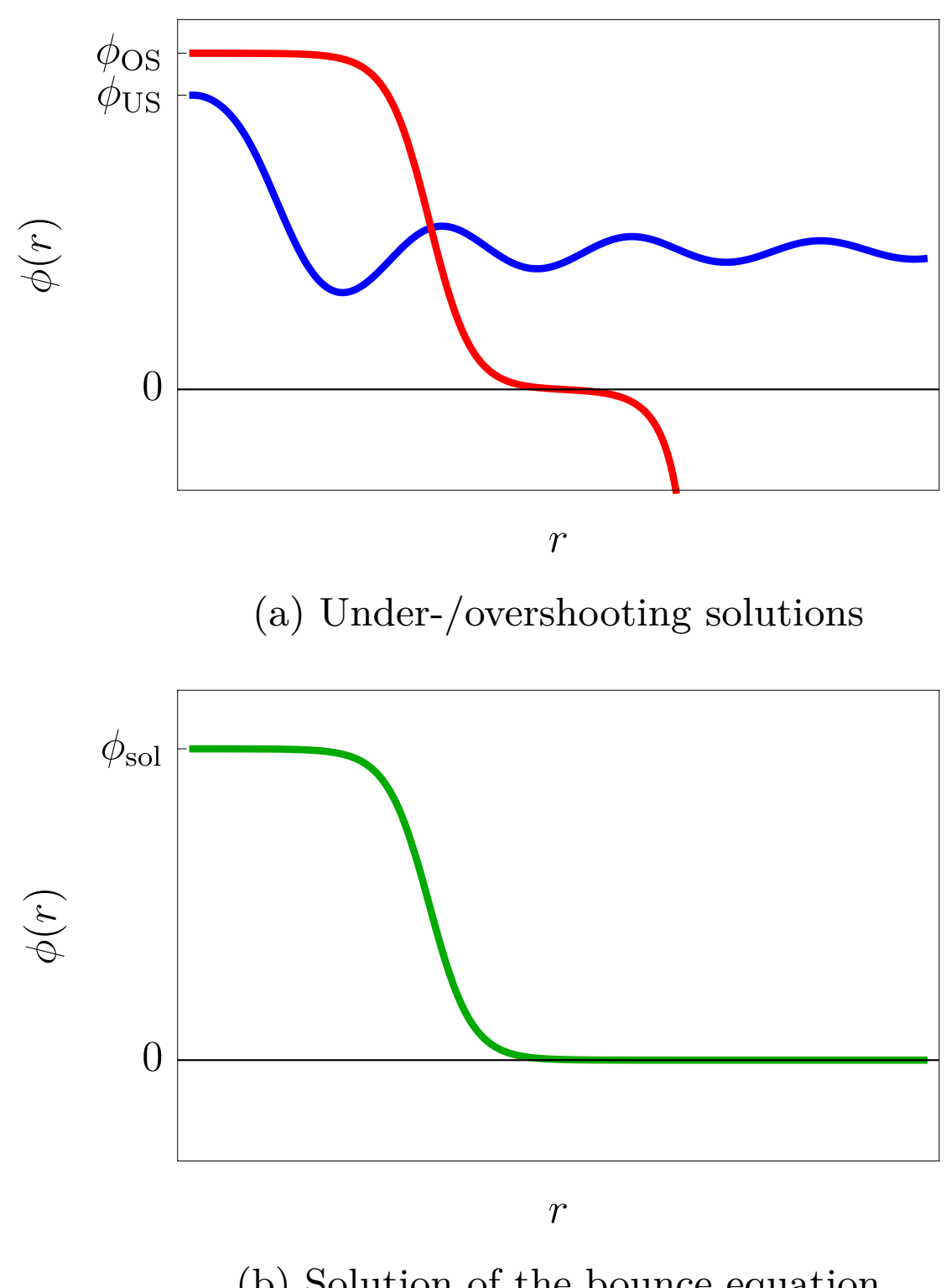

(a) Under-/overshooting solutions

(b) Solution of the bounce equation

Figure A.4: Bubble wall profiles $\phi(r)$ for different initial values with the hierarchy $\phi_{\mathrm{US}} < \phi_{\mathrm{sol}} < \phi_{\mathrm{OS}} < \phi_{\mathrm{min}}$.

# Acknowledgments

I would firstly like to thank my advisors Joachim Kopp and Pedro Schwaller for giving me the opportunity to contribute towards this fascinating field of physics. Furthermore, I am very grateful for the enjoyable and fostering environment provided by the THEP group. In particular, the many illuminating discussions with Lukas Mittnacht, Toby Opferkuch, Michael J. Baker and Eric Madge, amongst many others, have assisted me to greatly improve my understanding of a large range of both related and unrelated topics. Finally, a heartfelt thanks goes to my friends, to my parents Iris and Rainer and to my partner Alexandra for their unconditional support.

# List of Abbreviations

| | |
|---|---|
| $\Lambda$CDM | *Lambda cold dark matter* |
| BBN | *Big Bang nucleosynthesis* |
| BBO | *Big Bang Observer* |
| CMB | *cosmic microwave background* |
| DECIGO | *Deci-hertz Interferometer Gravitational-Wave Observatory* |
| DM | *dark matter* |
| DOF | *degree of freedom* |
| EPTA | *European Pulsar Timing Array* |
| EWBG | *electroweak baryogenesis* |
| EWPT | *electroweak phase transition* |
| EWSB | *electroweak symmetry breaking* |
| FRW | *Friedmann-Robertson-Walker metric* |
| GW | *gravitational wave* |
| IR | *infrared* |
| KMS | *Kubo-Martin-Schwinger relation* |
| LEP | *Large Electron-Positron Collider* |
| LHC | *Large Hadron Collider* |
| LIGO | *Laser Interferometer Gravitational-Wave Observatory* |
| LISA | *Laser Interferometer Space Antenna* |
| MACHO | *massive astrophysical compact halo object* |
| MHD | *magnetohydrodynamic turbulence* |
| NANOGrav | *North American Nanohertz Observatory for Gravitational Waves* |
| PT | *phase transition* |
| PTA | *pulsar timing array* |
| QFT | *quantum field theory* |
| RGE | *renormalization group equation* |
| SKA | *Square Kilometre Array* |
| SM | *Standard Model* |
| SNR | *signal-to-noise ratio* |
| SUSY | *supersymmetry* |
| TT | *transverse-traceless gauge* |
| UV | *ultraviolet* |
| VEV | *vacuum expectation value* |
| WIMP | *weakly interacting massive particle* |

# Bibliography

[1] A. Einstein, *Approximative Integration of the Field Equations of Gravitation*, *Sitzungsber. Preuss. Akad. Wiss. Berlin (Math. Phys.)* **1916** (1916) 688.

[2] A. Einstein, *Über Gravitationswellen*, *Sitzungsber. Preuss. Akad. Wiss. Berlin (Math. Phys.)* **1918** (1918) 154.

[3] Virgo, LIGO Scientific collaboration, *Observation of Gravitational Waves from a Binary Black Hole Merger*, *Phys. Rev. Lett.* **116** (2016) 061102 [1602.03837].

[4] E.W. Kolb, and M.S. Turner, *The Early Universe*, *Front. Phys.* **69** (1990) 1.

[5] D. Baumann, *Cosmology - Part III Mathematical Tripos*, Lecture notes.

[6] E. Hubble, *A relation between distance and radial velocity among extra-galactic nebulae*, *Proc. Nat. Acad. Sci.* **15** (1929) 168.

[7] A. Einstein, *Cosmological Considerations in the General Theory of Relativity*, *Sitzungsber. Preuss. Akad. Wiss. Berlin (Math. Phys.)* **1917** (1917) 142.

[8] Supernova Search Team collaboration, *The farthest known supernova: support for an accelerating universe and a glimpse of the epoch of deceleration*, *Astrophys. J.* **560** (2001) 49 [astro-ph/0104455].

[9] Planck collaboration, *Planck 2015 results. XIII. Cosmological parameters*, *Astron. Astrophys.* **594** (2016) A13 [1502.01589].

[10] Planck collaboration, *Planck 2013 results. I. Overview of products and scientific results*, *Astron. Astrophys.* **571** (2014) A1 [1303.5062].

[11] L. Husdal, *On Effective Degrees of Freedom in the Early Universe*, *Galaxies* **4** (2016) 78 [1609.04979].

[12] N. Palanque-Delabrouille et al., *Neutrino masses and cosmology with Lyman-alpha forest power spectrum*, *JCAP* **1511** (2015) 011 [1506.05976].

[13] A.D. Sakharov, *Violation of CP Invariance, C asymmetry, and baryon asymmetry of the universe*, *Pisma Zh. Eksp. Teor. Fiz.* **5** (1967) 32 [Usp. Fiz. Nauk161,no.5,61(1991)].

[14] P. Huet, *Electroweak baryogenesis and the standard model*, in *Phenomenology of Unification from Present to Future: Proceedings of the 1st International*

*Conference, March 23-26, 1994, Rome*, pp. 77–91, 1994, http://www-public.slac.stanford.edu/sciDoc/docMeta.aspx?slacPubNumber=SLAC-PUB-6492 [hep-ph/9406301].

[15] D.E. Morrissey, and M.J. Ramsey-Musolf, *Electroweak baryogenesis*, *New J. Phys.* **14** (2012) 125003 [1206.2942].

[16] D. Clowe, M. Bradac, A.H. Gonzalez, M. Markevitch, S.W. Randall, C. Jones et al., *A direct empirical proof of the existence of dark matter*, *Astrophys. J.* **648** (2006) L109 [astro-ph/0608407].

[17] T.D. Brandt, *Constraints on MACHO Dark Matter from Compact Stellar Systems in Ultra-Faint Dwarf Galaxies*, *Astrophys. J.* **824** (2016) L31 [1605.03665].

[18] N. Menci, A. Merle, M. Totzauer, A. Schneider, A. Grazian, M. Castellano et al., *Fundamental physics with the Hubble Frontier Fields: constraining Dark Matter models with the abundance of extremely faint and distant galaxies*, *Astrophys. J.* **836** (2017) 61 [1701.01339].

[19] C. Di Paolo, F. Nesti, and F.L. Villante, *Phase space mass bound for fermionic dark matter from dwarf spheroidal galaxies*, [1704.06644].

[20] S.D.M. White, C.S. Frenk, and M. Davis, *Clustering in a Neutrino Dominated Universe*, *Astrophys. J.* **274** (1983) L1 [,80(1984)].

[21] K. Griest, and M. Kamionkowski, *Unitarity Limits on the Mass and Radius of Dark Matter Particles*, *Phys. Rev. Lett.* **64** (1990) 615.

[22] C.M. Ho, and R.J. Scherrer, *Limits on MeV Dark Matter from the Effective Number of Neutrinos*, *Phys. Rev.* **D87** (2013) 023505 [1208.4347].

[23] G. Steigman, B. Dasgupta, and J.F. Beacom, *Precise Relic WIMP Abundance and its Impact on Searches for Dark Matter Annihilation*, *Phys. Rev.* **D86** (2012) 023506 [1204.3622].

[24] T. Marrodán Undagoitia, and L. Rauch, *Dark matter direct-detection experiments*, *J. Phys.* **G43** (2016) 013001 [1509.08767].

[25] T.R. Slatyer, *TASI Lectures on Indirect Detection of Dark Matter*, in *Theoretical Advanced Study Institute in Elementary Particle Physics: Anticipating the Next Discoveries in Particle Physics (TASI 2016) Boulder, CO, USA, June 6-July 1, 2016*, 2017, https://inspirehep.net/record/1630762/files/arXiv:1710.05137.pdf [1710.05137].

[26] T.M. Hong, *Dark matter searches at the LHC*, in *5th Large Hadron Collider Physics Conference (LHCP 2017) Shanghai, China, May 15-20, 2017*, 2017, https://inspirehep.net/record/1622267/files/arXiv:1709.02304.pdf [1709.02304].

[27] E.E. Flanagan, and S.A. Hughes, *The Basics of gravitational wave theory*, *New J. Phys.* **7** (2005) 204 [gr-qc/0501041].

[28] S. Weinberg, *The quantum theory of fields. Vol. 2: Modern applications*, Cambridge University Press (2013).

[29] S.R. Coleman, and E.J. Weinberg, *Radiative Corrections as the Origin of Spontaneous Symmetry Breaking*, *Phys. Rev.* **D7** (1973) 1888.

[30] M. Quiros, *Finite temperature field theory and phase transitions*, in *Proceedings, Summer School in High-energy physics and cosmology: Trieste, Italy, June 29-July 17, 1998*, pp. 187–259, 1999, http://alice.cern.ch/format/showfull?sysnb=0302087 [hep-ph/9901312].

[31] G. 't Hooft, and M.J.G. Veltman, *Regularization and Renormalization of Gauge Fields*, *Nucl. Phys.* **B44** (1972) 189.

[32] C. Delaunay, C. Grojean, and J.D. Wells, *Dynamics of Non-renormalizable Electroweak Symmetry Breaking*, *JHEP* **04** (2008) 029 [0711.2511].

[33] Y. Yang, *An introduction to thermal field theory*, Master's thesis, Imperial College London, 2011.

[34] D. Curtin, P. Meade, and H. Ramani, *Thermal Resummation and Phase Transitions*, [1612.00466].

[35] D. Comelli, and J.R. Espinosa, *Bosonic thermal masses in supersymmetry*, *Phys. Rev.* **D55** (1997) 6253 [hep-ph/9606438].

[36] M.E. Carrington, *The Effective potential at finite temperature in the Standard Model*, *Phys. Rev.* **D45** (1992) 2933.

[37] M. Carena, A. Megevand, M. Quiros, and C.E.M. Wagner, *Electroweak baryogenesis and new TeV fermions*, *Nucl. Phys.* **B716** (2005) 319 [hep-ph/0410352].

[38] R. Jackiw, *Functional evaluation of the effective potential*, *Phys. Rev.* **D9** (1974) 1686.

[39] A. Andreassen, W. Frost, and M.D. Schwartz, *Consistent Use of Effective Potentials*, *Phys. Rev.* **D91** (2015) 016009 [1408.0287].

[40] S.R. Coleman, *The Fate of the False Vacuum. 1. Semiclassical Theory*, *Phys. Rev.* **D15** (1977) 2929.

[41] H. Widyan, and M. Al-Wardat, *Classical Solution for the Bounce Up to Second Order*, *Chin. J. Phys.* **48** (2010) 736 [1206.2734].

[42] A.D. Linde, *Decay of the False Vacuum at Finite Temperature*, *Nucl. Phys.* **B216** (1983) 421.

[43] C.L. Wainwright, *CosmoTransitions: Computing Cosmological Phase Transition Temperatures and Bubble Profiles with Multiple Fields*, *Comput. Phys. Commun.* **183** (2012) 2006 [1109.4189].

[44] F. Staub, *SARAH 4 : A tool for (not only SUSY) model builders*, *Comput. Phys. Commun.* **185** (2014) 1773 [1309.7223].

[45] J.M. Moreno, M. Quiros, and M. Seco, *Bubbles in the supersymmetric standard model*, *Nucl. Phys.* **B526** (1998) 489 [hep-ph/9801272].

[46] C. Caprini et al., *Science with the space-based interferometer eLISA. II: Gravitational waves from cosmological phase transitions*, *JCAP* **1604** (2016) 001 [1512.06239].

[47] J.R. Espinosa, T. Konstandin, J.M. No, and G. Servant, *Energy Budget of Cosmological First-order Phase Transitions*, *JCAP* **1006** (2010) 028 [1004.4187].

[48] C. Grojean, and G. Servant, *Gravitational Waves from Phase Transitions at the Electroweak Scale and Beyond*, *Phys. Rev.* **D75** (2007) 043507 [hep-ph/0607107].

[49] S. Weinberg, *Gravitation and Cosmology*, John Wiley and Sons, New York (1972).

[50] P.J. Steinhardt, *Relativistic Detonation Waves and Bubble Growth in False Vacuum Decay*, *Phys. Rev.* **D25** (1982) 2074.

[51] M. Laine, *Bubble growth as a detonation*, *Phys. Rev.* **D49** (1994) 3847 [hep-ph/9309242].

[52] H. Kurki-Suonio, and M. Laine, *Supersonic deflagrations in cosmological phase transitions*, *Phys. Rev.* **D51** (1995) 5431 [hep-ph/9501216].

[53] D. Bodeker, and G.D. Moore, *Electroweak Bubble Wall Speed Limit*, *JCAP* **1705** (2017) 025 [1703.08215].

[54] D. Bodeker, and G.D. Moore, *Can electroweak bubble walls run away?*, *JCAP* **0905** (2009) 009 [0903.4099].

[55] P. Schwaller, *Gravitational Waves from a Dark Phase Transition*, *Phys. Rev. Lett.* **115** (2015) 181101 [1504.07263].

[56] S.J. Huber, and T. Konstandin, *Gravitational Wave Production by Collisions: More Bubbles*, *JCAP* **0809** (2008) 022 [0806.1828].

[57] M. Hindmarsh, S.J. Huber, K. Rummukainen, and D.J. Weir, *Numerical simulations of acoustically generated gravitational waves at a first order phase transition*, *Phys. Rev.* **D92** (2015) 123009 [1504.03291].

[58] C. Caprini, R. Durrer, and G. Servant, *The stochastic gravitational wave background from turbulence and magnetic fields generated by a first-order phase transition*, *JCAP* **0912** (2009) 024 [0909.0622].

[59] D.J. Weir, *Gravitational waves from a first order electroweak phase transition: a brief review*, *Phil. Trans. Roy. Soc. Lond.* **A376** (2018) 20170126 [1705.01783].

[60] Z. Kang, P. Ko, and T. Matsui, *Strong first order EWPT & strong gravitational waves in $Z_3$-symmetric singlet scalar extension*, *JHEP* **02** (2018) 115 [1706.09721].

[61] W. Chao, W.-F. Cui, H.-K. Guo, and J. Shu, *Gravitational Wave Imprint of New Symmetry Breaking*, [1707.09759].

[62] Y. Chen, M. Huang, and Q.-S. Yan, *Gravitation waves from QCD and electroweak phase transitions*, [1712.03470].

[63] LIGO Scientific collaboration, *Advanced LIGO: The next generation of gravitational wave detectors*, *Class. Quant. Grav.* **27** (2010) 084006.

[64] VIRGO collaboration, *Advanced Virgo: a second-generation interferometric gravitational wave detector*, *Class. Quant. Grav.* **32** (2015) 024001 [1408.3978].

[65] S. Babak, J. Gair, A. Sesana, E. Barausse, C.F. Sopuerta, C.P.L. Berry et al., *Science with the space-based interferometer LISA. V: Extreme mass-ratio inspirals*, *Phys. Rev.* **D95** (2017) 103012 [1703.09722].

[66] G.M. Harry, P. Fritschel, D.A. Shaddock, W. Folkner, and E.S. Phinney, *Laser interferometry for the big bang observer*, *Class. Quant. Grav.* **23** (2006) 4887 [Erratum: Class. Quant. Grav.23,7361(2006)].

[67] S. Kawamura et al., *The Japanese space gravitational wave antenna DECIGO*, *Class. Quant. Grav.* **23** (2006) S125.

[68] S. Sato et al., *The status of DECIGO*, *J. Phys. Conf. Ser.* **840** (2017) 012010.

[69] F. Jenet et al., *The North American Nanohertz Observatory for Gravitational Waves*, [0909.1058].

[70] R. van Haasteren et al., *Placing limits on the stochastic gravitational-wave background using European Pulsar Timing Array data*, *Mon. Not. Roy. Astron. Soc.* **414** (2011) 3117 [1103.0576], [Erratum: Mon. Not. Roy. Astron. Soc.425,no.2,1597(2012)].

[71] C.L. Carilli, and S. Rawlings, *Science with the Square Kilometer Array: Motivation, key science projects, standards and assumptions*, *New Astron. Rev.* **48** (2004) 979 [astro-ph/0409274].

[72] L. Lentati et al., *European Pulsar Timing Array Limits On An Isotropic Stochastic Gravitational-Wave Background*, *Mon. Not. Roy. Astron. Soc.* **453** (2015) 2576 [1504.03692].

[73] NANOGRAV collaboration, *The NANOGrav 11-year Data Set: Pulsar-timing Constraints On The Stochastic Gravitational-wave Background*, [1801.02617].

[74] E. Thrane, and J.D. Romano, *Sensitivity curves for searches for gravitational-wave backgrounds*, *Phys. Rev.* **D88** (2013) 124032 [1310.5300].

[75] C.J. Moore, S.R. Taylor, and J.R. Gair, *Estimating the sensitivity of pulsar timing arrays*, *Class. Quant. Grav.* **32** (2015) 055004 [1406.5199].

[76] G. Janssen et al., *Gravitational wave astronomy with the SKA*, *PoS* **AASKA14** (2015) 037 [1501.00127].

[77] N. Cornish, and T. Robson, *The construction and use of LISA sensitivity curves*, [1803.01944].

[78] S. Isoyama, H. Nakano, and T. Nakamura, *Multiband Gravitational-Wave Astronomy: Observing binary inspirals with a decihertz detector, B-DECIGO*, [1802.06977].

[79] K. Yagi, and N. Seto, *Detector configuration of DECIGO/BBO and identification of cosmological neutron-star binaries*, *Phys. Rev.* **D83** (2011) 044011 [1101.3940], [Erratum: Phys. Rev.D95,no.10,109901(2017)].

[80] H. Kudoh, A. Taruya, T. Hiramatsu, and Y. Himemoto, *Detecting a gravitational-wave background with next-generation space interferometers*, *Phys. Rev.* **D73** (2006) 064006 [gr-qc/0511145].

[81] S. Kuroyanagi, S. Tsujikawa, T. Chiba, and N. Sugiyama, *Implications of the B-mode Polarization Measurement for Direct Detection of Inflationary Gravitational Waves*, *Phys. Rev.* **D90** (2014) 063513 [1406.1369].

[82] C.J. Moore, R.H. Cole, and C.P.L. Berry, *Gravitational-wave sensitivity curves*, *Class. Quant. Grav.* **32** (2015) 015014 [1408.0740].

[83] S.L. Glashow, *Partial Symmetries of Weak Interactions*, *Nucl. Phys.* **22** (1961) 579.

[84] S. Weinberg, *A Model of Leptons*, *Phys. Rev. Lett.* **19** (1967) 1264.

[85] A. Salam, *Weak and Electromagnetic Interactions*, *Conf. Proc.* **C680519** (1968) 367.

[86] M. D'Onofrio, and K. Rummukainen, *Standard model cross-over on the lattice*, *Phys. Rev.* **D93** (2016) 025003 [1508.07161].

[87] K. Kajantie, M. Laine, K. Rummukainen, and M.E. Shaposhnikov, *Is there a hot electroweak phase transition at m(H) larger or equal to m(W)?*, *Phys. Rev. Lett.* **77** (1996) 2887 [hep-ph/9605288].

[88] K. Rummukainen, M. Tsypin, K. Kajantie, M. Laine, and M.E. Shaposhnikov, *The Universality class of the electroweak theory*, *Nucl. Phys.* **B532** (1998) 283 [hep-lat/9805013].

[89] F. Csikor, Z. Fodor, and J. Heitger, *Endpoint of the hot electroweak phase transition*, *Phys. Rev. Lett.* **82** (1999) 21 [hep-ph/9809291].

[90] J.M. Cline, *Baryogenesis*, in *Les Houches Summer School - Session 86: Particle Physics and Cosmology: The Fabric of Spacetime Les Houches, France, July 31-August 25, 2006*, 2006 [hep-ph/0609145].

[91] M.J. Baker, and J. Kopp, *Dark Matter Decay between Phase Transitions at the Weak Scale*, *Phys. Rev. Lett.* **119** (2017) 061801 [1608.07578].

[92] M.J. Baker, M. Breitbach, J. Kopp, and L. Mittnacht, *Dynamic Freeze-In: Impact of Thermal Masses and Cosmological Phase Transitions on Dark Matter Production*, *JHEP* **03** (2018) 114 [1712.03962].

[93] G. 't Hooft, *Naturalness, chiral symmetry, and spontaneous chiral symmetry breaking*, *NATO Sci. Ser. B* **59** (1980) 135.

[94] K. Kadota, T. Sekiguchi, and H. Tashiro, *A new constraint on millicharged dark matter from galaxy clusters*, [1602.04009].

[95] P.V. Dong, and H.N. Long, *The Economical SU(3)(C) X SU(3)(L) X U(1)(X) model*, *Adv. High Energy Phys.* **2008** (2008) 739492 [0804.3239].

[96] F.P. Huang, and X. Zhang, *Probing the hidden gauge symmetry breaking through the phase transition gravitational waves*, [1701.04338].

[97] B. Holdom, *Two U(1)'s and Epsilon Charge Shifts*, *Phys. Lett.* **166B** (1986) 196.

[98] A. de Gouvêa, and D. Hernández, *New Chiral Fermions, a New Gauge Interaction, Dirac Neutrinos, and Dark Matter*, *JHEP* **10** (2015) 046 [1507.00916].

[99] R. Essig et al., *Working Group Report: New Light Weakly Coupled Particles*, in *Proceedings, 2013 Community Summer Study on the Future of U.S. Particle Physics: Snowmass on the Mississippi (CSS2013): Minneapolis, MN, USA, July 29-August 6, 2013*, 2013, http://inspirehep.net/record/1263039/files/arXiv:1311.0029.pdf [1311.0029].

[100] S.L. Glashow, *Positronium Versus the Mirror Universe*, *Phys. Lett.* **167B** (1986) 35.

[101] E.C.G. Stueckelberg, *Interaction energy in electrodynamics and in the field theory of nuclear forces*, *Helv. Phys. Acta* **11** (1938) 225.

[102] J. Liu, X.-P. Wang, and F. Yu, *A Tale of Two Portals: Testing Light, Hidden New Physics at Future $e^+e^-$ Colliders*, *JHEP* **06** (2017) 077 [1704.00730].

[103] A. Addazi, and A. Marciano, *Gravitational waves from dark first order phase transitions and dark photons*, *Chin. Phys.* **C42** (2018) 023107 [1703.03248].

[104] K. Hashino, M. Kakizaki, S. Kanemura, P. Ko, and T. Matsui, *Gravitational waves from first order electroweak phase transition in models with the $U(1)_X$ gauge symmetry*, [1802.02947].

[105] A. Berlin, and N. Blinov, *Thermal Dark Matter Below an MeV*, *Phys. Rev. Lett.* **120** (2018) 021801 [1706.07046].

[106] R.H. Cyburt, B.D. Fields, K.A. Olive, and T.-H. Yeh, *Big Bang Nucleosynthesis: 2015*, *Rev. Mod. Phys.* **88** (2016) 015004 [1505.01076].

[107] G. Mangano, G. Miele, S. Pastor, T. Pinto, O. Pisanti, and P.D. Serpico, *Relic neutrino decoupling including flavor oscillations*, *Nucl. Phys.* **B729** (2005) 221 [hep-ph/0506164].

[108] A. Nieto, *Evaluating sums over the Matsubara frequencies*, *Comput. Phys. Commun.* **92** (1995) 54 [hep-ph/9311210].

[109] C. Manuel, *On the thermal gauge boson masses of the electroweak theory in the broken phase*, *Phys. Rev.* **D58** (1998) 016001 [hep-ph/9801364].